\documentclass[aps,prd,10pt,twocolumn,superscriptaddress,floatfix,nofootinbib,amsmath,amssymb,altaffilletter,preprintnumbers,longbibliography,eprint]{revtex4-2}

\usepackage[normalem]{ulem}
\usepackage{natbib}
\usepackage[english]{babel}
\usepackage[utf8]{inputenc}
\DeclareUnicodeCharacter{2009}{\,}

\usepackage{enumitem}
\usepackage{braket}
\usepackage{footnote}
\usepackage{tablefootnote}
\usepackage{amsmath,amssymb,amsfonts,mathrsfs}
\usepackage{booktabs}
\usepackage{mathtools}
\usepackage{bm}
\usepackage{dcolumn}
\usepackage{graphicx}   
\usepackage{latexsym}
\usepackage[acronym]{glossaries}
\usepackage{physics}
\usepackage{siunitx}
\usepackage{comment}
\DeclareSIUnit{\persqrthz}{\ensuremath{\text{Hz}^{-1/2}}}
\usepackage[svgnames]{xcolor}
\usepackage[final]{hyperref}
\hypersetup{%
    colorlinks=true,
    citecolor=DarkOliveGreen,
    }%

\newcommand{\new}[1]{{\color{Black}{#1}}}
\newcommand{\newJZ}[1]{{\color{Black}{#1}}}
\newcommand{\newHI}[1]{{\color{Black}{#1}}}

\newcommand{\myhyperref}[1]{\hyperref[#1]{\ref{#1}}}
\newcommand{\orcid}[1]{\href{https://orcid.org/#1}{\includegraphics[width=8pt]{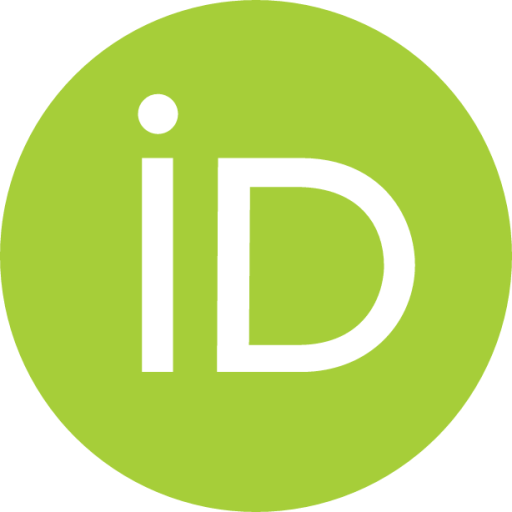}}}

\begin{document}

\title{Measuring gravitational wave memory with LISA\vspace{-.2cm}}

\author{Henri Inchauspé \orcid{0000-0002-4664-6451}}
\email{henri.inchauspe@kuleuven.be}
\affiliation{Institute for Theoretical Physics, KU Leuven,
Celestijnenlaan 200D, B-3001 Leuven, Belgium}
\affiliation{Leuven Gravity Institute, KU Leuven,
Celestijnenlaan 200D box 2415, 3001 Leuven, Belgium}
\affiliation{Institut f\"ur Theoretische Physik, Universit\"at Heidelberg, Philosophenweg 16, 69120 Heidelberg, Germany}

% \author{Henri Inchauspé}
% \email{inchauspe@tphys.uni-heidelberg.de}
% \affiliation{Institut f\"ur Theoretische Physik, Universit\"at Heidelberg, Philosophenweg 16, 69120 Heidelberg, Germany}

\author{Silvia Gasparotto \orcid{0000-0001-7586-1786}}
\email{sgasparotto@ifae.es}
\affiliation{Grup de F\'{i}sica Te\`{o}rica, Departament de F\'{i}sica, Universitat Aut\`{o}noma de Barcelona, 08193 Bellaterra (Barcelona), Spain}
\affiliation{Institut de F\'isica d'Altes Energies (IFAE), The Barcelona Institute of
Science and Technology (BIST), Campus UAB, 08193 Bellaterra, Barcelona}

\author{Diego Blas \orcid{0000-0003-2646-0112}}
\affiliation{Institut de F\'isica d'Altes Energies (IFAE), The Barcelona Institute of
Science and Technology (BIST), Campus UAB, 08193 Bellaterra, Barcelona}
\affiliation{Instituci\'o Catalana de Recerca i Estudis Avan\c cats (ICREA), Passeig Llu\'is Companys 23, 08010 Barcelona, Spain}

\author{Lavinia Heisenberg}
\affiliation{Institut f\"ur Theoretische Physik, Universit\"at Heidelberg, Philosophenweg 16, 69120 Heidelberg, Germany}

\author{Jann Zosso \orcid{0000-0002-2671-7531}}
\affiliation{Institute for Theoretical Physics, ETH Zurich, Wolfgang-Pauli-Strasse 27, CH-8093 Zurich, Switzerland}
\affiliation{Albert Einstein Center, Institute for Theoretical Physics, University of Bern,
Sidlerstrasse 5, CH-3012 Bern, Switzerland}
\author{Shubhanshu Tiwari}
\affiliation{Physik-Institut, Universität Zürich, Winterthurerstrasse 190, 8057 Zürich, Switzerland}
%\vspace{5pt}

\begin{abstract}
Gravitational wave (GW) astronomy has revolutionized our capacity to explore nature. The next generation of observatories, among which the space-borne detector Laser Interferometer Space Antenna   LISA, is expected to yield orders of magnitude of signal-to-noise ratio improvement, and reach fainter and novel features of General Relativity. Among them, an exciting possibility is the detection of GW memory. Interpreted as a permanent deformation of the background spacetime after a GW perturbation has passed through the detector, GW memory offers a novel avenue to proof-test General Relativity, access the non-linear nature of gravity, and provide complementary information to better characterize the GW source. Previous studies have shown that GW memory detection from individual mergers of massive black hole binaries is expected with LISA. However, these works have not simulated the proper time domain response of the detector to the GW memory. This work is filling this gap and presents the detection prospects of LISA regarding GW memory and the expected signature of GW memory on the data-streams using the most up-to-date LISA consortium simulations of the response. We focus on the GW memory of massive black hole binary mergers and use state-of-the-art population models to assess the likelihood of detecting the GW memory within the LISA lifetime. We conclude that GW memory will be a key feature of several events detected by LISA, and will help to exploit the scientific potential of the mission fully.
\end{abstract}

%%%%%%%%%%%%%%%%%%%%%%%%%%%%%%%%%%%%%%%%%%%%%%%%%%%%%%%%%%%%%%%%%%%%%%%%%%%%%%%%%%%%%%%%%%%%%%
    
%Term definitions
\newacronym{e2e}{E2E}{End-To-End}
\newacronym{inrep}{INREP}{Initial Noise REduction Pipeline}
\newacronym{tdi}{TDI}{Time Delay Interferometry}
\newacronym{ttl}{TTL}{Tilt-To-Length couplings}
\newacronym{dfacs}{DFACS}{Drag-Free and Attitude Control System}
\newacronym{ldc}{LDC}{LISA Data Challenge}
\newacronym{lisa}{LISA}{the Laser Interferometer Space Antenna}
\newacronym{emri}{EMRI}{Extreme Mass Ratio Inspiral}
\newacronym{ifo}{IFO}{Interferometry System}
\newacronym{grs}{GRS}{Gravitational Reference Sensor}
\newacronym{tmdws}{TM-DWS}{Test-Mass Differential Wavefront Sensing}
\newacronym{ldws}{LDWS}{Long-arm Differential Wavefront Sensing}
\newacronym[	plural={MOSAs},
		        first={Moving Optical Sub-Assembly},
		        firstplural={Moving Optical Sub-Assemblies}
            ]{mosa}{MOSA}{Moving Optical Sub-Assembly}
\newacronym{siso}{SISO}{Single-Input Single-Output}
\newacronym{mimo}{MIMO}{Multiple-Input Multiple-Output}
\newacronym[plural=MBHB's, firstplural=Massive Black Holes Binaries (MBHB's)]{mbhb}{MBHB}{Massive Black Holes Binary}
\newacronym{cmb}{CMB}{Cosmic Microwave Background}
\newacronym{sgwb}{SGWB}{Stochastic Gravitational Waves Background}
\newacronym{pta}{PTA}{Pulsar Timing Arrays}
\newacronym{gw}{GW}{Gravitational Wave}
\newacronym{snr}{SNR}{Signal-to-Noise Ratio}
\newacronym{pbh}{PBH}{Primordial Black Holes}
\newacronym{psd}{PSD}{Power Spectral Density}
\newacronym{tcb}{TCB}{Barycentric Coordinate Time}
\newacronym{bcrs}{BCRS}{Barycentric Celestial Reference System}
\newacronym{lhs}{LHS}{Left-Hand Side}
\newacronym{rhs}{RHS}{Right-Hand Side}
\newacronym{mcmc}{MCMC}{Monte-Carlo Markov Chains}
\newacronym{cs}{CS}{Cosmic Strings}
\newacronym{ssb}{SSB}{Solar System Barycentric}
\newacronym{oms}{OMS}{Optical Metrology System}
\newacronym{dof}{DoF}{Degree of Freedom}
\newacronym{eob}{EOB}{Effective One-Body}
\newacronym{pn}{PN}{Post-Newtonian}
\newacronym{cce}{CCE}{Cauchy-Characteristic Evolution}
\newacronym{imr}{IMR}{Inspiral-Merger-Ringdown}
\newacronym{scird}{SciRD}{LISA Science Requirement Document}

%%%%%%%%%%%%%%%%%%%%%%%%%%%%%%%%%%%%%%%%%%%%%%%%%%%%%%%%%%%%%%%%%%%%%%%%%%%%%%%%%%%%%%%%%%%%%%%%%

%
\maketitle

%%%%%%%%%%%%%%%%%%%%%%%%%%%%%%%%%%%%%%%%%
\section{\label{Intro}Introduction}
%%%%%%%%%%%%%%%%%%%%%%%%%%%%%%%%%%%%%%%%%

When \new{an unbound flux of} matter or radiation \new{is released from a localized event}, the final metric of \new{the asymptotic region does not return} to its original form, but the relative \new{proper} distances \new{between} freely falling observers are permanently modified with respect to their original ones.  This \new{phenomenon} is called \emph{gravitational wave memory} effect, see e.g.~\cite{Zeldovich:1974gvh,Braginsky:1985vlg,Braginsky:1987,Christodoulou:1991cr}, or, in recent language, \emph{displacement} memory, and is the dominant of several effects related to the changes of asymptotic states \new{after} the passage of radiation~\cite{Flanagan:2018yzh,Grant:2021hga,Grant:2022bla,Siddhant:2024nft,Pasterski:2015tva,Nichols:2018qac}. Remarkably, these effects are intimately connected to the asymptotic structure of space-time in General Relativity, namely with the symmetries of the BMS  group~\cite{Strominger:2014pwa,Ashtekar:2019viz,Mitman:2024uss}, \new{as well as the} soft theorems~\cite{Strominger:2014pwa,Strominger:2017zoo} \new{in scattering theory}.  

In this work, we will focus on the GW memory generated by the passage of \emph{gravitational flux}, \new{known as \emph{non-linear} memory to differentiate it from the \emph{linear} memory, which is related to the unbound energy flux of other fields such as matter radiation or matter components.}\footnote{\new{Motivated by the BMS \emph{balance laws}~\cite{Ashtekar:2019viz}, \gls{gw} memory is sometimes also categorized into so called \emph{null} memory that represents memory that is sourced by any unbound null energy flux of radiation, and \emph{ordinary} memory, associated to unbound energy-momentum of massive components that do not reach null infinity \cite{Bieri:2013ada}. The non-linear memory of GR is therefore also part of the null memory.}}
Moreover, we will restrict to the \emph{displacement} memory and not consider other subdominant effects, such as the \emph{spin} \cite{Pasterski:2015tva} or the \emph{center-of-mass} \cite{Nichols:2018qac,bieri_gravitational_2024} memories.

\new{As the name suggests,} the origin of \new{such non-linear displacement} memory stems from the non-linear nature of General Relativity:  gravitational \new{waves themselves carry energy-momentum that} generates \new{additional gravitational radiation, which precisely induces a displacement} memory ~\cite{Christodoulou:1991cr,Wiseman:1991ss,Blanchet:1992br,Thorne:1992sdb,Favata:2008yd,Blanchet:2008je,Favata:2010zu,Garfinkle:2022dnm,Zosso:2024xgy}. \new{Interestingly, \gls{gw} memory is however no mere second order effect, but can be understood as a fundamental component of any gravitational radiation that reaches null infinity at leading order~\cite{Christodoulou:1991cr}.}
The \new{relation of non-linear memory to the oscillatory \emph{primary} \glspl{gw}} explains some of the features of the effect since its magnitude and its typical time scale, \new{as well as} its angular dependence are related to the flux emitted by the primary signal. \new{However, memory nevertheless represents a unique non-linear effect that accumulates over time, whose non-oscillatory nature marks a clear distinction with} the oscillatory \new{primary} waveform \cite{favata_nonlinear_2009,Favata:2010zu}.
As a result, the properties of the \gls{gw} memory strongly differ from those of the \new{primary} waves and, despite \new{only having a small imprint on typical interferometric \gls{gw} detectors}, memory may carry key complementary information from the primary signal.
For instance, it may allow one to break the degeneracy between luminosity distance and inclination~\cite{gasparotto_can_2023,Xu:2024ybt} or between the merger of binary black holes or neutron stars~\cite{Tiwari:2021gfl,Lopez:2023aja}. The use of \gls{gw} memory has also been proposed to test the asymptotic symmetries of space-time~\cite{Goncharov:2023woe}, or as a part of consistency tests of waveform models exploiting balance laws \cite{heisenberg_balance_2023, dambrosio_testing_2024}. Furthermore, since \gls{gw} memory probes the non-linear nature of gravitation, it is natural to expect changes in modified theories of gravity~\cite{Du:2016hww,Tahura:2021hbk,Heisenberg:2023prj,Heisenberg:2024cjk}.
Finally, the different frequency span of the \gls{gw} memory as compared to the primary signal allows looking for merger events
%\new{\sout{of primordial}
at frequencies higher than the detector band~\cite{McNeill:2017uvq} and even search for the same events in ground-base and space-borne detectors~\cite{Ghosh:2023rbe}. In conclusion, the detection of \gls{gw} memory will open several possibilities to probe astrophysics and fundamental physics with \glspl{gw}. 

\gls{gw} observations have so far not detected \gls{gw} memory. This is mainly due to the smallness of the \new{signal within current \gls{gw} observatories} and is true both for binary black hole coalescences observed in LIGO/Virgo data~\cite{Hubner:2019sly,Ebersold:2020zah,Zhao:2021hmx,Hubner:2021amk} and supermassive black hole coalescences that may explain Pulsar Timing Arrays data~\cite{Seto:2009nv,Cordes:2012zz,NANOGrav:2023vfo}. In fact, it is not expected for current ground-based interferometers to detect the memory from the coalescence of \emph{individual} binary compact objects. Instead, it has been suggested that the combination of events with \gls{gw} memory \gls{snr} below threshold over 2 to 5 years may be enough to reach detection in these set-ups~\cite{Grant:2022bla}. The prospects of detecting the memory from individual events are more promising for the next generation of detectors, either ground-based, such as the ET~\cite{Punturo:2010zz} and CE~\cite{Reitze:2019iox}, and space-based interferometers like \gls{lisa} and TianQin~\cite{Grant:2022bla,gasparotto_can_2023,Favata:2009ii,Johnson:2018xly,Islo:2019qht,Islam:2021old,Sun:2022pvh,Sun:2024nut} or SKA  \cite{vanHaasteren:2009fy,Janssen:2014dka}.

In this paper, we focus on the characteristic imprint of the memory on the LISA detector from the coalescence of Massive Black Holes (MBHs) that merge in the frequency band between $10^{-4}\,$Hz and $10^{-1}\,\si{\hertz}$. These events will have an extraordinarily high \gls{snr}, especially during the merger, when most of the memory is created. Consequently,  these objects are the most promising source of detectable \gls{gw} memory. Recent works~\cite{gasparotto_can_2023,Sun:2022pvh,Sun:2024nut} have confirmed that several MBH mergers will have an \gls{snr} of the \gls{gw} memory sufficiently high to claim detection, though the expected number varies considerably depending on the astrophysical population considered. Building on these results, one of our main goals is to improve them by investigating the \gls{gw} memory with the \emph{full time-domain response of LISA}, based on \gls{tdi}  and its most updated noise characterization.  
Addressing the robustness of the detection of \gls{gw} memory when TDI is considered is essential to build a solid understanding of the imprint of this effect, especially if one aims at using it as an additional source of information complementary to the leading oscillatory signal. We devote the rest of the paper to the first investigation in this direction for the LISA mission.  

% Compared to previous work, we also consider new waveforms that directly contain the \gls{gw} memory~\cite{yoo_numerical_2023} which, for non-precessing binaries, is mainly contained in the (2,0) mode~\cite{Favata:2009ii}. We also investigate the overlap of the memory and additional oscillatory features excited during the merger and the ringdown, both present in the (2,0), and we discuss potential strategies to distinguish the two.
\newHI{For the calculation of \gls{gw} memory, we consider the \texttt{NRHybSur3dq8} waveform~\cite{varma_surrogate_2019}, which is one of the most accurate waveforms available to date, trained on numerical relativity simulations and including higher modes up to $\ell\leq 4$. This model has recently been updated to capture the \gls{gw} memory \texttt{NRHybSur3dq8\_}\texttt{CCE}~\cite{yoo_numerical_2023}, which for non-preprocessing binaries is mainly in the (2,0) mode~\cite{Favata:2009ii}. This helps to perform consistency checks and to investigate the mixing of the memory and additional oscillatory features excited during the merger and ringdown, both of which are present in the (2,0) mode.}
We further perform a broad study of the parameter space of the binary merger by varying the mass ratio of the two black holes and the amplitude of the aligned spin of the sources, though we will not consider the presence of precession or non-trivial eccentricities. The detectability of \gls{gw} memory by LISA will be assessed with state-of-the-art population models of MBHs described in Refs.~\cite{Barausse:2020gbp, Barausse:2020mdt} (and based on previous work presented in~\cite{EB12,Sesana:2014bea,Antonini:2015sza}).   

The paper is organized as follows. In Sec.~\ref{sec:mem_lisa} we describe the necessary tools for the subsequent analysis, such as the TDI processing, the LISA noise characterization and the waveform models. In Sec.~\ref{subsection: lisa signature}, we comprehensively describe  how the displacement memory impacts the signal of LISA. In Sec.~\ref{tdi} we discuss the signal-to-noise ratio and the detectability predictions. Sec.~\ref{sec:populations} is devoted to the study of MBHs mergers and the prospect of detecting memory from them in LISA. In Sec.~\ref{sec:q_and_spin}, we explore the particular dependence of the memory \gls{snr} w.r.t. mass ratio and spin. Finally, we discuss our conclusions and outlook in Sec.~\ref{discussion}.

%%%%%%%%%%%%%%%%%%%%%%%%%%
\section{\label{sec:mem_lisa}Displacement Memory imprints on LISA interferometric data}
%%%%%%%%%%%%%%%%%%%%%%%%%%

\gls{gw} interferometers, either ground or space-based, are not designed to observe a permanent shift of the strain because they are sensitive to a limited frequency band. As a result, they cannot detect the permanent offset from the \gls{gw} memory. However, the time-dependent transition in strain induced by the \gls{gw} memory at the detector location does exhibit significant spectral content at low frequencies, which may be detected. 
%LISA's sensitivity band spans \glspl{gw} between $10^{-4}$\,Hz and $10^{-1} \si{\hertz}$  \cite{colpi_lisa_2024}.
\new{
% \sout{Its better sensitivity at low frequencies as compared to current ground-based interferometers implies that it is a better detector of \gls{gw} memory for the (MBH) mergers that have support at its frequency sensitivity peak (mHz). }
Most of the MBH mergers in the LISA band, sensitive between $10^{-4}\si{\hertz}$ and $10^{-1}\si{\hertz}$~\cite{colpi_lisa_2024}, will have a much higher SNR than events currently detected by ground-based interferometers, which allow the fine details of the waveform, such as the GW memory, to be distinguished.  }
The proper determination of LISA's sensitivity to \gls{gw} memory requires a full time-domain simulation of the projection of the signal onto the laser antenna response  (down to TDI data streams). 
In this section, we show a comprehensive {\it end-to-end} time-domain simulation of the LISA response to the \gls{gw} memory using the most up-to-date LISA consortium simulations (\texttt{LISAGWResponse}~\cite{bayle_lisa_2023, bayle_lisa_2022}, \texttt{LISAInstrument} \cite{bayle_unified_2023, lisa_instrument}) and post-processing software (\texttt{PyTDI}~\cite{staab_pytdi_2023}). \\
%%%%%%%%%%%%%%%%%%%%%%%%%%%%%%%%%%%%%%%%%%%%%%%%%%%%%%%%%%%%%%%
\new{
\subsection{GW memory model and simulation}
\label{subsection: mem model}
}

\new{
\subsubsection{Definition and computation of non-linear memory}
\label{subsection: thorne formula}
}

\begin{figure}[t]
    \centering
    % trim={<left> <lower> <right> <upper>}
    \includegraphics[width=1.0\columnwidth, trim={0.2cm 0.0cm 0.2cm 0.0cm}, clip]{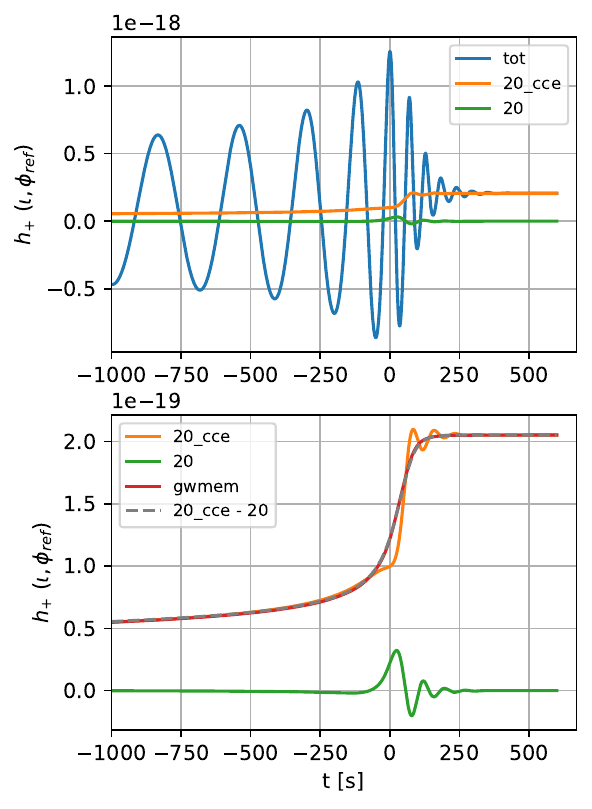}
    \caption{Breakdown of the oscillatory and memory component of the waveform from a $M_{tot} = 10^{6} M_{\odot}$ \acrshort{mbhb} merger.
    %, showing the different components in Eq.~(\myhyperref{eq: surr_20}) \new{as well as a comparison to Eq.~\eqref{eq:memorymodes}}. 
    On the top plot: the total waveform (blue), the $(l,m)=(2,0)$ component from \texttt{NRHybSur3dq8\_CCE} (orange) and \texttt{NRHybSur3dq8} (green) models. On the bottom plot: a zoom-in on the \texttt{NRHybSur3dq8\_CCE} model (orange) and the estimated contribution from \gls{gw} memory \new{(grey-dashed)} according to Eq.~(\myhyperref{eq: surr_20}), \new{together with a comparison to the $(2,0)$ mode computed with the \texttt{GWMemory} package~\cite{Talbot:2018sgr} (red)}.}
    \label{fig: surr_cce}
\end{figure}

To assess the detectability of the memory, it is important \new{in a first step to provide a pertinent theoretical definition that allows} to isolate it in the total \gls{gw} strain. \new{Focusing on non-linear memory, such a definition is given by the following integral over the energy flux emitted in primary \glspl{gw} \cite{Wiseman:1991ss,Thorne:1992sdb,Favata:2010zu,Heisenberg:2023prj}\footnote{This \new{formula} explicitly does not capture any \emph{linear} memory. \new{A more general formula involving any type of unbound energy-momentum content can for instance be found in \cite{Heisenberg:2024cjk}.} However, \new{for binary black hole coalescence the only linear memory comes from a potential final black hole kick, which} is a \new{very} subdominant contribution~\cite{Kick_Favata2009}.}
\begin{equation}\label{eq:memoryequation}
    \left[h^{mem}_{ij}\right]^\text{TT}=\frac{4G}{Rc^4}\int_{-\infty}^u \mathrm{d}u'\int \mathrm{d}\Omega' \frac{\mathrm{d}E}{\mathrm{d}u'\mathrm{d}\Omega'}\left[\frac{n'_in'_j}{1-n'_k N^k}\right]^\text{TT},
\end{equation}
where $u$ is the retarded time, $R$ the luminosity distance \newJZ{of the detector} to the source, $n'\equiv n(\Omega')$ are unit radial vectors centered at the source, $N\equiv n(\Omega)$ is the unit line-of-sight vector, and TT denotes a projection onto the TT-gauge along the relevant direction $N$ \cite{maggiore2008gravitational,Heisenberg:2023prj}. 
\newJZ{This formula describes a non-linear displacement memory component of the asymptotic radiation that is caused by any spherically asymmetric release of \gls{gw} energy flux from a localized source. }
%\JZ{I was not sure whether the following sentence is needed or is already too much:} 
\newJZ{The characteristic angular dependence within the square brackets in Eq.~\eqref{eq:memoryequation} arises as a combination of the universal structure of any asymptotic energy-momentum tensor and the source-to-detector vector within the Green's function of the D’Alembert operator \cite{Garfinkle:2022dnm,Heisenberg:2024cjk}.} 
Moreover, since the energy flux of GWs is integrated over the entire binary evolution, the GW memory is a \textit{hereditary} effect. More precisely, it accumulates over time leaving an increasing off-set of the initial zero mean of the strain, leading to a non-zero geodesic deviation between any freely falling test-masses after the \glspl{gw} have passed and therefore by definition induces a displacement memory effect. However, most of the effect comes from the merger stage, where about $5\%$ of the total mass is released in GWs~\cite{Barausse:2012qz} which explains the characteristic step-like behaviour of the memory as seen in Figure~\ref{fig: surr_cce}.}

%\JZ{I never thought of this before, but I don't think that in this case we can talk about a ``quadrupole moment'' here, at lest it is not evident. The classic quadrupole moment formula in GR arises in a non-relativistic expansion of the source in a $v/c$ expansion, showing that in this limit the moltipole that dominates is the quadrupole one. For the primary radiation such a non-relativistic expansion does of course not make sense?}

\new{
Fundamentally, the gauge-invariant energy flux of primary waves reads 
\begin{equation}\label{eq:energyflux}
    \frac{\mathrm{d}E}{\mathrm{d}u'\mathrm{d}\Omega'}=\frac{R^2 c^3}{16\pi G}\langle \dot{h}_{0+}^2+\dot{h}_{0\times}^2\rangle,
\end{equation}
where $h_{0+}$ and $h_{0\times}$ are the amplitudes of the corresponding polarization states and the spacetime average $\langle...\rangle$ is to ensure a well-defined energy content of \glspl{gw} \cite{PhysRev.121.1556,PhysRev.166.1272,maggiore2008gravitational,Favata:2011qi,Heisenberg:2023prj}.
Such an average over the scales of variation of the \glspl{gw} naturally arises within an Isaacson approach to defining \glspl{gw} \cite{Isaacson_PhysRev.166.1263,Isaacson_PhysRev.166.1272}, and in the context of memory, it allows for a clear distinction between so-called {\it primary waves}, defined as the oscillatory part of asymptotic radiation that we denote as $h_0$, and the non-linear memory that is fundamentally understood as a corresponding low-frequency correction \cite{Heisenberg:2023prj}.
% Such a sharp separation in scales of variation between the memory signal and the primary waves is important in order to be able to claim a detection of memory, as we elaborate in App.~\ref{App:memorymodeldetails}. Indeed, as we will see in Section~\myhyperref{subsection: lisa signature} the differentiation between memory and the primary oscillatory signal is evident within a time-frequency analysis of the signal.}
However, for quasi-circular and non-precessing compact binary black hole coalescence, a special case which this work is restricted to, the spacetime average in \eqref{eq:energyflux} can effectively be dropped, as we explain in detail in App.~\ref{App:memorymodeldetails}. It will therefore be ignored throughout this study. Nevertheless, is worth emphasizing that, as soon as eccentric or precessing binaries are considered, averaging over high-frequency scales becomes inevitable to define a memory signal clearly distinguishable from the primary wave.}
In the following, we therefore call the {\it memory signal} the part of the \gls{gw} that is generated merely by the energy flux emitted in primary \glspl{gw}.

\new{It is useful to explicitly take into account the freedom in rotating the polarization basis of any outgoing radiation by defining the complex scalar 
\begin{equation}
    h\equiv h_+-ih_\times,
\end{equation}
with explicit spin-weight of $s=-2$ \cite{DAmbrosio:2022clk}, that can therefore be expanded in terms of spin-weighted spherical harmonics
\begin{equation}\label{eq:mode_decomp}
    h(u,\Omega)= \sum_{\ell\geq 2}\,\sum_{|m|\leq \ell} h^{\ell m}(u)\, \phantom{}_{-2}Y_{\ell m}(\iota,\phi) .
\end{equation}
Projecting both the memory signal as well as the primary wave into the spin-weighted spherical harmonics basis, Eq.~\eqref{eq:memoryequation} can efficiently be evaluated, \newHI{as the expansion helps to disentangle the angular and the time dependencies and yields a more compact representation of the angular dependence of the flux}. Then,  the angular integration in Eq.~\eqref{eq:memoryequation} can be performed either numerically as done by the \texttt{GWMemory} software (see Eq.~(8) and~(9) of ~\cite{Talbot:2018sgr}), or even analytically (see Eq.~\eqref{eq:memorymodes} in the Appendix).
% \begin{equation}
%     \delta h (u, \Omega) = \frac{R}{4 \pi c} \Lambda_{\ell_1, \ell_2, m_1, m_2} \int_{-\infty}^{u} \dot{h}_{\ell_1, m_1} \dot{h}_{\ell_2, m_2}^* du',
% \end{equation}
% with:
% \begin{multline}
%         \Lambda_{\ell_1, \ell_2, m_1, m_2} = \frac{1}{2} \left( e_{jk}^+ - e_{jk}^\times \right) \times
%     \\
%     \int_{S^2} {}_{-2} Y_{\ell_1, m_1}(\Omega')\ {}_{-2} Y_{\ell_2, m_2}(\Omega') \left[\frac{n'_in'_j}{1-n'_ln^l}\right]^\text{TT} d\Omega'
% \end{multline}
% ~\footnote{\new{Additionally, in the case of quasi-circular and non-precessing compact binary black hole coalescence's considered in this work, the spacetime average in \eqref{eq:energyflux} can effectively be dropped as we explain in detail in App.~\ref{App:memorymodeldetails}. However, as soon as eccentric or precessing binaries are considered, averaging over high-frequency scales is expected to become inevitable to define a clearly distinguishable memory signal.}}
From the analytical derivation, it can be seen that the memory is essentially showing up in the $(2,0)$ mode, since most of the primary \glspl{gw} energy flux is released in the $(2,2)$ mode. More details are given in the App.~\ref{App:memorymodeldetails}.}

Throughout this work, we will therefore use \new{the $(2,0)$ memory mode} computed through the \new{publicly available and numerically optimized} \texttt{GWMemory} package \new{with an input of primary} waveforms from the surrogate model \texttt{NRHybSur3dq8}~\cite{varma_surrogate_2019}. \newHI{This surrogate model is the most accurate available for the parameter space of interest in this study (low mass-ratio $q$, focus on the merger)}. It is trained on numerical waveforms limited to mass-ratio $q < 8$ and spins aligned to the angular momentum of the binary with amplitude $\chi_{1,z},\chi_{2,z}\in [-0.8,0.8]$ (the model does not include precession or eccentricity). 

\new{
\subsubsection{Alternative modeling approach}
}
%\newHI{Moreover, a recent update of this model \cite{yoo_numerical_2023} which uses a novel waveform extraction scheme called \gls{cce} \cite{Mitman:2020pbt,Mitman:2024uss}, naturally captures the non-linear memory, hence comparison and consistency checks will be explored and discussed throughout this study.} \JZ{Perhaps this part is not necessary here as we introduce the CCE waveforms in the next subsection?}

\new{The surrogate waveform model \texttt{NRHybSur3dq8}~\cite{varma_surrogate_2019} has recently been updated with \texttt{NRHybSur3dq8\_CCE}~\cite{yoo_numerical_2023}, which is the first waveform model to include the full $(2,0)$ memory mode. These waveforms are calibrated to Numerical Relativity (NR) simulations that use a novel waveform extraction scheme called \gls{cce}~\cite{Mitman:2020pbt,Mitman:2024uss}, which directly captures the full low-frequency memory contribution.  However, \new{it is important to note that} the $(2,0)$ mode \new{does not purely capture the memory signal, but} also includes oscillatory contributions excited during the ringdown, \new{which are present in both the \texttt{NRHybSur3dq8} and \texttt{NRHybSur3dq8\_CCE} waveforms (see also App.~\ref{appendix: comp surrmem gwmem} for further discussion).}}
%\new{With the use of} a \gls{cce} extraction procedure of the strain from numerical relativity simulations \cite{Handmer:2016mls}, \new{it is possible today to generate numerical waveforms from binary coalescence's that directly capture the full low-frequency memory contribution. These numerical waveforms are} now implemented in the SXS catalogue~\cite{Mitman:2020pbt,Mitman:2024uss}. \new{Based on these results, the surrogate model \texttt{NRHybSur3dq8}~\cite{varma_surrogate_2019} was recently updated to a version \texttt{NRHybSur3dq8\_CCE}~\cite{yoo_numerical_2023} that represents} the first waveform model that includes the full memory mode $(2,0)$ calibrated to the \new{\gls{cce}} Numerical Relativity simulations. However, \new{it is important to know that} the $(2,0)$ mode \new{does not purely capture the memory signal, but} also includes oscillating contributions excited during the ringdown, \new{which are present in both the \texttt{NRHybSur3dq8} and \texttt{NRHybSur3dq8\_CCE} waveforms (see also App. for further discussion).}

\new{Yet, with these two surrogate waveform models at hand, an alternative modelling approach for the memory within the $(2,0)$ mode can be envisaged by simply subtracting the oscillatory part present in \texttt{NRHybSur3dq8} from the full mode in \texttt{NRHybSur3dq8\_CCE} that contains the memory}
\begin{equation}
h_{+}^{mem}(t) \equiv h_{+}^{(20)|cce}(t) - h_{+}^{(20)}(t). 
\label{eq: surr_20}
\end{equation}
This separation is shown in Figure~\myhyperref{fig: surr_cce}, where we show the total waveform (blue) and the $(2,0)$ mode from the \texttt{NRHybSur3dq8\_CCE} model (orange) and from the \texttt{NRHybSur3dq8} (green) for a merger event with parameters [$M_{tot} = 10^{6}\ M_{\odot}$, $z = 1.0$, $q = 1.0$, $\iota = \SI{1.047}{\radian}$, $S=0.0$]. \new{From Figure~\myhyperref{fig: surr_cce}, it becomes} clear that Eq.~(\myhyperref{eq: surr_20}) \new{(grey-dashed) indeed} allows us to isolate the memory from the oscillatory features of the ringdown in the $(2, 0)$ mode.

\new{In Figure~\myhyperref{fig: surr_cce} we also compare the memory model defined in Eq.~\eqref{eq: surr_20} with the direct theoretical prediction in Eq.~\eqref{eq:memoryequation} computed by the \texttt{GWMemory} package (red). While the two memory models appear to be in good agreement at first sight, we find significant discrepancies, which are significantly accentuated for high mass ratio systems. \newHI{The origin of this discrepancy is not entirely elucidated. However, it is suspected that different gauge fixing options taken by \texttt{NRHybSur3dq8} and \texttt{NRHybSur3dq8\_CCE}, related to the different strain extraction at null infinity strategies \cite{yoo_numerical_2023}, make the direct comparison in Eq.~\eqref{eq: surr_20} not well defined in general and especially for higher mass ratios $q$. This discrepancy will be investigated empirically, and the results of the comparison between Eq.~\eqref{eq: surr_20} and~\texttt{GWMemory} calculations will be presented in detail in App.~\myhyperref{appendix: comp surrmem gwmem}, as a by-product of this work.}}

\new{Because of these uncertainties in isolating the memory model based on NR simulations, as opposed to the clear understanding of memory from Eq.~\eqref{eq:memoryequation}, we will use the memory computed by the \texttt{GWMemory} package~\cite{Talbot:2018sgr} throughout this work.
This choice is also justified because the memory computed by Eq.~\eqref{eq:memoryequation} gives more conservative estimates in terms of SNR and smooth behaviour over the parameter space.} \new{It is emphasized that we are here focusing on the detectability of the memory only, as opposed to that of the entire $(2,0)$ mode, to which the oscillatory contribution from the ringdown adds significantly. We note that a phenomenological waveform model for the $(2,0)$ mode that includes both the memory and ringdown components has recently been implemented as an extension of the computationally efficient \texttt{IMRPhenomTHM} model~\cite{Rossello-Sastre:2024zlr} for the similar parameter space covered by \texttt{NRHybSur3dq8}, though without restriction on the length of the waveform. Once published, this can be used in the future for extensive Bayesian parameter estimation, which is currently very computationally expensive.}

\subsection{Detector response simulation}
\label{subsection: detector model}
%%%%%%%%%%%%%%%%%%%%%%%%%%%%%%%%%%%%%%%%%%%%%%%%%%%%

The simulation of the detector response requires the projection of the strain time-series computed in Section~\myhyperref{subsection: mem model}  onto the space antenna response. For this, one first computes the interferometer's six single-link time-domain responses to the memory strain time-series using  \texttt{LISAGWResponse} \cite{bayle_lisa_2023, bayle_lisa_2022}. This yields $6$ time-series of relative (Doppler) frequency modulation $y(t) = \frac{\delta\nu}{\bar{\nu}}(t)$ of the laser link frequency caused by the \gls{gw} crossing over the laser beam path:
\begin{equation}
\label{eq: response}
    %y_{ij}^{\mathrm{GW}} 
    y_{ij}= \frac{1}{2} \frac{\epsilon^a \epsilon^b}{1 - \vec{\epsilon}\cdot\vec{k}} \left[ h_{ab}\left(c t_{r} - \vec{k} \cdot \vec{x}_{r}\right) - h_{ab}\left(c t_{e} - \vec{k} \cdot \vec{x}_{e}\right) \right].
\end{equation}
\newHI{ Eq.~\eqref{eq: response} is called the two-pulse response \cite{vallisneri_geometric_2005} for the variable $y_{ij}(t)$, which is the relative Doppler shift time series observed from a vertex $i$ of LISA between the local, emitted frequency $\nu_{ij}$ and the received distant laser frequency $\nu_{ji}$. The latter frequency is subject to space-time geometry fluctuations as it travels along $\vec{\epsilon} \equiv \vec{\epsilon}_{j \rightarrow i}$ direction across the optical arm. The relative frequency fluctuation $y_{ij}(t)$ is reduced to the difference between the \gls{gw} strain $h_{ab}$ evaluated at the time of emission $(t_e, \vec{x}_e)$ and reception $(t_r, \vec{x}_r)$.  Consequently, for \gls{gw} wavelengths much larger than the constellation arm length $L$, $y \propto \dot{h}$ and the response in Eq.~\eqref{eq: response} behaves as a first order time differentiator. A comprehensive derivation of the time-domain link response can be found in App. A of Ref.~\cite{baghi_uncovering_2023}.}

The single-link frequency fluctuation data streams $y(t)$ are then injected into the \texttt{PyTDI} software \cite{staab_pytdi_2023} to simulate the virtual interferometer (i.e. suppressing primary noises, such as laser \cite{bayle_lisa_2023} and spacecraft jitter noise \cite{lisa_dynamics}). \newHI{This generates the ultimate $3$ interferometer time-series, selecting the A, E and T (second generation) variables \cite{tinto_time-delay_2020, vallisneri_geometric_2005} to work with nearly independent channels and ignoring residual cross-correlations, a sufficient set-up for \acrshort{snr} computations (more details about this are given in Section \myhyperref{subsection: noise}). Second-generation is needed to account for constellation flexing during the laser beam propagation time in the virtual interferometers, and then to suppress effectively laser noise below minimum requirements \cite{lisa_scird}. However, this second generation \gls{tdi} impacts LISA response to \gls{gw}, and typically acts as a second order time differentiator again at low frequency, i.e. for $\lambda_{\mathrm{GW}} \gg L$~\cite{babak_lisa_2021}.} Keplerian orbits are considered for the constellation spacecraft, both for the single-link response and the time delays in \gls{tdi} post-processing.

In addition, we performed several processing adjustments to facilitate the computation of the \gls{snr}  around the merger time. First, the time reception shift of the wave at the detector location is ignored (setting \texttt{sun\_shift~=~False}). This is a parameter specific to \texttt{LISAGWResponse}, where the timing frame for the input strain time series is understood from the location of the \gls{ssb} reference frame barycentre,  implying a propagation time until it reaches the space interferometer. This makes the merger time depend on the sky-localization of the source, adding unnecessary complications for our scope here. Second, the time origin was adapted so that the merger time --- defined as the time of the maximum $(l,m)=(2,2)$ amplitude as conventional in literature \cite{boyle_sxs_2019} --- is set to be $t_{\text{merger}} = 0$. The time window width around $t=0$ has been set to adapt dynamically to the total mass $M_{tot}$ of the \gls{mbhb} system, ensuring to capture most of the oscillatory \gls{snr} for events with $M_{tot}>10^5 M_{\odot}$, and to fully capture the memory \gls{snr}  for all relevant \acrshort{mbhb} masses ($10^4 M_{\odot}-10^8 M_{\odot}$). Finally, to avoid biases at low frequencies when representing the waveform in the frequency domain,  especially occurring due to edge effects from the restricted time window and the \gls{gw} memory step-like shape, we apply the Planck-taper window centred on the merger time, as extensively used in \gls{gw} data analysis \cite{mckechan_tapering_2010}. 

%%%%%%%%%%%%%%%%%%%%%%%%%%%%%%%%%%%%%%%%%
\subsection{Noise settings}
\label{subsection: noise}
%%%%%%%%%%%%%%%%%%%%%%%%%%%%%%%%%%%%%%%%%

For the \gls{snr} determinations, LISA noise spectra are computed from single-link noise models, merely containing the two dominant secondary noises, i.e. the test mass (TM) acceleration noise and the \gls{oms} noise, and in agreement with the \gls{scird} \cite{lisa_scird},
\begin{widetext}
\begin{align}
    & S_{n|\text{TM}}^{1/2} = 3 \times 10^{-15} \sqrt{\left[ 1 + \left( \frac{\SI{0.4}{\milli\hertz}}{f} \right)^2 \right] \left[ 1 + \left( \frac{f}{\SI{8}{\milli\hertz}} \right)^4 \right]} \  \si{\frac{\metre }{\second^{2}}
 \persqrthz}\label{eq: TM noise}, \\
    & S_{n|\text{OMS}}^{1/2} = 15 \times 10^{-12} \sqrt{\left[ 1 + \left( \frac{\SI{2}{\milli\hertz}}{f} \right)^4 \right]} \ \si{\metre\persqrthz} \label{eq: OMS noise}.
\end{align}
\end{widetext}
\clearpage
We then apply the \gls{tdi} transfer function matrices on the single-link data streams based on the \gls{tdi}-2 combinations, defined as
\begin{align}
\label{equ:X2}
X_2 =&\, X_{1.5} + \vb{D}_{13121}y_{12}+\vb{D}_{131212}y_{21}+\vb{D}_{1312121}y_{13} \notag
\\
& + \vb{D}_{13121213}y_{31} - \big[\vb{D}_{12131}y_{13}+\vb{D}_{121313}y_{31}
\\
& +\vb{D}_{1213131}y_{12} + \vb{D}_{12131312}y_{21}\big]\notag,
\end{align}
with
\begin{align}
\label{equ:X1}
X_{1.5} =& \,y_{13} + \vb{D}_{13}y_{31} + \vb{D}_{131}y_{12}+ \vb{D}_{1312}y_{21} \notag
\\
& - ( y_{12} + \vb{D}_{12}y_{21}+\vb{D}_{121}y_{13}+\vb{D}_{1213}y_{31}).
\end{align}
\new{and where $Y_2$ and $Z_2$ can be deduced by circular permutation of the spacecraft indices $1\xrightarrow{}2\xrightarrow{}3\xrightarrow{}1$.}
There, for example, $y_{12}$ is the frequency modulation of the laser beam emitted by spacecraft $2$ towards spacecraft $1$, and as measured on spacecraft $1$, then accumulating \gls{gw} modulation along its path.
$\vb{D}_{ij}$ is a delay operator which takes into account the light time travelled from one spacecraft $i$ to the other $j$ separated by the arm distance $L_{ij}$. In the frequency domain, the delay operator is a simple phase operator (Eq.~(\myhyperref{eq: delay operator})),
\begin{equation}
    \vb{D}_{ij}x(t)= x(t-L_{ij}(t)) \, \xrightarrow{FT} \, \vb{\Tilde{D}}_{ij} \Tilde{x}(f) = \Tilde{x}(f)\ e^{-2\pi if L_{ij}(t)}
\label{eq: delay operator}
\end{equation}
and $\vb{D}_{ijk...}$ are nested time-delay operators. For example, the second term of Eq.~(\myhyperref{equ:X2}), $\vb{D}_{13121} = \vb{D}_{13} \vb{D}_{31} \vb{D}_{12} \vb{D}_{21}$, is the time-domain operator used to delay the single-link data $y_{12}$ by a time shift equal to the light travel time $L_{13121}(t)$ of a laser beam circulating across the spacecraft constellation, starting at spacecraft $1$, making a round trip along arm $21$, then going for a second round trip along $31$ arm to finally ending back at spacecraft $1$. \\
%From there, we move to the frequency domain, where we approximate the delay operators as simple phase operators (Eq.~(\myhyperref{eq: delay operator})),
%\begin{equation}
%    \vb{D}_{ij}x(t)= x(t-L_{ij}(t)) \, \xrightarrow{FT} \, \vb{\Tilde{D}}_{ij} \Tilde{x}(f) = \Tilde{x}(f)\ e^{-2\pi if L_{ij}(t)}.
%\label{eq: delay operator}
%\end{equation}
\new{Arranging the \gls{tdi} operators as a matrix applied on the single-link noise spectra as in \cite{baghi_uncovering_2023}, one gets the \gls{tdi}-XYZ noise cross spectral density $3 \times 3$ matrix $S_{XYZ}$:
\begin{equation}
    S_{XYZ}(f) = \mathbf M^{\text{TDI}}(f) \ S_{ij}(f)\ \mathbf M^{\text{TDI}, \dag}(f).
\end{equation}
Converting to \gls{tdi}-AET channels by a simple change of basis \cite{quang_nam_time-delay_2023},
% \begin{align}
%     \begin{bmatrix}
%         A \\
%         E \\
%         T
%     \end{bmatrix}
%     =
%     \begin{pmatrix}
%         -\frac{1}{\sqrt{2}} & 0 & \frac{1}{\sqrt{2}} \\
%         \frac{1}{\sqrt{6}} & -\frac{2}{\sqrt{6}} & \frac{1}{\sqrt{6}}\\
%         \frac{1}{\sqrt{3}} & \frac{1}{\sqrt{3}} & \frac{1}{\sqrt{3}}
%     \end{pmatrix}
%     \begin{bmatrix}
%         X \\
%         Y \\
%         Z
%     \end{bmatrix},
% \label{eq: AET}
%  \end{align}
one can get the A, E, T noise spectra:
\begin{align}
S_{AET}
=
\begin{pmatrix}
    -\frac{1}{\sqrt{2}} & 0 & \frac{1}{\sqrt{2}} \\
    \frac{1}{\sqrt{6}} & -\frac{2}{\sqrt{6}} & \frac{1}{\sqrt{6}}\\
    \frac{1}{\sqrt{3}} & \frac{1}{\sqrt{3}} & \frac{1}{\sqrt{3}}
\end{pmatrix}
S_{XYZ}.
\label{eq: AET noise}
\end{align}
% \begin{align}
% & S_{AA} = S_{XX} - S_{XY}, \\
% & S_{EE} = S_{XX} - S_{XY}, \\
% & S_{TT} = S_{XX} + 2\ S_{XY},
% \label{eq: AET noise}
% \end{align}
We assume $S_{AET}$ to be diagonal to first approximation, as rotating \gls{tdi} from $XYZ$ to $AET$ basis makes the noise to be nearly uncorrelated. Such property is exact for symmetric performances of $X$, $Y$ and $Z$ channels only \footnote{the \gls{tdi}-AET basis diagonalizes the cross-correlation matrix $S_{XYZ}$ in case of a constellation symmetric over spacecraft cyclic permutation.}. However, we account for unequal arm constellation orbits in the modeling of $S_{AA}$, $S_{EE}$ and $S_{TT}$ diagonal elements for improved accuracy of the noise power spectrum models.} \\
In addition to instrumental noise, we add for completeness a $4$-years galactic confusion noise spectrum, equivalent to the residual power left by the unresolved (and non-extracted) \glspl{gw} from galactic binaries  after a $4$-years observation run (minimal expected mission duration) \cite{pitte_detectability_2023, lisa_scird}.
Finally, it is worth noting that the noise power spectrum required a smoothing treatment at the zero-response frequencies to counteract numerical instabilities when computing the \gls{snr}. Processing the single-link noise spectrum, we have considered multiple \gls{tdi} transfer functions evaluated at $20$ distinct epochs of the year and averaged the outputs to get the finally used \gls{tdi} noise power spectrum. The procedure then provides a year-averaged power spectrum, as well as smoothed-out resonances, which now have a negligible impact on the \gls{snr} values.\\

We finally define the \gls{snr} of the primary and memory signals in LISA data as \cite{babak_lisa_2021}:
\begin{equation}
    \rho_{C}^2 = 4 \Re \int_{\rm f_{\rm max}}^{f_{\rm max}} \frac{\Tilde{d}(f)\Tilde{d}^*(f)}{S_{CC}(f)} \mathrm{d}f,
\label{eq: snr per channel}
\end{equation}
\newHI{where $\Tilde{d}(f)$ is the Fourier transform of an interferometer output channel $C\in\{A, E, T\}$ and all the spectra are {\it one-sided}. We combine the 3 \gls{tdi} channels $C\in\{A, E, T\}$ summing the individual \gls{snr} $\rho_{C}$ quadratically:}
\begin{equation}
    \rho_{\text{tot}} = \sqrt{\sum_{C\in\{A, E, T\}} \rho_{C}^2},
\label{eq: snr tot}
\end{equation}
\new{using the property of $A$, $E$ and $T$ channels of being approximately independent.} Limited frequency range with $f_{\rm min}= 5 \times 10^{-5}\ \si{\hertz}$ and $f_{\rm max} =10^{-1}\ \si{\hertz}$ is considered in the computation of $\rho_{\text{tot}}$, making sure that all the memory \gls{snr} and most of the $(2,2)$ mode are captured, for any masses ranging from $10^4\, M_{\odot}$ to $10^8\, M_{\odot}$.

%%%%%%%%%%%%%%%%%%%%%%%%%%%%%%%%%%%%%%%%%%%%%%%%%%%%
\section{Memory signature in LISA data}
\label{subsection: lisa signature}
%%%%%%%%%%%%%%%%%%%%%%%%%%%%%%%%%%%%%%%%%%%%%%%%%%%%

Equipped with the above state-of-the-art simulation tools for the LISA TDI time-domain response to \gls{gw}, we can now study the projection of the wave strain, defined as the $(2,2)$ mode, and of the memory strain, defined as the memory component of the $(2,0)$ mode explained above.
\begin{figure}[t!]
    \centering
    % trim={<left> <lower> <right> <upper>}
    \includegraphics[width=1.0\columnwidth, trim={0.2cm 0.0cm 0.2cm 0.0cm}, clip]{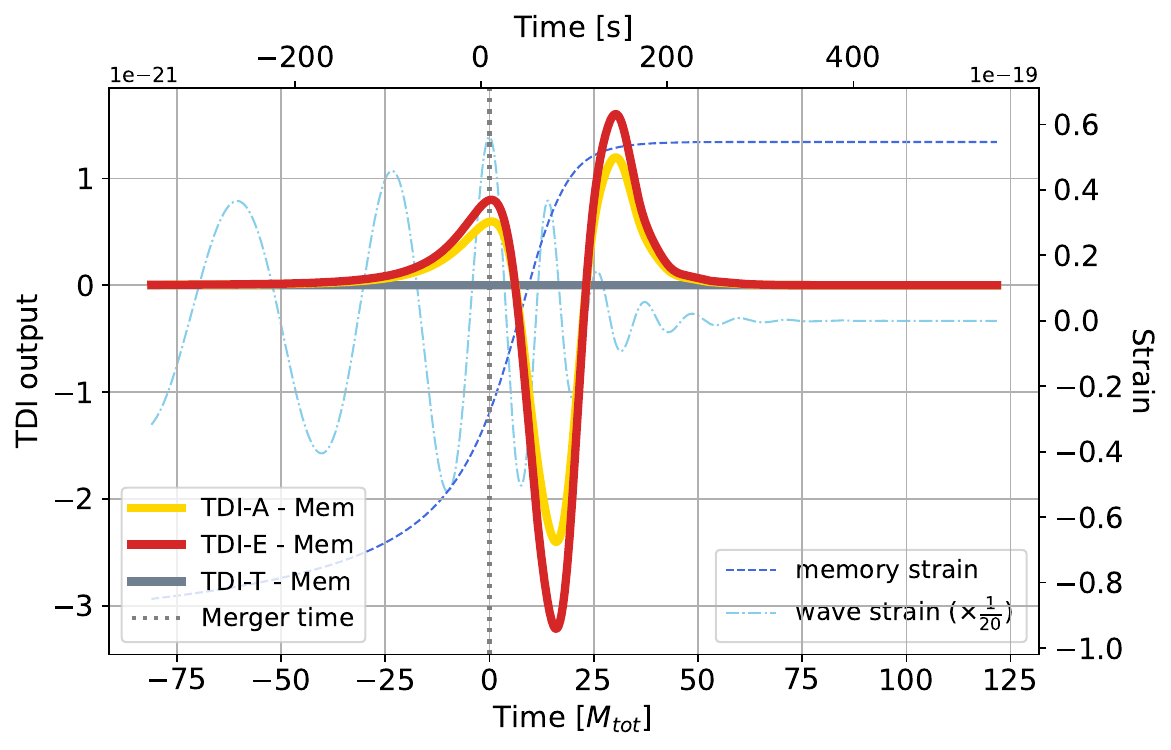}
    \caption{GW memory imprint of a binary merger with parameters [$M_{tot} = 10^{6}\ M_{\odot}$, $z = 1.0$, $q = 1.0$, $\iota = \SI{1.047}{\radian}$, $S=0.0$] \new{observed at sky latitude and longitude $[\beta = 0.52 \si{\radian}, \lambda = 3.24 \si{\radian}]$}, on TDI time series A, E and T (resp. yellow, red and blue plain traces).  The \gls{gw} memory burst time-series as well as the sourcing oscillatory waveform strain ($\times 0.05$ down-scaled in amplitude) are superimposed in dashed-lines for timing and shape reference. Left $Y$-axis provides units for the TDI A, E and T channels (plain lines) while right $Y$-axis units are attached to the strain atime-series (dashed-lines). The merger time ($t=0$) is indicated with a vertical dot-line.}
    \label{fig: mem_tdi}
\end{figure}
This is shown for the three different channels A, E and T in Figure~\ref{fig: mem_tdi} for the \gls{gw} of the same event considered in Figure~\myhyperref{fig: surr_cce}. In the background of the same figure, and attached to the additional right-hand $y$-axis, we trace the $(2,2)$ mode (down-scaled by $\times 0.05$) and the memory strain for a timing reference; the merger time is also indicated with a black dotted line. One observes from Figure~\myhyperref{fig: mem_tdi} that the instrument response function acts as a high-pass filter for the memory signal, significantly reducing the overall amplitude and exhibiting time oscillations. 
In fact, at low frequency, i.e. for wavelengths longer than the arm length, the instrument response behaves as a high order differentiator~\cite{Babak:2021mhe}. As a result, the TDI output for the memory is mostly determined by $\partial_t^3 h$, as we show in Appendix~\ref{sec:app}.

\begin{figure}[t!]
    \centering
    % trim={<left> <lower> <right> <upper>}
    \includegraphics[width=1\columnwidth, trim={0.0cm 0.0cm 0.0cm 0.0cm}, clip]{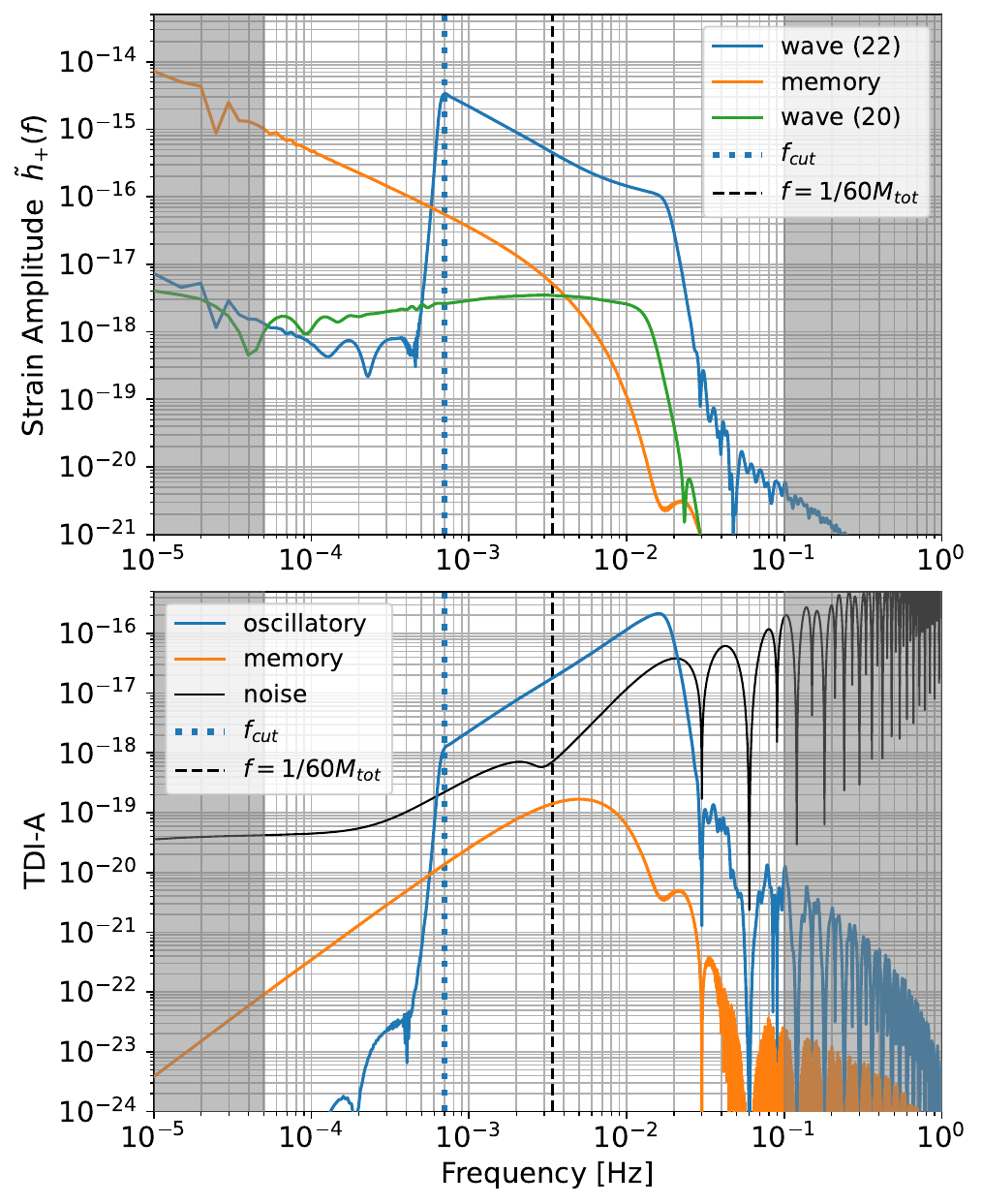}
    \caption{Up: Spectrum of the strain time-series $h_{+}(t)$, the wave $(l,m)=(2,2)$ mode in blue, the wave $(l,m)=(2,0)$ mode in green, and the memory in orange. Down: \gls{tdi}-X time-series response to the wave $(l,m)=(2,2)$ mode (in blue) and to the memory (in orange). The same system as in Figure \myhyperref{fig: mem_tdi}, i.e. [$M_{tot} = 10^{6}\ M_{\odot}$, $z = 1.0$, $q = 1.0$, $\iota = \SI{1.047}{\radian}$, $S=0.0$] is considered. The dashed gray lines on both figures mark the cut-off frequency applied at the inspiral due to computational time limitation (the oscillatory modes are invalid below that frequency. The gray frequency domains ($f < 5\times 10^{-5} \si{\hertz}$ and $f > 10^{-1} \si{\hertz}$) are excluded from the \gls{snr} computation.}
    \label{fig: spectra}
\end{figure}

This can also be seen in the frequency domain in Figure~\myhyperref{fig: spectra}, where we compare the spectra of the injected signals (up) and of the TDI outputs (down). The Fourier Transform (FT) of the memory on the upper panel (orange) scales as $1/f$ at low frequency and it decays at frequencies $f\gtrsim 1/60\, M_z$ where \new{$M_z=(1+z)M_{tot}$} is the redshifted total mass of the binary. This follows from the fact that, on time scales longer than its rising time, the memory is well approximated by a step-like function around the merger time, whose FT is $\sim1/f$. At high frequency, the cutoff frequency is given by the duration of the memory saturation~$\tau\sim 60\, M_z$, that corresponds to the time window around the merger, during which most of the \gls{gw} energy~$E_\mathrm{GW}$ is radiated~\cite{Buonanno:2006ui}.  

We also show in Figure~\myhyperref{fig: spectra} the oscillating component of the  $(2,0)$ mode (green). It is interesting to note that the frequency $f=(60 M_z)^{-1}$ (black dashed) represents a good transition above which this oscillatory component overcomes the memory.  Again, this is consistent with identifying the pure memory with the low-frequency signal of the  $(2,0)$ mode.
On the lower panel of Figure~\myhyperref{fig: spectra}, we show the spectra of the \gls{tdi} response to the (2,2) and memory modes of the waveform. Comparing the upper and lower subplots, focusing on the low-frequency {\it memory} spectra, we confirm the $\sim f^3(1/f)$ behavior of the LISA response which, in the long wavelength regime, manifestly approximates as a third-order time differentiator.

We stress that, to recover the expected spectrum of the memory strain shown in the upper panel of  Figure~\myhyperref{fig: spectra}, some processing of the waveform is needed that is different from the usual one applied to the oscillating wave.  In particular, to recover the low-frequency behavior of the memory, we pad the strain to its final value for a sufficiently long time after the end of the generated waveform and use a Planck-taper window function to put the final value gradually to zero. This last step is crucial: otherwise, spectral leaking would completely spoil the weak signal from the memory. The issue of correctly processing the memory has also recently been discussed in \cite{Chen:2024ieh},  where an alternative method to recover a clean spectrum is proposed. However, we found that this issue is not present in the \gls{tdi} output of the memory shown on the lower panel of Figure~\myhyperref{fig: mem_tdi}, since effectively the \gls{tdi} output is proportional to higher derivatives of the injected signal and the time series naturally goes to zero at the edges.

\begin{figure}[ht!]
    \centering
    % trim={<left> <lower> <right> <upper>}
    \includegraphics[width=\columnwidth, trim={0.0cm 0.0cm 0.0cm 0.0cm}, clip]{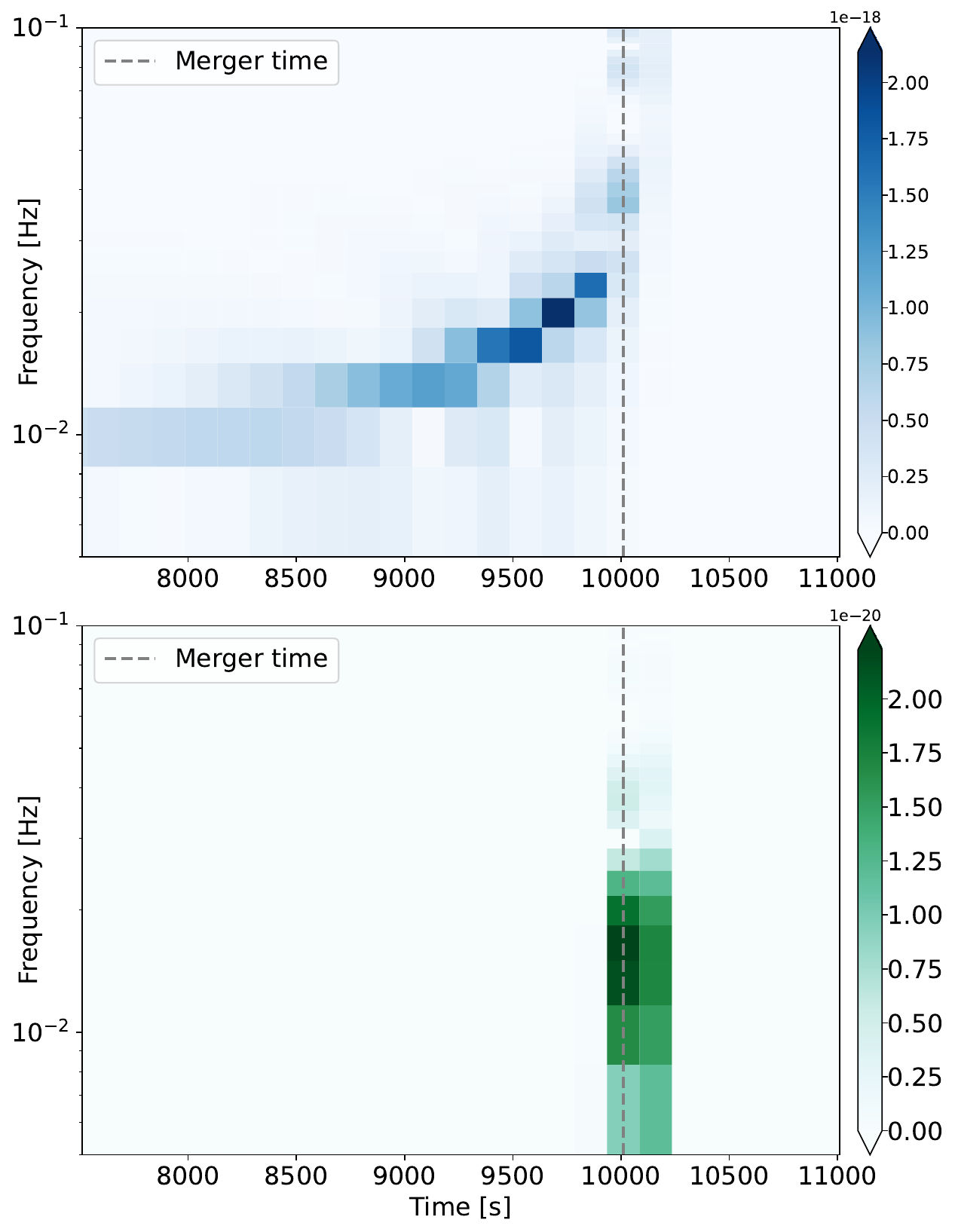}
    \caption{Spectrograms (short-time Fourier transforms) of the \gls{tdi}-A signals from the \gls{gw} (2,2) mode (\new{top}) and from the \gls{gw} memory component (\new{bottom}) of an equal-mass system with total mass $M_{tot} = 10^5 M_{\odot}$.  The two signals present a distinct localization in time-frequency representation, the (2,2) mode exhibiting the well-known {\it chirp} transient signal, increasing power towards high frequencies, whereas the memory is confined to merger time and relatively spread over frequencies, though with a maximum power frequency distinct from the (2,2) merger frequency. This representation provides helpful insight regarding strategies to identify memory from the main \gls{gw} modes.}
    \label{fig: spectrograms}
\end{figure}

Further insight into the characterization of the memory in the data is provided by the time-frequency plot of the TDI signals of the primary wave and the memory. As shown in Figure~\myhyperref{fig: spectrograms}, the two signals present a different localization in the time-frequency representation, with the (2,2) mode exhibiting the well-known {\it chirp} transient signal, increasing in power towards high frequencies, while the memory is confined to the merger time and relatively spread over frequencies, although with a maximum power frequency distinct from the (2,2) merger frequency.

\begin{table*}[t]
\setlength\tabcolsep{0.4cm}
\centering
\begin{tabular}{||l|c c c c c c||} 
 \hline
 Baseline & $q$ & $\chi$ & inclination $\iota\ [\si{\radian}]$ & lat. $\beta\ [\si{\radian}]$ & long. $\lambda\ [\si{\radian}]$ & pixel $p$ \\ [0.5ex] 
 \hline\hline
 1. Conservative & 2.5 & 0.0 & 1.047 & 0.62 & 0.20 & 145 \\ 
 \hline
 2. Optimistic & 1.0 & 0.0 & 1.571 & 0.52 & 3.24 & 192 \\
 \hline
 3. Opt. \& Spin. & 1.0 & 0.8 & 1.571 & 0.52 & 3.24 & 192 \\
 \hline
\end{tabular}
\caption{Table listing the three baselines used in this work regarding the choice of relevant merger parameters, such as the binary mass ratio $q$, the effective aligned spin $\chi$, the inclination of the source $\iota$, the sky-position (given as longitude and latitude angles $\beta$ and $\lambda$ in the \gls{ssb} frame, or equivalently, as a pixel index in the sky assuming a sky discretization with $N_{\text{side}}=8$ {\it RING} scheme via \texttt{healpy} \cite{healpy}).}
\label{table: baselines}
\end{table*}

%%%%%%%%%%%%%%%%%%%%%%%%%%%%%%%%%%%%%%%%%%%%%%%%%%%%%%%%%%%%%%%%%%%%%%
\section{\label{tdi} Signal-to-noise ratio and detectability predictions}
%%%%%%%%%%%%%%%%%%%%%%%%%%%%%%%%%%%%%%%%%%%%%%%%%%%%%%%%%%%%%%%%%%%%%%%%

We begin by examining the \gls{snr} of the primary wave and the memory as a function of sky position, which is important for two main reasons: first, to study the potential improvement in sky localization from the information encoded in the memory; second, to understand its effect on the calculation of \gls{snr}. LISA will have a different relative sensitivity for the memory and the primary wave depending on the sky localization, as the antenna response function is frequency-dependent\new{~\cite{Robson:2018ifk}}. This could help better characterize the source signal's direction, especially for short signals, similar to what is found for the inclination angle in~\cite{gasparotto_can_2023}.  In addition, the direction of the source in the sky will primarily affect the \gls{snr} of the memory, which is more of a burst-like event, whereas since the main signal will be in the band for longer, the motion of the detector will average out the effects of localization. 

\begin{figure*}[t!]
    \centering
    % trim={<left> <lower> <right> <upper>}
    \includegraphics[width=\textwidth, trim={0.0cm 0.0cm 0.0cm 0.0cm}, clip]{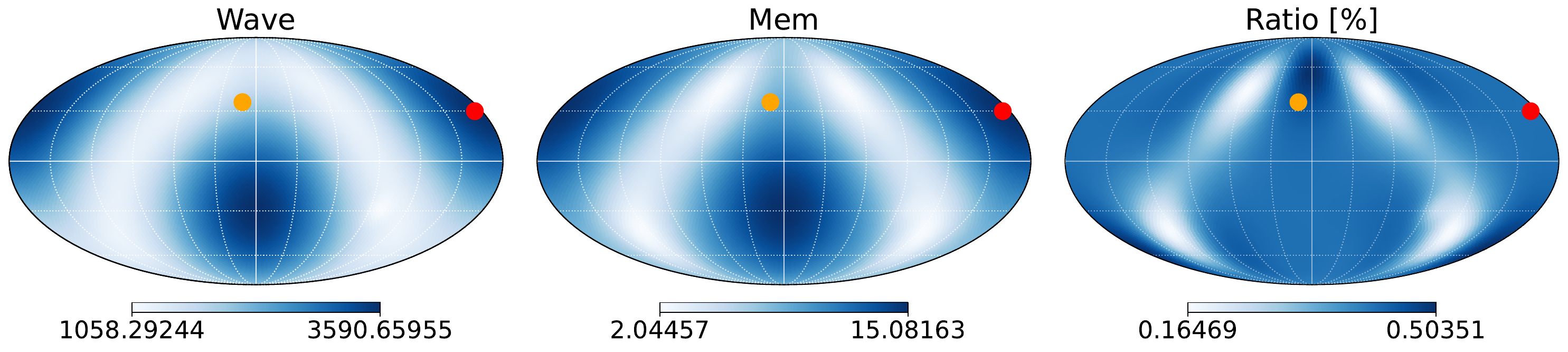}
    \caption{Sky-localization dependence of the \gls{gw} (2,2) and memory modes \gls{snr} (resp. left and centre), and their ratio shown in percentage (right). The system studied here is a non-spinning $[M_{tot} = 10^6 M_{\odot}, q = 2.5, z = 1.0, \iota=1.047 \si{rad}]$ merger. The celestial sphere is decomposed into equal area pixels via the tool \href{https://healpy.readthedocs.io/en/latest/}{\texttt{healpy}} \cite{healpy, healpix}), and the longitude $\lambda$ and latitude $\beta$ are converted into pixel indexes to be iterated over. We have sampled the sky into $N_{\text{pix}}=768$ pixels. The resulting sky map has been upsampled and smoothed spanning a Gaussian symmetry beam on the raw pixel data (see \texttt{smoothing} method from \href{https://healpy.readthedocs.io/en/latest/}{\texttt{healpy}} \cite{healpy, healpix}). The yellow dots point to a sky-direction with average memory \gls{snr} used in baseline $1$, while the red dot represents the optimal direction for memory \gls{snr}, which is considered in baseline $2$ (see Table \myhyperref{table: baselines}).}
    \label{fig: snr_skymap} 
\end{figure*}
We show an example of a sky-dependent map of the \gls{snr} for the primary wave, the memory, and their ratio, respectively, in Figure~\myhyperref{fig: snr_skymap} for a binary's merger with parameters $M_{tot}=10^{6} M_\odot$, $z = 1.0$, $q = 2.5$, zero spin and inclination $\iota=1.047$ ({\it conservative baseline $1$} in Table \myhyperref{table: baselines}).
As expected, we find differences in the sky pattern between the primary and memory signals, captured by the different \gls{snr} ratios. Furthermore, we confirm our expectation that the sky localization primarily affects the \gls{snr} of the memory, e.g. for this particular system we find a relative factor within sky locations of $\sim 7.5$ for the memory compared to $\sim 3$ for the primary. These values depend on the particular source we choose, and we expect this difference to be even greater for lighter binaries that stay longer in band.
The yellow dots in Figure~\myhyperref{fig: snr_skymap} indicate a sky direction with average \gls{snr}, while the red dots indicate the sky direction with maximum \gls{snr}. The first is considered for the {\it conservative baseline} for the detectability study of section~\myhyperref{sec:populations}, while the second is considered for the {\it optimistic baseline} for sky position. Since the sky maps are frequency-dependent, the sky directions for maximum and average \gls{snr} change slightly for different masses. The baseline set in our analysis (see Table \myhyperref{table: baselines}) is selected so that conservative and optimal assumptions remain valid across the examined mass range, \new{in other words we marginalize over the total mass. }

\begin{figure*}[t!]
   \centering
    \includegraphics[width=\textwidth, trim={0.25cm 0.0cm 0.25cm 0.0cm}, clip]
    {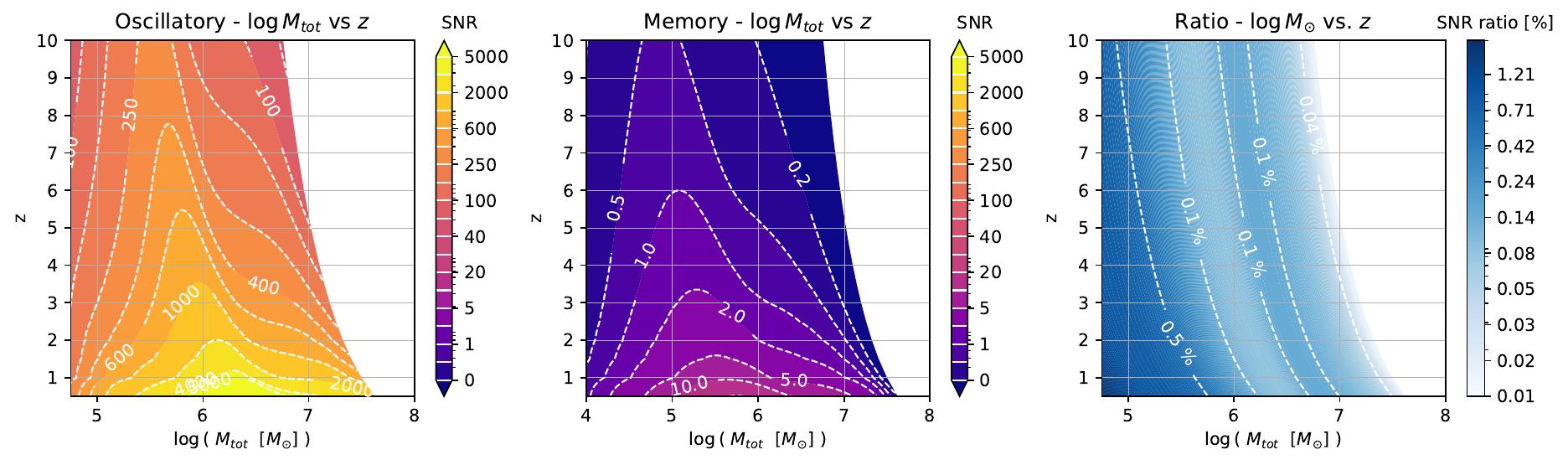}
    \caption{Contour plots of constant \gls{snr} lines for the primary wave \textit{(left)}, the memory \new{\textit{(middle)} and the ratio between the \gls{snr} of the memory and that of the primary \textit{(right)}} considering the average scenario (baseline $1$ described in Table \myhyperref{table: baselines}) for different \gls{mbhb} merger systems.}
    \label{fig: waterfall_MvsZ_Gal4y_Conservative}
 \includegraphics[width=\textwidth, trim={0.2cm 0.0cm 0.2cm 0.0cm}, clip]{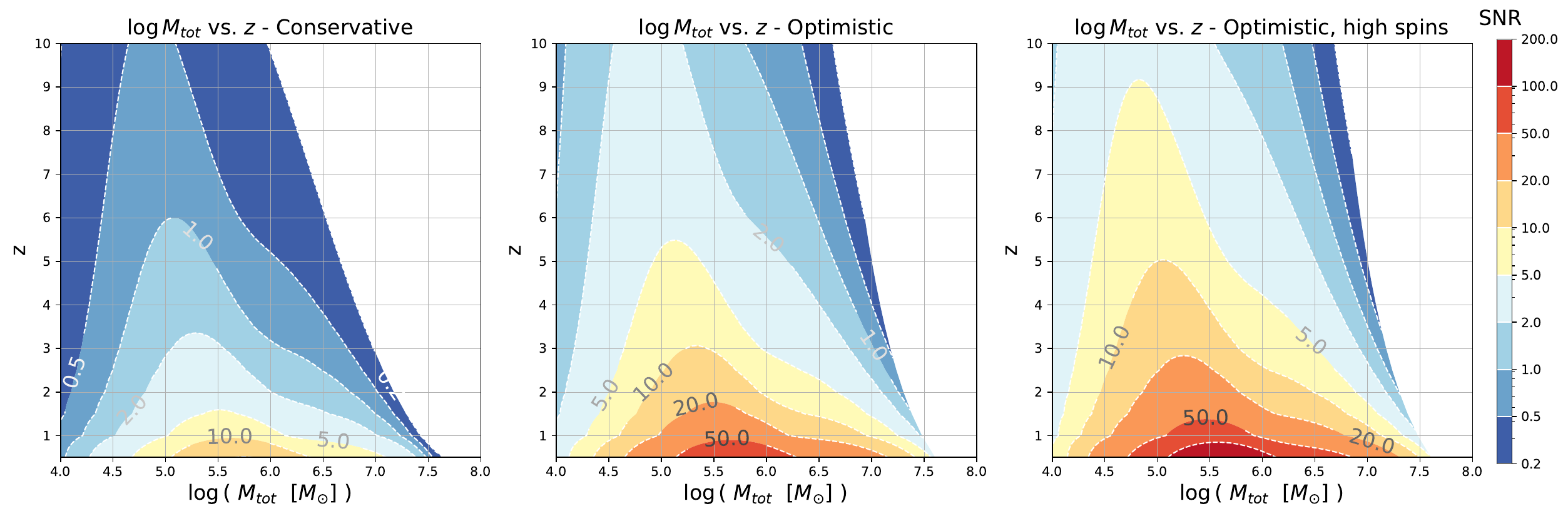}
    \caption{Contour plots for the constant \gls{snr} lines of the memory for the three baselines described in Table \myhyperref{table: baselines}, corresponding to different scenarios with increasing levels of optimism. \textit{Left:} median inclination and sky location, mass ratio $q=2.5$ and zero spin (baseline 1), \textit{Middle:} optimal inclination and sky-position, $q=1$ and zero spin (baseline 2), \textit{Right: }optimal inclination and sky-position, $q=1$ and $\chi_{1,z}=\chi_{2,z}=0.8$ (baseline 3). }
    \label{fig: waterfall_MvsZ_Gal4y_BaselineComp}
% \end{subfigure}%
\end{figure*}

We now examine the dependence on the mass and redshift of the binary black hole. The \gls{snr} results for the primary and memory signals as a function of \gls{mbhb} mass and redshift are shown in Figures~\myhyperref{fig: waterfall_MvsZ_Gal4y_Conservative} for an average scenario (baseline $1$ in Table \myhyperref{table: baselines}). Note that the peak of the \gls{snr} for the memory occurs at lower masses compared to that of the primary wave, due to the clear frequency separation of the two signals. This is reflected in the right panel of Figures~\myhyperref{fig: waterfall_MvsZ_Gal4y_Conservative}, where we see that the relative \gls{snr} of memory, which is less than the percentage level for the conservative baseline, increases for masses $M_{tot}\leq 10^{5.5}M_\odot$. The frequency content of the {\it memory signal} indeed lands closer to \gls{lisa}'s high sensitivity spot for such lighter masses, hence the expectation of a greater effect in these cases.

In Figure~\myhyperref{fig: waterfall_MvsZ_Gal4y_BaselineComp}, we show the same waterfall-type memory plots for the three different baselines of Table \myhyperref{table: baselines}: \emph{conservative}, \emph{optimistic}, and \emph{optimistic with spins}. \new{Note that the conservative and optimistic scenarios differ in sky location, inclination (median vs optimal), and also in the mass ratio ($q=2.5$ vs $q=1$). The optimal inclination for the memory is edge-on, since the $(2,0)$ mode depends on the inclination angle as $(\sin{\iota})^2$~\cite{Favata:2008yd}, as opposed to the $(2,2)$ mode, which is maximum for a face-on system. }
We will expand on the effect of the spin and the mass ratio in Section \ref{sec:q_and_spin}. Interestingly, the \gls{snr} can be enhanced by an order of magnitude for the same source which is relevant if one wants to use it to test General Relativity (which can be done with just a  few loud events). 
We cross-checked our \gls{snr} results and found very good agreement with those presented in~\cite{Pitte:2023ltw} for the primary wave and with those of~\cite{gasparotto_can_2023} for the memory, where the \gls{snr} was computed in the frequency domain (without computing the TDI projection of the signal) and the antenna's pattern functions were averaged over the sky.  

%%%%%%%%%%%%%%%%%%%%%%%%%%%%%%%%%%%%%%%%%%%%%%%%%%%%%%
\section{\label{sec:populations} MBHs populations and detectability prospects}
%%%%%%%%%%%%%%%%%%%%%%%%%%%%%%%%%%%%%%%%%%%%%%%%%%%%%%

\begin{figure*}[h!]
    \centering
    % trim={<left> <lower> <right> <upper>}
    \includegraphics[width=\textwidth, trim={0.0cm 0.0cm 0.0cm 0.0cm}, clip]{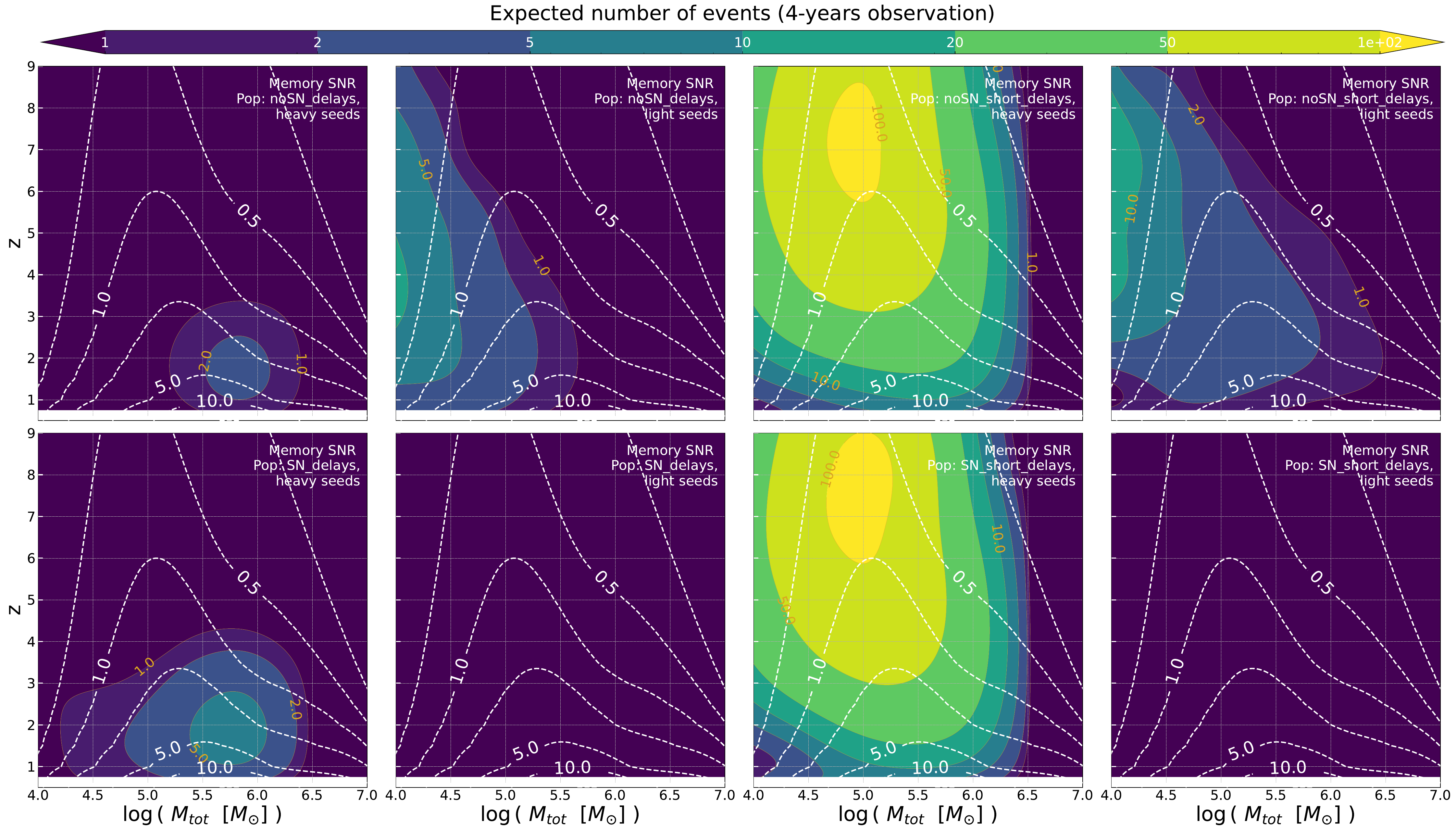}
    \caption{Overlay of the contours of the constant \gls{snr} of the memory for the \textit{conservative} scenario (baseline 1) and the average number of mergers expected in 4 years for the different astrophysical population models described in~\cite{Barausse:2020gbp, Barausse:2020mdt}. The first row shows the results for populations with SN feedback with ``delays'' or ``short delays'' for either the Heavy Seeds (HS) or the Light Seeds (LS). The second row is the same, but for populations without SN feedback.  }
    \label{fig: waterfall_MvsZ_Gal4y_Conservative_WithPop}
\end{figure*}

\begin{figure*}[h!]
    \centering
    % trim={<left> <lower> <right> <upper>}
    \includegraphics[width=\textwidth, trim={0.0cm 0.0cm 0.0cm 0.0cm}, clip]{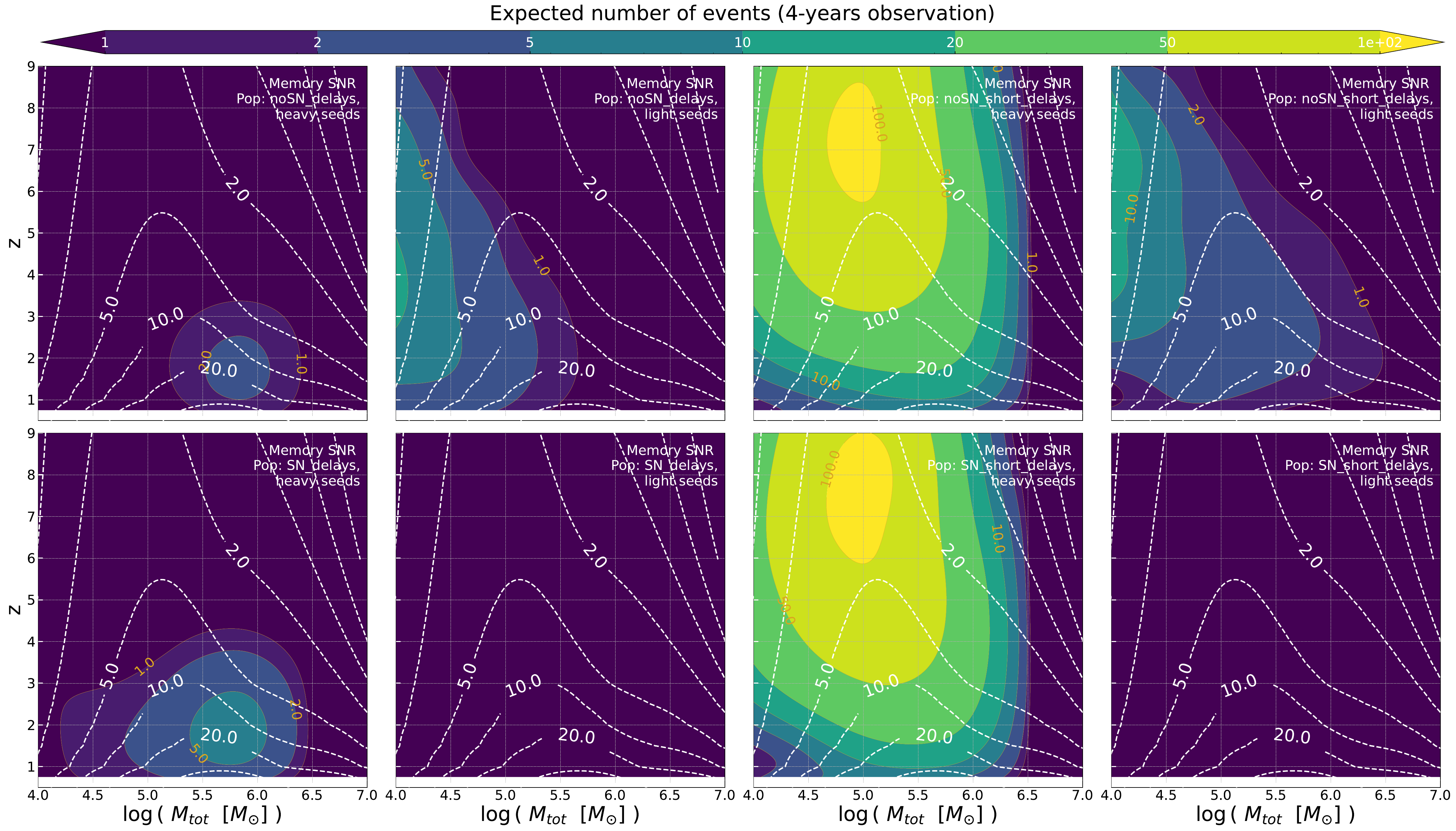}
    \caption{As Figure~\myhyperref{fig: waterfall_MvsZ_Gal4y_Conservative_WithPop}, but for the \emph{optimistic} scenario (baseline 2).  }
    \label{fig: waterfall_MvsZ_Gal4y_Optimistic_WithPop}
\end{figure*}

With the results described in the previous sections, we are now in a position to study the number of mergers with detectable memory expected for LISA. To do this,  we consider eight different astrophysical models of \glspl{mbhb} mergers described in Refs.~\cite{Barausse:2020gbp, Barausse:2020mdt}, each of them considering a different choice for one of the main astrophysical uncertainties affecting the evolution of MBHs, which leads to rather different LISA event rates. The first uncertainty concerns the initial mass function 
for the ``seeds'' of the MBH population [``light seeds'' (LS) of population III stars,  
or ``heavy seeds'' (HS) from the direct collapse of protogalactic gas disks], which primarily affects the final mass distribution of the population and thus the loudness of the mergers in the LISA band. Second, different time delays between the galaxy merger and the corresponding BBH mergers are considered: ``delayed'' models are more realistic and try to model processes at parsec distances (such as stellar hardening, triplet interactions, etc.); ``short delays'' neglect these and simply account for dynamical friction between halos. Since, for more realistic delays, the bulk of the mergers is shifted to lower redshifts, these models are the most promising ones in terms of detectability. Third, ``SN'' models include the effect of supernovae, which tends to inhibit the accretion of the MBH, so these models have lower final masses compared to ``noSN'' models. 

Our main results are presented in Figures~\myhyperref{fig: waterfall_MvsZ_Gal4y_Conservative_WithPop} and \myhyperref{fig: waterfall_MvsZ_Gal4y_Optimistic_WithPop} for the \emph{conservative} and \emph{optimistic} baselines of Table \myhyperref{table: baselines}, where we overlay the contour plots of the \gls{snr} to the contour plots of the number of mergers for the astrophysical populations in the mass-redshift plane, expected for $4$ years of LISA observations. 
In the \emph{conservative} scenario (cf. {\it baseline 1} in Table \myhyperref{table: baselines}), as in the previous section, we fix the sky direction corresponding to the average \gls{snr}, but for computing the waterfall-type plot we take the median of the distributions of the inclination $\iota$  and mass ratio $q$. The median is the middle value of the sample, so it tells us that half of the mergers will have a lower inclination (or mass ratio) and the other half a greater inclination (or mass ratio). If the direction of the angular momentum of the binary is random, then the inclination angle defined between $0\leq\iota\leq \pi/2$ follows a $\cos(\iota)$ distribution\footnote{A pedagogical explanation can be found in \url{https://keatonb.github.io/archivers/uniforminclination}.}  whose median is $\iota \approx \SI{1.047}{\radian}$. 
For the mass ratio, we find that, in all the different astrophysical populations considered, the median is $q\approx2.5$, as opposed to the average value which varies for the different populations, since the mass ratio distributions are skewed and different, especially between the ``heavy'' and ``light'' seeds. The spin distributions are instead symmetric around zero. Therefore, the mean and the median are close to each other, and we consider no spin in the average scenario. In the \emph{optimistic} scenario (cf. {\it baseline 2} in Table \myhyperref{table: baselines}), we consider an edge-on system with optimal sky-position, equal mass and zero spin.

As shown in Figures~\myhyperref{fig: waterfall_MvsZ_Gal4y_Conservative_WithPop} and \myhyperref{fig: waterfall_MvsZ_Gal4y_Optimistic_WithPop}, the prediction for the number of events with detectable memory depends strongly on the astrophysical model considered. We confirm previous results \cite{gasparotto_can_2023} that the most promising models are those starting with HS since, being the binaries more massive, the signals are louder. Among these HS models, the ones with short delays present many more events since the binaries start merging earlier, and we find no big effect of the SN feedback for those. On the other hand, the prospects are less promising for LS models, where the signals are weaker, and even less so for models with SN feedback, where the MBHs are lighter. 

%%%%%%%%%%%%%%%%%%%%%%%%%%%%%%%%%%%%%%%%%%%%%%%%%%%%%%%%%
%%%%%%%%%%%%%%%%%%%%%%%%%%%%%%%%%%%%%%%%%%%%%%%%5%%%%%%%%%
\section{\label{sec:q_and_spin} Mass ratio and spin dependence}
%%%%%%%%%%%%%%%%%%%%%%%%%%%%%%%%%%%%%%%%%%%%%%%%%%%%%%%%%
%%%%%%%%%%%%%%%%%%%%%%%%%%%%%%%%%%%%%%%%%%%%%%%%5%%%%%%%%%

\begin{figure}[h!]
    \centering
    \includegraphics[width=1.0\columnwidth, trim={0.2cm 0.0cm 0.2cm 0.0cm}, clip]{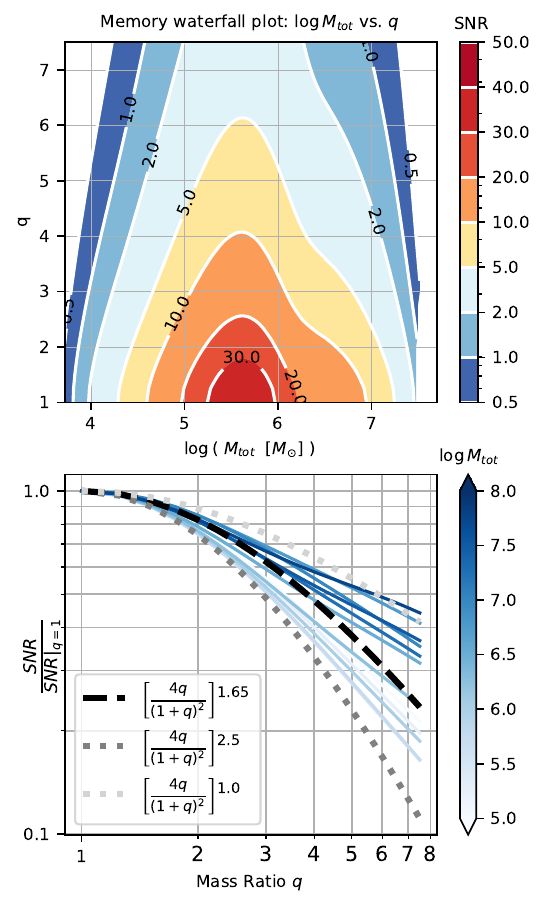}
    % trim={<left> <lower> <right> <upper>}
    \caption{Study of the mass ratio dependence of the memory \gls{snr}, using the baseline $2$ in Table \myhyperref{table: baselines} (apart from $q$ obviously variable here). 
    %\new{suggesting that the memory extraction strategy of Eq.~(\myhyperref{eq: surr_20}) is not valid for $q \gtrsim 2.0$\textbf{maybe change this part}}. 
    \new{On the top panel: contour plot of the memory \gls{snr} as a function of total mass $M_{tot}$ ($x$-axis) and mass ratio $q$ ($y$-axis)}.
    % We have noticed an excess of signal when computing memory from {CCE} surrogate waveforms and Eq.~(\myhyperref{eq: surr_20}) given by the bump visible at masses $M_{tot}\gtrsim 10^6 M_\odot$ and high mass ratio.
    On the lower panel: dependence of the \gls{snr} with the mass ratio $q$  for a set of system total masses. This is compared with the scaling $\sim [4q/(1+q)^2]^{\alpha}$ with $1\leq\alpha\leq2.5$, where the $\alpha=1$ is the Post-Newtonian expectation and $\alpha=1.65$ was found in Ref.~\cite{2021PhRvD.103d3005L}.}
\label{fig: waterfall_MvsQ}
\end{figure}

In this section, we investigate the dependence of the \gls{snr} on the mass ratio and the total spin for different MBHs masses. The dependence of the memory on the mass ratio has been previously studied in~\cite{2021PhRvD.103d3005L,Islam:2021old, Favata:2008yd,Talbot:2018sgr,Zhao:2021hmx}, where it was shown that the final amplitude of the memory strongly decreases for asymmetric configurations. \new{This trend is verified in Figure~\ref{fig: waterfall_MvsQ}, showing \gls{snr} contours as a function of binary total masses $M_{tot}$ and mass ratio $q$. The \gls{snr} peaks about $\log{M_{tot}}=5.7$ and $q=1$ and decreases monotonously with increasing values of $q$, while we recognize again in the horizontal behavior the frequency dependence of LISA sensitivity.}
% In Figure~\ref{fig: waterfall_MvsQ}, we present the \gls{snr} as a function of $(q,M_{tot})$ for the memory computed through Eq.~\eqref{eq:memorymodes} with \texttt{GWMemory} package, and we underlay in dashed lines the computation using Eq.~\eqref{eq: surr_20}. We found some discrepancy between the memory \gls{snr} calculated \new{with Eq.~\eqref{eq:memorymodes} and Eq.~\eqref{eq: surr_20} as it can be seen from the bump appearing at masses $M_{tot}\gtrsim 10^6 M_\odot$ and high mass ratio when the memory is computed with Eq.~\eqref{eq: surr_20}}. As explained in section~\ref{subsection: mem model}, this may be because the two waveforms \texttt{NRHybSur3dq8\_CCE} and \texttt{NRHybSur3dq8} are not calculated in the same way, so that subtraction of the two leaves a residual feature at high frequencies which is masked at low masses but becomes relevant at higher masses. However, this difference may also (partially) have a physical origin, as NR waveforms return slightly higher memory and high-frequency features in the merger/ringdown. 
% %\new{To be conservative, for mass ratio values of $q\geq2$ we compute the memory through Eq.~\eqref{eq:memorymodes}, whose overall \gls{snr} is smaller and does not present the bump present in Figure~\ref{fig: waterfall_MvsQ} \textbf{maybe change?}}.  
On the lower panel of Figure~\ref{fig: waterfall_MvsQ}, we show that the scaling of the memory \gls{snr} with the mass ratio $q$ is approximated by $\sim [4q/(1+q)^2]^{\alpha}$ with \new{$1\leq\alpha\leq 2.5$}, where $\alpha=1$ is the Post-Newtonian expectation and $\alpha=1.65$ was found in Ref.~\cite{2021PhRvD.103d3005L} for the final memory offset. We want to highlight that for light SMBHs, for which we expect a bigger impact of the memory on the total waveform, the memory decreases faster with mass ratio, strongly decreasing the possibility of detecting it far from the equal mass case. 
Instead, for higher total mass, the \gls{snr} scaling becomes less severe, because even if the amplitude of the signal decreases with $q$, the high-frequency cut-off of the memory shifts closer to the LISA high sensitivity frequency bins, hence mitigating the \gls{snr} loss.

\begin{figure}[t!]
    \centering
    % trim={<left> <lower> <right> <upper>}
    \includegraphics[width=1.0\columnwidth, trim={0.0cm 0.0cm 0.0cm 0.0cm}, clip]{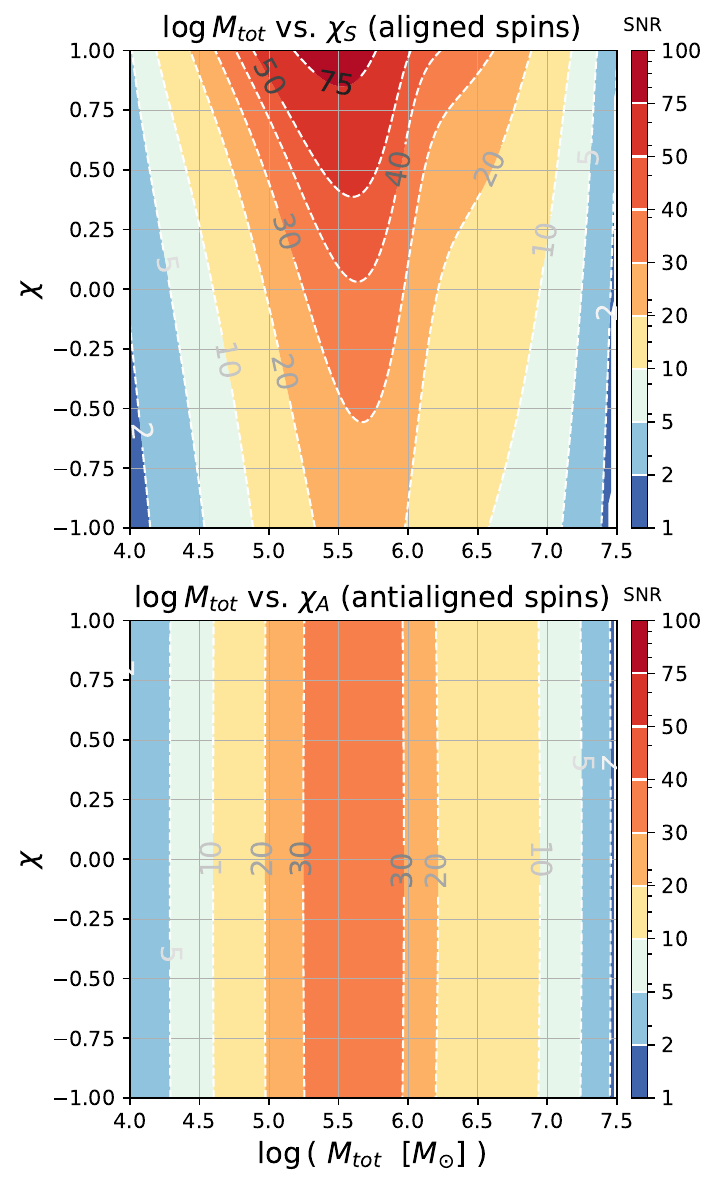}
    \caption{Study of the dependence of the memory \gls{snr} on the total amplitude of the spin for the aligned $\chi_{1,z}=\chi_{2,z}=\chi$ and anti-aligned $\chi_{1,z}=-\chi_{2,z}$ configurations. We use the baseline $2$ in Table \myhyperref{table: baselines} (apart from $\chi$ variable here).}
    \label{fig: waterfall_MvsS}
\end{figure}

The dependence of the memory on the spins was studied in \cite{Cao_2016,2021PhRvD.103d3005L,Pollney:2010hs,Islam:2021old}, where it was shown that the final amplitude of the memory increases monotonically with the symmetric spin $\chi_S=(m_1\chi_{1,z}+m_2\chi_{2,z})/M_{tot}$ and is independent of the antisymmetric combination $\chi_A=(m_1\chi_{1,z}-m_2\chi_{2,z})/2M_{tot}$, at least for the equal mass case. We confirm this dependence in Figure~\ref{fig: waterfall_MvsS}, where we compute the \gls{snr} of the memory for an equal mass case as a function of $M_{tot}$ for different values of the individual spin, more precisely the spin component along the angular momentum $|\chi_{1,z}|=|\chi_{2,z}|=\chi$. As expected, the \gls{snr} of the memory just depends on the symmetric spin configuration, i.e. $\chi_{1,z}=\chi_{2,z}=\chi$, and has a maximum when the spins are aligned with the binary angular momentum. In this case, the \gls{snr} is about three times bigger than in the anti-aligned case. On the other hand, the memory \gls{snr} does not change when $\chi_{1,z}=-\chi_{2,z}$ since the symmetric spin is always zero for equal mass binary.

%%%%%%%%%%%%%%%%%%%%%%%%%%%%%%%%%%%%%%%%%%%%%%%%%%%%%%%
\section{\label{discussion} Discussion and Conclusion}
%%%%%%%%%%%%%%%%%%%%%%%%%%%%%%%%%%%%%%%%%%%%%%%%%%%%%%%
% \new{\textbf{We have to update the text regarding the memory generation}}

In this paper, we have extensively studied the imprint of gravitational (displacement) memory on the LISA detector. For the first time, we have simulated the TDI response of the detector to the passage of the \gls{gw} memory contained in the (2,0) mode.
% \new{\sout{of the \texttt{NRHybSur3dq8\_CCE} surrogate model}}. 
As the \gls{lisa} \gls{tdi} response to the raw strain behaves as a third order differentiator, the memory signal appears as a burst-like event rather than a persistent offset. This burst-like behavior is evident in a time-frequency plot, where the memory signal becomes prominent just near the merger time. Therefore, a long waveform is not necessary to accurately calculate the memory \gls{snr}. Due to the shorter duration of the signal, the memory \gls{snr} is more dependent on the sensitivity of the detector to that particular direction of the sky, as opposed to longer signals where the effect of sky localization is averaged along the orbital motion of LISA. Indeed, we find that for \gls{mbhb} mergers with different sky localizations, the memory \gls{snr} can vary by a factor of $\mathcal{O}(10)$. We then study the dependence of the memory \gls{snr} on the intrinsic parameters of the binary, such as mass and redshift, for 
three different scenarios with increasing levels of optimism \emph{conservative}, \emph{optimistic} and \emph{optimistic with spin}, described in Table \ref{table: baselines}.

We also compare the \gls{snr} results of eight different astrophysical populations of SMBHs to estimate the number of expected detectable sources and their parameter space distribution. Our results indicate that the detection prospects vary significantly with the population model, with heavy seeds models showing promising results and light seeds models yielding fewer detectable sources. Future work should explore these results further, especially for populations that better match the pulsar timing array signal as interpreted from the inspiral of SMBHs \cite{Barausse:2023yrx}.

% In this work, we isolate the memory component of the waveform by subtracting the (2,0) mode of the \texttt{NRHybSur3dq8}, which does not capture the memory, from that of the \texttt{NRHybSur3dq8\_CCE} and compare it with the memory calculated from the \texttt{NRHybSur3dq8} waveform using the \texttt{GWMemory} package. We found that this procedure isolates the memory from the oscillatory contributions excited during the merger and ringdown present in the (2,0) mode of the \texttt{NRHybSur3dq8\_CCE} model close to the equal mass ratio, but leaves some oscillatory residuals at higher values of $q$ that affect the \gls{snr} calculation as shown in Section \ref{sec:q_and_spin}. Therefore, care should be taken when using this method to separate the memory from the rest of the waveform. We leave the discussion of whether these differences have numerical or physical origin for the future (they are practically irrelevant to the \gls{snr} studies of this work).

Data analysis issues regarding the extraction of \gls{gw} memory signal from LISA \gls{tdi} data will be an immediate extension of this work. The end-to-end time-domain simulation of the detection of the full \gls{imr} waveform, including memory, is readily available for designing and testing data analysis strategies. Two approaches can be pursued. First, one can
%\new{\sout{exploit Eq.~(\myhyperref{eq:memorymodes}) or the \texttt{GWMemory} package \cite{talbot_gravitational-wave_2018} to}}
jointly fit the memory component, \new{as computed from Eq.~\eqref{eq:memoryequation}}, together with the oscillatory component \cite{garcia-quiros_multimode_2020}, hence evaluating the evidence of a memory signal detection (Bayes factor and hypothesis test). On the other hand, one can instead use an agnostic approach, combining a parameterized waveform model \cite{garcia-quiros_multimode_2020} together with a template-free time-frequency representation (e.g. wavelets), optimized for capturing low-frequency burst-like components such as the {\it memory signal}.

Other possible extensions of this project include investigating the effect of memory on binary parameter estimation and identifying regions of parameter space where neglecting memory leads to biases, similar to the analysis performed in \cite{pitte_detectability_2023} for higher harmonics. In addition, we could extend our study to subdominant memory components, such as \textit{spin} memory. Our results are also relevant to the use of \gls{gw} memory to perform consistency checks complementary to those performed with ringdown. Since the memory depends on the energy flux at infinity, it could potentially test additional radiation channels excited during the merger. Indeed, modified theories beyond General Relativity can affect the memory in both tensor perturbations of the metric and additional polarizations, as shown in \cite{Heisenberg:2023prj,Heisenberg:2024cjk}. 
Finally, \emph{linear} memory from hyperbolic encounters may also leave an imprint in the LISA band, see e.g. \cite{Caldarola:2023ipo}. This low-frequency part of the waveform may be sensitive to the presence of dissipation
during the encounter \cite{Grobner:2020drr,Dandapat:2023zzn}. We leave for future work a more detailed study of this possibility.

Having derived the first realistic estimates of LISA's sensitivity to \gls{gw} memory, we are convinced that this subtle effect of General Relativity will play a relevant role in the scientific results of future \gls{gw} missions. 

\section{\label{acknowledgments} Acknowledgments}

The authors would like to thank M. Besançon, J. Garc\'ia Bellido, M. Maggiore, A. Petiteau, C. Pitte, and L. Magaña Zertuche for their precious insight and the fruitful discussions we had during this project. The authors acknowledge E. Barausse for his help regarding \gls{mbhb} population models and catalogs, which were critical inputs for this work. The authors also thank the LISA Simulation Working Group and the LISA Simulation Expert Group for the lively discussions on all simulation-related activities.

The research leading to these results has received funding from the Spanish Ministry of Science and Innovation (PID2020-115845GB-I00/AEI/10.13039/501100011033).
IFAE is partially funded by the CERCA program of the Generalitat de Catalunya. SG has the support of the predoctoral program AGAUR FI SDUR 2022 from the Departament de Recerca i Universitats from Generalitat de Catalunya and the European Social Plus Fund. D. Blas acknowledges the support from the Departament de Recerca i Universitats from Generalitat de Catalunya to the Grup de Recerca 00649 (Codi: 2021 SGR 00649).  

LH would like to acknowledge financial support from the European Research Council (ERC) under the European Union Horizon 2020 research and innovation programme grant agreement No 801781. LH further acknowledges support from the Deutsche Forschungsgemeinschaft (DFG, German Research Foundation) under Germany’s Excellence Strategy EXC 2181/1 - 390900948 (the Heidelberg STRUCTURES Excellence Cluster). The authors thank the Heidelberg STRUCTURES Excellence Cluster for financial support.

ST is supported by the Swiss National Science Foundation (SNSF) Ambizione Grant Number: PZ00P2-202204.

JZ is supported by the Swiss National Science Foundation (SNSF) through a Postdoc.Mobility Fellowship (Grant No. P500PT-222346).

Computations were performed on the DANTE platform, APC, France.

%\clearpage

\appendix

%%%%%%%%%%%%%%%%%%%%%%%%%%%%%%%%%%%%%%%%%%%%%%%%%
\new{
\section{Details on the definition and modeling of memory}\label{App:memorymodeldetails}
}

\new{In this appendix, we provide useful details on the practical evaluation of Eq.~\eqref{eq:memoryequation}, as well as its use for a viable definition of memory to claim a detection of the phenomenon.}\\

\new{
\subsubsection{Memory in the $(2,0)$ mode}
\label{App:meory is in the 20 mode}
}

\new{The claim that non-linear memory of non-precessing and non-eccentric binary black hole coalescence's is predominantly found in the $(2,0)$ mode follows directly from its formula in terms of a spin-weighted mode decomposition by explicitly evaluating the angular integral within Eq.~\eqref{eq:memoryequation}. Such an analytic computation of the angular integral is rendered possible through a translation of the TT projected factor into spherical harmonics as for instance shown in paragraph IV (C) of Ref.~\cite{Heisenberg:2023prj}, resulting in the explicit formula \cite{Favata:2008yd,Zosso:2024xgy}
\begin{equation}
    h^{\ell m}_{mem}=\frac{R}{c}\sum_{\ell',\ell''\geq 2}\, \sum_{m',m''} \Gamma^{l'm'm''l''}_{lm} \int_{-\infty}^u \mathrm{d}u'\langle\dot{h}_0^{\ell' m'} \dot{h}_0^{*\ell'' m''}\rangle,
\label{eq:memorymodes}
\end{equation}
with
\begin{widetext}
\begin{equation}
    \Gamma^{l'm'm''l''}_{lm}\equiv (-1)^{m+m''}\sqrt{\frac{(l-2)!}{(l+2)!}} \sqrt{\frac{(2l'+1)(2l''+1)(2l+1)}{4\pi}}
    \begin{pmatrix}
        l' & l'' & l\\
         m' & -m'' & -m
    \end{pmatrix}
    \begin{pmatrix}
        l' & l'' & l\\
        2 & -2 & 0
    \end{pmatrix},
\end{equation}
\end{widetext}
where the big parenthesis represent $3-j$ symbols, which in this case are only non zero if $m=m'-m''$ and $|l'-l''|\leq l\leq l'+l''$.

\new{While Eq.~\eqref{eq:memorymodes} provides the formula for the memory modes with the least amount of numerical computation needed, it is equivalent to the output of the publicly available and numerically optimized \texttt{GWMemory} package~\cite{Talbot:2018sgr}, as both correspond to a spin-weighted expansion of Eq.~\eqref{eq:memoryequation}.}

From the selection rule above it becomes evident that for a quasi-circular and non-precessing signal, for which} most of the energy in \glspl{gw} is released into the $(2,2)$ mode \new{of the primary wave}, the memory is predominantly found in \new{the $m=0$ modes with $2\leq l \leq 4$. Moreover, compared to the $(2,0)$ memory mode, the $(4,0)$ mode is suppressed by two orders of magnitude while the $(3,0)$ memory mode vanishes entirely due to the symmetry over the orbital plane \cite{Zosso:2024xgy}, such that in the end the memory of non-precessing binary black hole coalescence's predominantly} appears in the $(2,0)$ mode~\cite{Favata:2009ii,dambrosio_testing_2024}.

\new{To be more precise, the conclusion that all modes other than the $(2,0)$ mode can be disregarded in Eq.~\eqref{eq:memorymodes} is more complex, since with increasing accuracy also other modes than the $(2,2)$ mode of the primary waves become important. However, due to the presence of the spacetime averaging over the scales of variation of primary waves, this conclusion will not change, as we will further discuss below.}  \\

\new{
\subsubsection{Evaluation of the spacetime averaging}
\label{App:Evaluation of the averaging}
}

\new{But first, we want to discuss the implications of the averaging for the dominant $(2,0)$ mode of memory. It turns out that for $m=0$ modes the average can effectively be dropped in the case of quasi-circular and non-precessing binaries\footnote{\new{See also the discussion in App. C of \cite{Favata:2011qi}.}}

To understand this statement it is first necessary to know that for a typical setting relevant for \gls{gw} detection and after having identified the gauge-invariant contribution to the \gls{gw} energy-flux, the spacetime average $\langle ...\rangle$ can be reduced to a purely temporal average over orbital time-scales (see Sec. 1.4.3 in \cite{maggiore2008gravitational}). If in addition the integrand in Eq.~\eqref{eq:memoryequation} does not vary significantly over these scales of the primary waves, the average can effectively be disregarded. Notice that this is precisely the case for memory sourced by the primary $(2,2)$ mode only, since the dependence on the orbital phase of the primary modes cancels within the integrand in Eq.~\eqref{eq:memorymodes} for $m=0$ modes. Thus, for quasi-circular and non-precessing compact binary coalescence's any explicit spacetime average over orbital scales within the computation of the $(2,0)$ mode of memory through Eq.~\eqref{eq:memorymodes} can be dropped, as it is common practice in the community.
}\\

\new{
\subsubsection{Definition of memory and claim of detection}
\label{App:Averaging and claiming detection}
}
\new{However, one should nevertheless keep in mind that in defining non-linear memory, the average in Eqs.~\eqref{eq:memoryequation} and \eqref{eq:memorymodes} is in fact essential. \newJZ{This is because the automatic cancellation of the oscillations on orbital time-scales within the memory modes is only approximate and can not be relied on in general, in particular in the case of eccentric and precessing binaries for instance.} Yet, a sharp separation in scales of variation between non-linear memory and the primary waves is important in order to be able to talk about memory as a clearly distinct component of the total radiation that can individually be searched for.}

\newJZ{In fact, even for the quasi-circular and non-precessing binaries addressed in the present work, the average is necessary to identify the $(2,0)$ mode as the only relevant mode of the memory signal. This is because with increasing accuracy of detectors it is actually not true that all modes beyond the dominant $(2,2)$ mode of the primary wave can simply be discarded. Indeed, when computing the memory through the \texttt{GWMemory} package we ought to consider all the modes available within the \texttt{NRHybSur3dq8} surrogate model, as otherwise the resulting memory would be significantly off. This in turn implies that also additional memory modes in Eq.~\eqref{eq:memorymodes} are present that would contain oscillatory features if the averaging would simply be dropped. What is more, the oscillatory features within these additional modes would actually impact our SNR computations for certain regions of the parameter space. Thus, the modes computed through Eq.~\eqref{eq:memorymodes} with $m\neq0$ are only truly insignificant if an explicit averaging is considered that cancels such oscillations on orbital timescales. In other words, the statement that it is sufficient to only compute the $(2,0)$ mode within Eq.~\eqref{eq:memorymodes} actually necessitates the presence of the space-time averaging. In the case of precessing and eccentric binaries, such a distinction between oscillatory features and memory within Eq.~\eqref{eq:memorymodes} will be even more pressing, since also the $m\neq 0$ modes will contain non-negligible memory components.

We want to stress, however, that such oscillatory features within the asymptotic radiation that are sourced by the primary waves are not un-physical. In the language of balance laws~\cite{Ashtekar:2019viz,Khera:2020mcz,Mitman:2020bjf,Zhao:2021hmx,Zhao:2021hmx,Sun:2022pvh}, they belong to the \textit{null} part of the total strain, in contrast to the familiar oscillatory contributions that are sourced by the movement of masses, such as the dominant $(2,2)$ mode or the oscillations within the total $(2,0)$ mode shown in Fig.~\ref{fig: surr_cce} associated to the ring-down of the final black-hole, which represent the \textit{ordinary} part of the radiation. Yet, none of the oscillations on orbital and ring-down time-scales are to be considered as part of the \gls{gw} displacement memory, regardless whether they appear within the null part or the ordinary part of the waveform.}\footnote{\newJZ{It is important to note that strictly speaking, the BMS balance laws do not allow for an isolation of a displacement memory component as an independent entity within the time-varying waveform. The balance-laws only provide an unambiguous identification of the total final offset of memory. A separation into the null and ordinary parts of the total time-dependent strain through the balance laws only provide an approximate isolation of the displacement memory in the case of non-precessing and non-eccentric binaries, since in this case almost all the memory is excursively present within the null part of the asymptotic radiation, while the oscillations on orbital and ring-down time-scales are mostly found in the ordinary piece. In more general cases, an averaging procedure would however be inevitable to isolate a time-dependent non-linear memory signal (see also App. C in \cite{Heisenberg:2023prj}).}}

\new{Indeed, a true measurement of non-linear displacement memory that is associated with the profound theoretical consequences outlined in the introduction\footnote{\new{In particular, \gls{gw} memory provides an empirical window into fascinating questions on fundamental physics through its relation to asymptotic symmetries as well as the soft theorems of scattering theory.}} requires the claim of detecting its defining property of a permanent distortion of spacetime after the passage of gravitational radiation. However, of course current GW detectors designed to measure changes in relative distance at a restricted frequency band are not capable of measuring such a permanent modification of proper distances but rather have only access to the sharp raise of non-linear memory with a characteristic time scale of the merger, as clearly shown if Fig.~\ref{fig: mem_tdi}. In this context, for a claim of detection of memory it is crucial to ensure that the memory signal, which appears as an additional oscillatory feature within the detector response of LISA, can unambiguously be associated to an offset in proper distances between freely falling test masses. Thus, in a search for memory, any oscillatory feature on orbital or ringdown time-scales that are not associated with the raise of displacement memory, need to be disregarded by demanding a sharp separation in scales of variation between memory and the primary waves. This is precisely ensured through an explicit averaging within Eqs.~\eqref{eq:memoryequation} and \eqref{eq:memorymodes}.}\\

 %%%%%%%%%%%%%%%%%%%%%%%%%%%%%%%%%%%%%%%%%%%%%%%%%%%%%%%%%%%%%%%%%%%%%%%%%%%%%%%%%%%%%%%%%%%%%%%

\section{TDI memory response in the long wavelength limit}

In this section, we want to provide a brief analytical understanding of why the TDI output of the memory signal in Figure \ref{fig: mem_tdi} looks like a burst-like event compared to the naively expected step-like function. For this, we need to understand how the TDI $X$ variables are related to the original \gls{gw} strain $h$. This is given by a complicated combination of the single-link data streams shifted by several delay operators, as given in Eq. \eqref{equ:X2}. In the low-frequency limit $\omega L\ll1$, i.e. for frequencies $\omega$ much smaller than the inverse of the detector arm $L$, one can show the following relationship between the output of the first generation TDI combination $X_{1.5}$ and the \gls{gw} strain $h$ (see Eq. (51) of Ref.~\cite{Babak:2021mhe})
\begin{equation}
    X_{1.5}\approx 2L^2(\epsilon_{12}^a \epsilon_{12}^b-\epsilon_{13}^a \epsilon_{13}^b)\partial_t^2h_{ab},
\end{equation}
\new{where $\epsilon_{ij}$ is the unit vector from spacecraft $j$ towards spacecraft $i$.}
The second generation of TDI combination $X_{2}$ is related to the first generation by
\begin{equation}
    |X_{2}|=2\sin(2\omega L)|X_{1.5}|
\end{equation}
so, in the low-frequency limit $|X_{2}|\approx4\omega L|X_{1.5}|$, and we recognize that in frequency space this is simply the derivative of $X_{1.5}$. Therefore, $X_{2}$ is proportional to an additional derivative of the strain $X_{2}\propto \partial^3_th$.  

\begin{figure}[h!]
    \centering
    \includegraphics[scale=0.35]{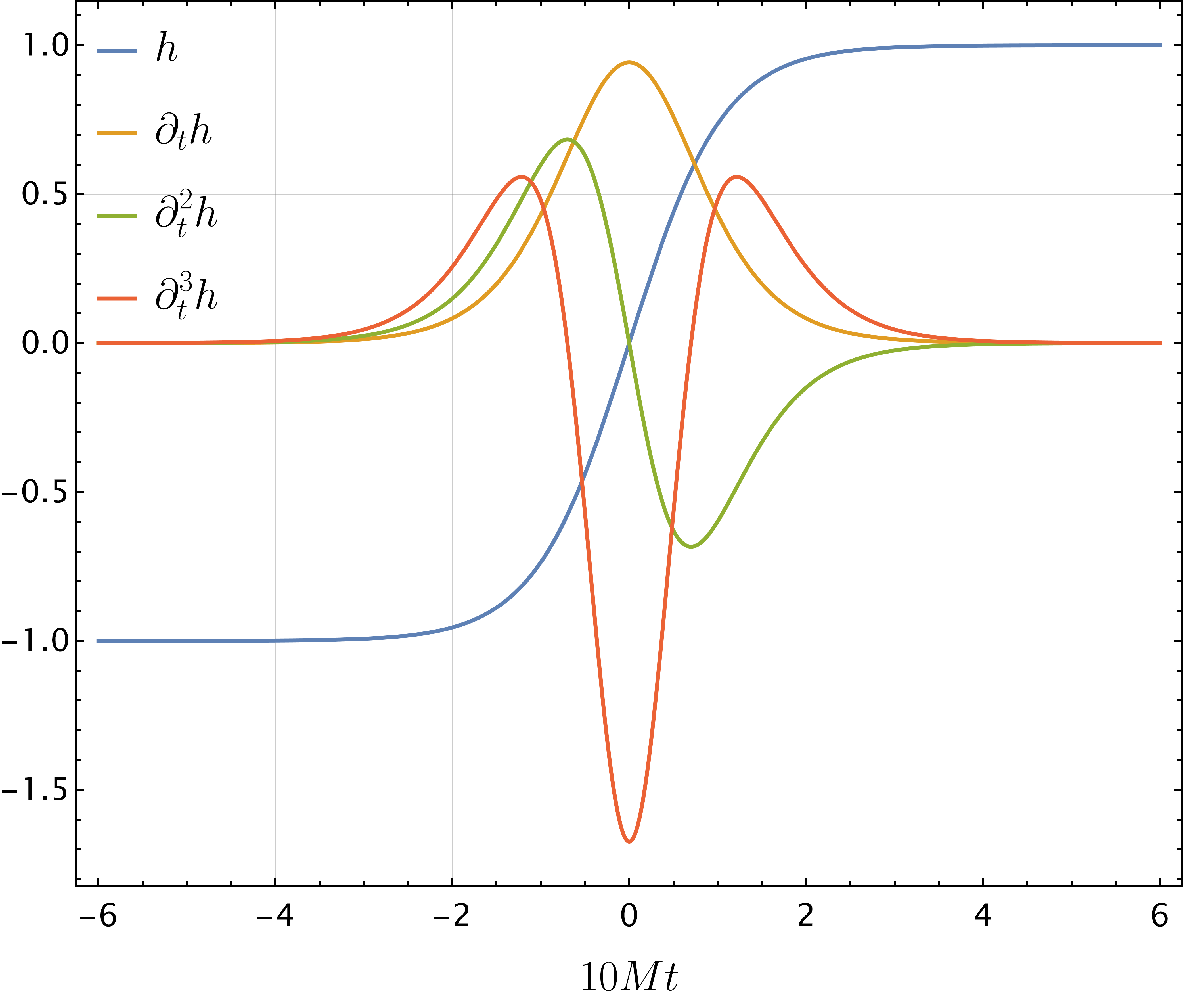}
    \caption{Visualisation of the step-like behaviour of the memory approximated by the hyperbolic tangent, Eq.~\eqref{eq:tanh}, and its first, second and third derivatives. We take $t_c=0$ and $\Delta T= 60M_{tot}$ as explained in the main text. }
    \label{fig:ploApp}
\end{figure}

Going back to the step-like behavior of the memory signal, this can be approximated by a hyperbolic function with time raise $\Delta T= 60 M_{tot}$, as explained in Section \ref{sec:mem_lisa},
\begin{equation}\label{eq:tanh}
    h(t)=\tanh\left(\frac{2\pi(t-t_c)}{\Delta T}\right),
\end{equation}
with $t_c$ the merger time. We show this function in Figure \ref{fig:ploApp} with its higher derivatives up to the third, noting that they all go to zero sufficiently far from the merger. In particular, note how the third derivative (red line) resembles the TDI result of the memory signal shown in Figure \ref{fig: mem_tdi}.   

\label{sec:app}

\section{Comparison of the memory between Surrogate\_CCE and \texttt{GWMemory} package}
\label{appendix: comp surrmem gwmem}

\begin{figure}
	\centering
	\includegraphics[width=\columnwidth]{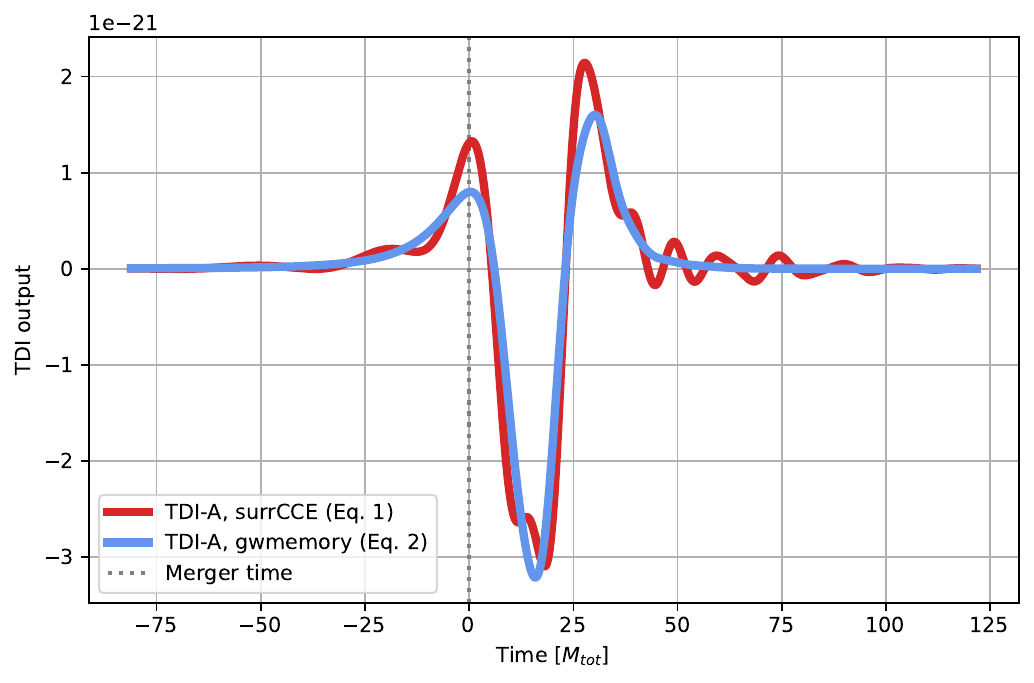} 
	\caption{Comparison of the TDI-A response for the same source of Figure \ref{fig: spectra}, for the memory computed through Eq.\eqref{eq: surr_20} (red) and the \texttt{GWMemory} package estimate from the \texttt{NRHybSur3dq8} waveforms (blue). }
	\label{fig:comparison2}
\end{figure}

\begin{figure}[h!]
	\centering
   % \hspace{-1cm}
	\includegraphics[width=0.9\columnwidth]{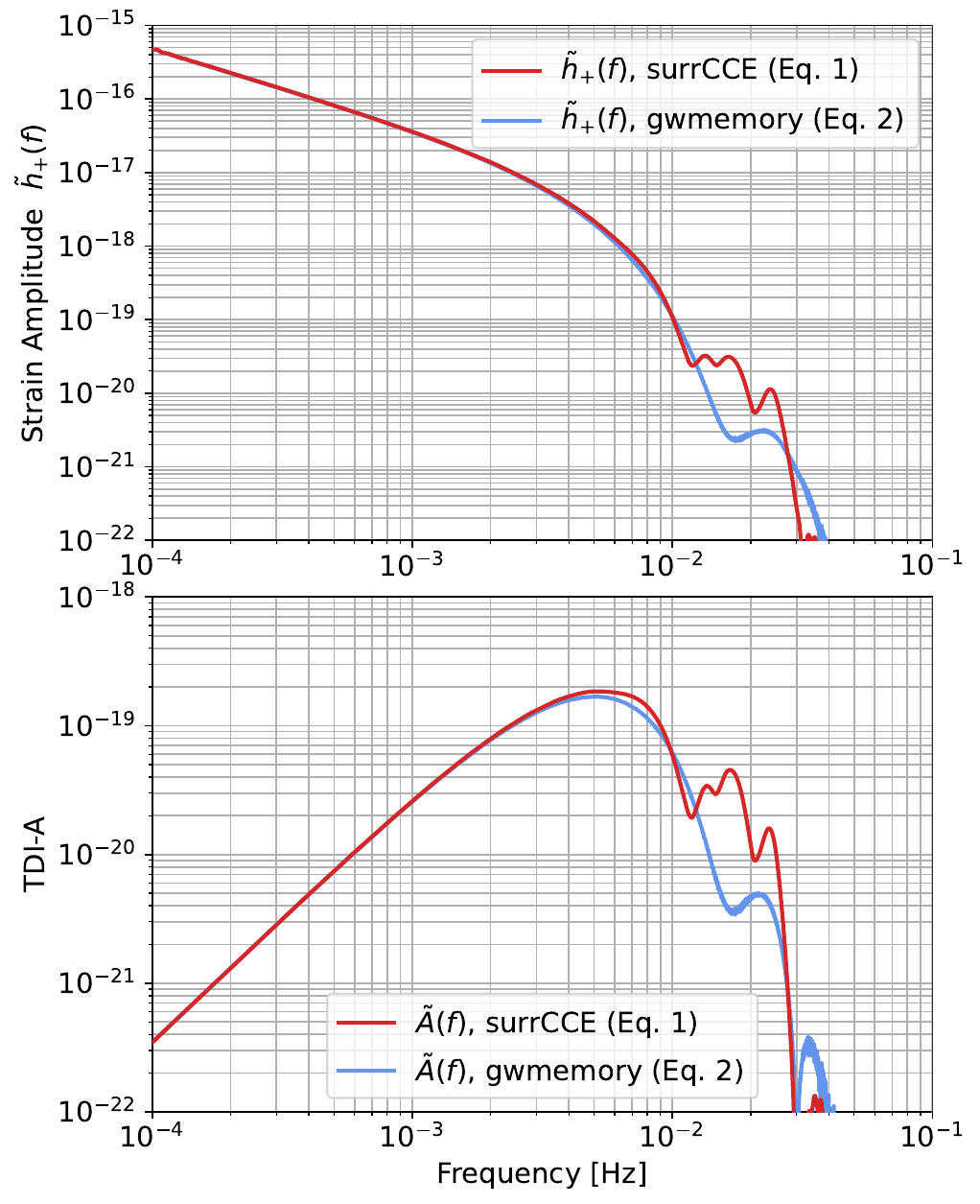} 
	\caption{As in Figure \ref{fig:comparison2} for the comparison of the strain amplitude and TDI-A response, here in the frequency domain. In the first method, the FT of the memory presents additional peaks at high frequencies, about the peak frequency of the (2,2) mode of the primary signal. At the level of computing the \gls{snr}, for this source, the mismatch between these two waveforms is smaller than $\mathcal{O}(1)$, thus not affecting the prediction on detectability. }
	\label{fig:comparison1}
\end{figure}

As mentioned in the main text, we found some differences between the \gls{gw} memory computed through Eq.~\eqref{eq: surr_20}, which subtracts the (2,0) mode of \texttt{NRHybSur3dq8\_CCE} to that of  \texttt{NRHybSur3dq8\_CCE}, and through Eq.~\eqref{eq:memoryequation} from the \texttt{NRHybSur3dq8} waveforms using the \texttt{GWMemory} package. The two differ by some oscillating features present in the first method compared to the second, as can be seen in Figure \ref{fig:comparison2}, where the smooth blue curve is the TDI of the memory calculated through the \texttt{GWMemory} package, while the red curve is calculated through Eq.~\eqref{eq: surr_20}. These oscillations are visible as small peaks in the power spectra of the signals shown in Figure \ref{fig:comparison1}, at frequencies higher than the typical frequencies of the memory. By comparing them with Figure \ref{fig: spectra} in the main text, one can see that they are just below the maximum frequency of the (2,2) mode of the primary signal. This might suggest that the discrepancy between the two has a physical origin that is better captured in the NR simulations.  However, as mentioned in the text, the two waveforms used in Eq.~\eqref{eq: surr_20} are not constructed exactly in the same way, then the comparison in Eq.~\eqref{eq: surr_20} might not be robust in general, and can give rise to some numerical error, such as residual coming from the ringdown.

\begin{figure}[h!]
    \centering
    \includegraphics[width=1.0\columnwidth, trim={0.2cm 0.0cm 0.2cm 0.0cm}, clip]{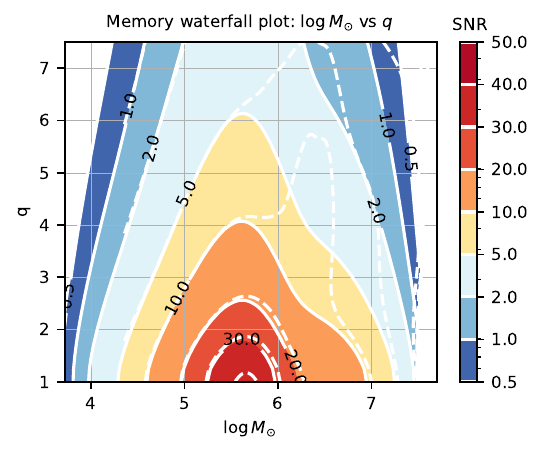}
    % trim={<left> <lower> <right> <upper>}
    \caption{Study of the $q$-dependence of the memory \gls{snr}, using the baseline $2$ in Table \myhyperref{table: baselines} (apart from $q$ obviously variable here). 
    %\new{suggesting that the memory extraction strategy of Eq.~(\myhyperref{eq: surr_20}) is not valid for $q \gtrsim 2.0$\textbf{maybe change this part}}. 
    We are comparing the \gls{snr} computed using \texttt{GWMemory} \cite{talbot_gravitational-wave_2018} (color-filled and plain-line contours) and Eq.~(\myhyperref{eq: surr_20}) (dashed-line contours). We have noticed an excess of signal when computing memory from {CCE} surrogate waveforms and Eq.~(\myhyperref{eq: surr_20}) given by the bump visible at masses $M_{tot}\gtrsim 10^6 M_\odot$ and high mass ratio.
    On the lower panel, we draw the evolution of the \gls{snr} for a set of system total masses as the mass ratio $q$ increases and we compare it with the scaling $\sim [4q/(1+q)^2]^{\alpha}$ with $1\leq\alpha\leq2.5$, where the $\alpha=1$ is the Post-Newtonian expectation and $\alpha=1.65$ was found in Ref.~\cite{2021PhRvD.103d3005L}.}
\label{fig: waterfall_MvsQ_GwMemVsSurrCCE}
\end{figure}

While rather small in the strain time-series (see left-hand side of \myhyperref{fig:comparison1}), these high-frequency features are significantly magnified by the \acrshort{lisa} response function, acting as a third-order high-pass filter (see right-hand plot of Figure \myhyperref{fig:comparison2}). We find, however, that these discrepancies do not impact sensibly \gls{snr} prospects for equal-mass sources, but have a stronger impact for higher $q$ as visible in Figure \myhyperref{fig: waterfall_MvsQ_GwMemVsSurrCCE}.
% \new{\sout{Since, at this point, we cannot robustly identify the origin of such mismatch, we take a conservative approach and compute the \gls{snr} from the  \texttt{GWMemory} when the mass ratio is $q\gtrsim2$ which returns a more conservative value, and keep numerical relativity inherited \texttt{NRHybSur3dq8\_CCE} evaluations for $q=1$ sources.}}

In Figure~\ref{fig: waterfall_MvsQ_GwMemVsSurrCCE}, we present the \gls{snr} as a function of $(q,M_{tot})$ for the memory computed through the \texttt{GWMemory} package, and we underlay in dashed lines the computation using Eq.~\eqref{eq: surr_20}. The discrepancy between the memory \gls{snr} calculated \new{with these two methods becomes obvious in this plot. In particular, we found a strange bump appearing at masses $M_{tot}\gtrsim 10^6 M_\odot$ and high mass ratio when the memory is computed with Eq.~\eqref{eq: surr_20}}. Again, we believe this is due to the two waveforms \texttt{NRHybSur3dq8\_CCE} and \texttt{NRHybSur3dq8} not being calculated in the same way, with same gauge fixing strategies \cite{Mitman:2024uss}, so that the two are not directly comparable in general. As a consequence, the subtraction of the two can leave a substantial residual at high frequencies, more pronounced for higher mass-ratio, and with brighter signatures through LISA response for higher total masses. However, this difference may also (partially) have a physical origin, as NR waveforms return slightly higher memory and high-frequency features in the merger/ringdown. 
% %\new{To be conservative, for mass ratio values of $q\geq2$ we compute the memory through Eq.~\eqref{eq:energyflux}, whose overall \gls{snr} is smaller and does not present the bump present in Figure~\ref{fig: waterfall_MvsQ} \textbf{maybe change?}}. 

\newpage

\bibliographystyle{apsrev4-1}
\bibliography{exported}

\begin{thebibliography}{108}%
\makeatletter
\providecommand \@ifxundefined [1]{%
 \@ifx{#1\undefined}
}%
\providecommand \@ifnum [1]{%
 \ifnum #1\expandafter \@firstoftwo
 \else \expandafter \@secondoftwo
 \fi
}%
\providecommand \@ifx [1]{%
 \ifx #1\expandafter \@firstoftwo
 \else \expandafter \@secondoftwo
 \fi
}%
\providecommand \natexlab [1]{#1}%
\providecommand \enquote  [1]{``#1''}%
\providecommand \bibnamefont  [1]{#1}%
\providecommand \bibfnamefont [1]{#1}%
\providecommand \citenamefont [1]{#1}%
\providecommand \href@noop [0]{\@secondoftwo}%
\providecommand \href [0]{\begingroup \@sanitize@url \@href}%
\providecommand \@href[1]{\@@startlink{#1}\@@href}%
\providecommand \@@href[1]{\endgroup#1\@@endlink}%
\providecommand \@sanitize@url [0]{\catcode `\\12\catcode `\$12\catcode
  `\&12\catcode `\#12\catcode `\^12\catcode `\_12\catcode `\%12\relax}%
\providecommand \@@startlink[1]{}%
\providecommand \@@endlink[0]{}%
\providecommand \url  [0]{\begingroup\@sanitize@url \@url }%
\providecommand \@url [1]{\endgroup\@href {#1}{\urlprefix }}%
\providecommand \urlprefix  [0]{URL }%
\providecommand \Eprint [0]{\href }%
\providecommand \doibase [0]{http://dx.doi.org/}%
\providecommand \selectlanguage [0]{\@gobble}%
\providecommand \bibinfo  [0]{\@secondoftwo}%
\providecommand \bibfield  [0]{\@secondoftwo}%
\providecommand \translation [1]{[#1]}%
\providecommand \BibitemOpen [0]{}%
\providecommand \bibitemStop [0]{}%
\providecommand \bibitemNoStop [0]{.\EOS\space}%
\providecommand \EOS [0]{\spacefactor3000\relax}%
\providecommand \BibitemShut  [1]{\csname bibitem#1\endcsname}%
\let\auto@bib@innerbib\@empty
%</preamble>
\bibitem [{\citenamefont {Zel'dovich}\ and\ \citenamefont
  {Polnarev}(1974)}]{Zeldovich:1974gvh}%
  \BibitemOpen
  \bibfield  {author} {\bibinfo {author} {\bibfnamefont {Y.~B.}\ \bibnamefont
  {Zel'dovich}}\ and\ \bibinfo {author} {\bibfnamefont {A.~G.}\ \bibnamefont
  {Polnarev}},\ }\href@noop {} {\bibfield  {journal} {\bibinfo  {journal} {Sov.
  Astron.}\ }\textbf {\bibinfo {volume} {18}},\ \bibinfo {pages} {17} (\bibinfo
  {year} {1974})}\BibitemShut {NoStop}%
\bibitem [{\citenamefont {Braginsky}\ and\ \citenamefont
  {Grishchuk}(1985)}]{Braginsky:1985vlg}%
  \BibitemOpen
  \bibfield  {author} {\bibinfo {author} {\bibfnamefont {V.~B.}\ \bibnamefont
  {Braginsky}}\ and\ \bibinfo {author} {\bibfnamefont {L.~P.}\ \bibnamefont
  {Grishchuk}},\ }\href@noop {} {\bibfield  {journal} {\bibinfo  {journal}
  {Sov. Phys. JETP}\ }\textbf {\bibinfo {volume} {62}},\ \bibinfo {pages} {427}
  (\bibinfo {year} {1985})}\BibitemShut {NoStop}%
\bibitem [{\citenamefont {{Braginsky}}\ and\ \citenamefont
  {{Thorne}}(1987)}]{Braginsky:1987}%
  \BibitemOpen
  \bibfield  {author} {\bibinfo {author} {\bibfnamefont {V.~B.}\ \bibnamefont
  {{Braginsky}}}\ and\ \bibinfo {author} {\bibfnamefont {K.~S.}\ \bibnamefont
  {{Thorne}}},\ }\href {\doibase 10.1038/327123a0} {\bibfield  {journal}
  {\bibinfo  {journal} {\nat}\ }\textbf {\bibinfo {volume} {327}},\ \bibinfo
  {pages} {123} (\bibinfo {year} {1987})}\BibitemShut {NoStop}%
\bibitem [{\citenamefont {Christodoulou}(1991)}]{Christodoulou:1991cr}%
  \BibitemOpen
  \bibfield  {author} {\bibinfo {author} {\bibfnamefont {D.}~\bibnamefont
  {Christodoulou}},\ }\href {\doibase 10.1103/PhysRevLett.67.1486} {\bibfield
  {journal} {\bibinfo  {journal} {Phys. Rev. Lett.}\ }\textbf {\bibinfo
  {volume} {67}},\ \bibinfo {pages} {1486} (\bibinfo {year}
  {1991})}\BibitemShut {NoStop}%
\bibitem [{\citenamefont {Flanagan}\ \emph {et~al.}(2019)\citenamefont
  {Flanagan}, \citenamefont {Grant}, \citenamefont {Harte},\ and\ \citenamefont
  {Nichols}}]{Flanagan:2018yzh}%
  \BibitemOpen
  \bibfield  {author} {\bibinfo {author} {\bibfnamefont {E.~E.}\ \bibnamefont
  {Flanagan}}, \bibinfo {author} {\bibfnamefont {A.~M.}\ \bibnamefont {Grant}},
  \bibinfo {author} {\bibfnamefont {A.~I.}\ \bibnamefont {Harte}}, \ and\
  \bibinfo {author} {\bibfnamefont {D.~A.}\ \bibnamefont {Nichols}},\ }\href
  {\doibase 10.1103/PhysRevD.99.084044} {\bibfield  {journal} {\bibinfo
  {journal} {Phys. Rev. D}\ }\textbf {\bibinfo {volume} {99}},\ \bibinfo
  {pages} {084044} (\bibinfo {year} {2019})}\BibitemShut {NoStop}%
\bibitem [{\citenamefont {Grant}\ and\ \citenamefont
  {Nichols}(2022)}]{Grant:2021hga}%
  \BibitemOpen
  \bibfield  {author} {\bibinfo {author} {\bibfnamefont {A.~M.}\ \bibnamefont
  {Grant}}\ and\ \bibinfo {author} {\bibfnamefont {D.~A.}\ \bibnamefont
  {Nichols}},\ }\href {\doibase 10.1103/PhysRevD.105.024056} {\bibfield
  {journal} {\bibinfo  {journal} {Phys. Rev. D}\ }\textbf {\bibinfo {volume}
  {105}},\ \bibinfo {pages} {024056} (\bibinfo {year} {2022})},\ \bibinfo
  {note} {[Erratum: Phys.Rev.D 107, 109902 (2023)]}\BibitemShut {NoStop}%
\bibitem [{\citenamefont {Grant}\ and\ \citenamefont
  {Nichols}(2023)}]{Grant:2022bla}%
  \BibitemOpen
  \bibfield  {author} {\bibinfo {author} {\bibfnamefont {A.~M.}\ \bibnamefont
  {Grant}}\ and\ \bibinfo {author} {\bibfnamefont {D.~A.}\ \bibnamefont
  {Nichols}},\ }\href {\doibase 10.1103/PhysRevD.107.064056} {\bibfield
  {journal} {\bibinfo  {journal} {Phys. Rev. D}\ }\textbf {\bibinfo {volume}
  {107}},\ \bibinfo {pages} {064056} (\bibinfo {year} {2023})},\ \bibinfo
  {note} {[Erratum: Phys.Rev.D 108, 029901 (2023)]}\BibitemShut {NoStop}%
\bibitem [{\citenamefont {Siddhant}\ \emph {et~al.}(2024)\citenamefont
  {Siddhant}, \citenamefont {Grant},\ and\ \citenamefont
  {Nichols}}]{Siddhant:2024nft}%
  \BibitemOpen
  \bibfield  {author} {\bibinfo {author} {\bibfnamefont {S.}~\bibnamefont
  {Siddhant}}, \bibinfo {author} {\bibfnamefont {A.~M.}\ \bibnamefont {Grant}},
  \ and\ \bibinfo {author} {\bibfnamefont {D.~A.}\ \bibnamefont {Nichols}},\
  }\href@noop {} {\  (\bibinfo {year} {2024})}\BibitemShut {NoStop}%
\bibitem [{\citenamefont {Pasterski}\ \emph {et~al.}(2016)\citenamefont
  {Pasterski}, \citenamefont {Strominger},\ and\ \citenamefont
  {Zhiboedov}}]{Pasterski:2015tva}%
  \BibitemOpen
  \bibfield  {author} {\bibinfo {author} {\bibfnamefont {S.}~\bibnamefont
  {Pasterski}}, \bibinfo {author} {\bibfnamefont {A.}~\bibnamefont
  {Strominger}}, \ and\ \bibinfo {author} {\bibfnamefont {A.}~\bibnamefont
  {Zhiboedov}},\ }\href {\doibase 10.1007/JHEP12(2016)053} {\bibfield
  {journal} {\bibinfo  {journal} {JHEP}\ }\textbf {\bibinfo {volume} {12}},\
  \bibinfo {pages} {053} (\bibinfo {year} {2016})}\BibitemShut {NoStop}%
\bibitem [{\citenamefont {Nichols}(2018)}]{Nichols:2018qac}%
  \BibitemOpen
  \bibfield  {author} {\bibinfo {author} {\bibfnamefont {D.~A.}\ \bibnamefont
  {Nichols}},\ }\href {\doibase 10.1103/PhysRevD.98.064032} {\bibfield
  {journal} {\bibinfo  {journal} {Phys. Rev. D}\ }\textbf {\bibinfo {volume}
  {98}},\ \bibinfo {pages} {064032} (\bibinfo {year} {2018})}\BibitemShut
  {NoStop}%
\bibitem [{\citenamefont {Strominger}\ and\ \citenamefont
  {Zhiboedov}(2016)}]{Strominger:2014pwa}%
  \BibitemOpen
  \bibfield  {author} {\bibinfo {author} {\bibfnamefont {A.}~\bibnamefont
  {Strominger}}\ and\ \bibinfo {author} {\bibfnamefont {A.}~\bibnamefont
  {Zhiboedov}},\ }\href {\doibase 10.1007/JHEP01(2016)086} {\bibfield
  {journal} {\bibinfo  {journal} {JHEP}\ }\textbf {\bibinfo {volume} {01}},\
  \bibinfo {pages} {086} (\bibinfo {year} {2016})}\BibitemShut {NoStop}%
\bibitem [{\citenamefont {Ashtekar}\ \emph {et~al.}(2020)\citenamefont
  {Ashtekar}, \citenamefont {De~Lorenzo},\ and\ \citenamefont
  {Khera}}]{Ashtekar:2019viz}%
  \BibitemOpen
  \bibfield  {author} {\bibinfo {author} {\bibfnamefont {A.}~\bibnamefont
  {Ashtekar}}, \bibinfo {author} {\bibfnamefont {T.}~\bibnamefont
  {De~Lorenzo}}, \ and\ \bibinfo {author} {\bibfnamefont {N.}~\bibnamefont
  {Khera}},\ }\href {\doibase 10.1007/s10714-020-02764-1} {\bibfield  {journal}
  {\bibinfo  {journal} {Gen. Rel. Grav.}\ }\textbf {\bibinfo {volume} {52}},\
  \bibinfo {pages} {107} (\bibinfo {year} {2020})}\BibitemShut {NoStop}%
\bibitem [{\citenamefont {Mitman}\ \emph {et~al.}(2024)\citenamefont {Mitman}
  \emph {et~al.}}]{Mitman:2024uss}%
  \BibitemOpen
  \bibfield  {author} {\bibinfo {author} {\bibfnamefont {K.}~\bibnamefont
  {Mitman}} \emph {et~al.},\ }\href@noop {} {\  (\bibinfo {year}
  {2024})}\BibitemShut {NoStop}%
\bibitem [{\citenamefont {Strominger}(2017)}]{Strominger:2017zoo}%
  \BibitemOpen
  \bibfield  {author} {\bibinfo {author} {\bibfnamefont {A.}~\bibnamefont
  {Strominger}},\ }\href@noop {} {\emph {\bibinfo {title} {{Lectures on the
  Infrared Structure of Gravity and Gauge Theory}}}}\ (\bibinfo {year}
  {2017})\BibitemShut {NoStop}%
\bibitem [{\citenamefont {Bieri}\ and\ \citenamefont
  {Garfinkle}(2014)}]{Bieri:2013ada}%
  \BibitemOpen
  \bibfield  {author} {\bibinfo {author} {\bibfnamefont {L.}~\bibnamefont
  {Bieri}}\ and\ \bibinfo {author} {\bibfnamefont {D.}~\bibnamefont
  {Garfinkle}},\ }\href {\doibase 10.1103/PhysRevD.89.084039} {\bibfield
  {journal} {\bibinfo  {journal} {Phys. Rev. D}\ }\textbf {\bibinfo {volume}
  {89}},\ \bibinfo {pages} {084039} (\bibinfo {year} {2014})},\ \Eprint
  {http://arxiv.org/abs/1312.6871} {arXiv:1312.6871 [gr-qc]} \BibitemShut
  {NoStop}%
\bibitem [{\citenamefont {Bieri}\ and\ \citenamefont
  {Polnarev}(2024)}]{bieri_gravitational_2024}%
  \BibitemOpen
  \bibfield  {author} {\bibinfo {author} {\bibfnamefont {L.}~\bibnamefont
  {Bieri}}\ and\ \bibinfo {author} {\bibfnamefont {A.}~\bibnamefont
  {Polnarev}},\ }\href {\doibase 10.1088/1361-6382/ad4dfe} {\bibfield
  {journal} {\bibinfo  {journal} {Class. Quant. Grav.}\ }\textbf {\bibinfo
  {volume} {41}},\ \bibinfo {pages} {135012} (\bibinfo {year} {2024})},\
  \Eprint {http://arxiv.org/abs/2402.02594} {arXiv:2402.02594 [gr-qc]}
  \BibitemShut {NoStop}%
\bibitem [{\citenamefont {Wiseman}\ and\ \citenamefont
  {Will}(1991)}]{Wiseman:1991ss}%
  \BibitemOpen
  \bibfield  {author} {\bibinfo {author} {\bibfnamefont {A.~G.}\ \bibnamefont
  {Wiseman}}\ and\ \bibinfo {author} {\bibfnamefont {C.~M.}\ \bibnamefont
  {Will}},\ }\href {\doibase 10.1103/PhysRevD.44.R2945} {\bibfield  {journal}
  {\bibinfo  {journal} {Phys. Rev. D}\ }\textbf {\bibinfo {volume} {44}},\
  \bibinfo {pages} {R2945} (\bibinfo {year} {1991})}\BibitemShut {NoStop}%
\bibitem [{\citenamefont {Blanchet}\ and\ \citenamefont
  {Damour}(1992)}]{Blanchet:1992br}%
  \BibitemOpen
  \bibfield  {author} {\bibinfo {author} {\bibfnamefont {L.}~\bibnamefont
  {Blanchet}}\ and\ \bibinfo {author} {\bibfnamefont {T.}~\bibnamefont
  {Damour}},\ }\href {\doibase 10.1103/PhysRevD.46.4304} {\bibfield  {journal}
  {\bibinfo  {journal} {Phys. Rev. D}\ }\textbf {\bibinfo {volume} {46}},\
  \bibinfo {pages} {4304} (\bibinfo {year} {1992})}\BibitemShut {NoStop}%
\bibitem [{\citenamefont {Thorne}(1992)}]{Thorne:1992sdb}%
  \BibitemOpen
  \bibfield  {author} {\bibinfo {author} {\bibfnamefont {K.~S.}\ \bibnamefont
  {Thorne}},\ }\href {\doibase 10.1103/PhysRevD.45.520} {\bibfield  {journal}
  {\bibinfo  {journal} {Phys. Rev. D}\ }\textbf {\bibinfo {volume} {45}},\
  \bibinfo {pages} {520} (\bibinfo {year} {1992})}\BibitemShut {NoStop}%
\bibitem [{\citenamefont {Favata}(2009{\natexlab{a}})}]{Favata:2008yd}%
  \BibitemOpen
  \bibfield  {author} {\bibinfo {author} {\bibfnamefont {M.}~\bibnamefont
  {Favata}},\ }\href {\doibase 10.1103/PhysRevD.80.024002} {\bibfield
  {journal} {\bibinfo  {journal} {Phys. Rev. D}\ }\textbf {\bibinfo {volume}
  {80}},\ \bibinfo {pages} {024002} (\bibinfo {year}
  {2009}{\natexlab{a}})}\BibitemShut {NoStop}%
\bibitem [{\citenamefont {Blanchet}\ \emph {et~al.}(2008)\citenamefont
  {Blanchet}, \citenamefont {Faye}, \citenamefont {Iyer},\ and\ \citenamefont
  {Sinha}}]{Blanchet:2008je}%
  \BibitemOpen
  \bibfield  {author} {\bibinfo {author} {\bibfnamefont {L.}~\bibnamefont
  {Blanchet}}, \bibinfo {author} {\bibfnamefont {G.}~\bibnamefont {Faye}},
  \bibinfo {author} {\bibfnamefont {B.~R.}\ \bibnamefont {Iyer}}, \ and\
  \bibinfo {author} {\bibfnamefont {S.}~\bibnamefont {Sinha}},\ }\href
  {\doibase 10.1088/0264-9381/25/16/165003} {\bibfield  {journal} {\bibinfo
  {journal} {Class. Quant. Grav.}\ }\textbf {\bibinfo {volume} {25}},\ \bibinfo
  {pages} {165003} (\bibinfo {year} {2008})},\ \bibinfo {note} {[Erratum:
  Class.Quant.Grav. 29, 239501 (2012)]}\BibitemShut {NoStop}%
\bibitem [{\citenamefont {Favata}(2010)}]{Favata:2010zu}%
  \BibitemOpen
  \bibfield  {author} {\bibinfo {author} {\bibfnamefont {M.}~\bibnamefont
  {Favata}},\ }\href {\doibase 10.1088/0264-9381/27/8/084036} {\bibfield
  {journal} {\bibinfo  {journal} {Class. Quantum Gravity}\ }\textbf {\bibinfo
  {volume} {27}},\ \bibinfo {pages} {084036} (\bibinfo {year}
  {2010})}\BibitemShut {NoStop}%
\bibitem [{\citenamefont {Garfinkle}(2022)}]{Garfinkle:2022dnm}%
  \BibitemOpen
  \bibfield  {author} {\bibinfo {author} {\bibfnamefont {D.}~\bibnamefont
  {Garfinkle}},\ }\href {\doibase 10.1088/1361-6382/ac7203} {\bibfield
  {journal} {\bibinfo  {journal} {Class. Quant. Grav.}\ }\textbf {\bibinfo
  {volume} {39}},\ \bibinfo {pages} {135010} (\bibinfo {year} {2022})},\
  \Eprint {http://arxiv.org/abs/2201.05543} {arXiv:2201.05543 [gr-qc]}
  \BibitemShut {NoStop}%
\bibitem [{\citenamefont {Zosso}(2024)}]{Zosso:2024xgy}%
  \BibitemOpen
  \bibfield  {author} {\bibinfo {author} {\bibfnamefont {J.}~\bibnamefont
  {Zosso}},\ }\emph {\bibinfo {title} {{Probing Gravity - Fundamental Aspects
  of Metric Theories and their Implications for Tests of General
  Relativity}}},\ \href {\doibase 10.3929/ethz-b-000675938} {Ph.D. thesis},\
  \bibinfo  {school} {Zurich, ETH} (\bibinfo {year} {2024})\BibitemShut
  {NoStop}%
\bibitem [{\citenamefont {Favata}(2009{\natexlab{b}})}]{favata_nonlinear_2009}%
  \BibitemOpen
  \bibfield  {author} {\bibinfo {author} {\bibfnamefont {M.}~\bibnamefont
  {Favata}},\ }\href {\doibase 10.1088/0004-637X/696/2/L159} {\bibfield
  {journal} {\bibinfo  {journal} {The Astrophysical Journal}\ }\textbf
  {\bibinfo {volume} {696}},\ \bibinfo {pages} {L159} (\bibinfo {year}
  {2009}{\natexlab{b}})},\ \bibinfo {note} {publisher: American Astronomical
  Society}\BibitemShut {NoStop}%
\bibitem [{\citenamefont {Gasparotto}\ \emph {et~al.}(2023)\citenamefont
  {Gasparotto}, \citenamefont {Vicente}, \citenamefont {Blas}, \citenamefont
  {Jenkins},\ and\ \citenamefont {Barausse}}]{gasparotto_can_2023}%
  \BibitemOpen
  \bibfield  {author} {\bibinfo {author} {\bibfnamefont {S.}~\bibnamefont
  {Gasparotto}}, \bibinfo {author} {\bibfnamefont {R.}~\bibnamefont {Vicente}},
  \bibinfo {author} {\bibfnamefont {D.}~\bibnamefont {Blas}}, \bibinfo {author}
  {\bibfnamefont {A.~C.}\ \bibnamefont {Jenkins}}, \ and\ \bibinfo {author}
  {\bibfnamefont {E.}~\bibnamefont {Barausse}},\ }\href {\doibase
  10.1103/PhysRevD.107.124033} {\bibfield  {journal} {\bibinfo  {journal}
  {Physical Review D}\ }\textbf {\bibinfo {volume} {107}},\ \bibinfo {pages}
  {124033} (\bibinfo {year} {2023})},\ \bibinfo {note} {publisher: American
  Physical Society}\BibitemShut {NoStop}%
\bibitem [{\citenamefont {Xu}\ \emph {et~al.}(2024)\citenamefont {Xu},
  \citenamefont {Rossell\'o-Sastre}, \citenamefont {Tiwari}, \citenamefont
  {Ebersold}, \citenamefont {Hamilton}, \citenamefont {Garc\'\i{}a-Quir\'os},
  \citenamefont {Estell\'es},\ and\ \citenamefont {Husa}}]{Xu:2024ybt}%
  \BibitemOpen
  \bibfield  {author} {\bibinfo {author} {\bibfnamefont {Y.}~\bibnamefont
  {Xu}}, \bibinfo {author} {\bibfnamefont {M.}~\bibnamefont
  {Rossell\'o-Sastre}}, \bibinfo {author} {\bibfnamefont {S.}~\bibnamefont
  {Tiwari}}, \bibinfo {author} {\bibfnamefont {M.}~\bibnamefont {Ebersold}},
  \bibinfo {author} {\bibfnamefont {E.~Z.}\ \bibnamefont {Hamilton}}, \bibinfo
  {author} {\bibfnamefont {C.}~\bibnamefont {Garc\'\i{}a-Quir\'os}}, \bibinfo
  {author} {\bibfnamefont {H.}~\bibnamefont {Estell\'es}}, \ and\ \bibinfo
  {author} {\bibfnamefont {S.}~\bibnamefont {Husa}},\ }\href@noop {} {\
  (\bibinfo {year} {2024})}\BibitemShut {NoStop}%
\bibitem [{\citenamefont {Tiwari}\ \emph {et~al.}(2021)\citenamefont {Tiwari},
  \citenamefont {Ebersold},\ and\ \citenamefont {Hamilton}}]{Tiwari:2021gfl}%
  \BibitemOpen
  \bibfield  {author} {\bibinfo {author} {\bibfnamefont {S.}~\bibnamefont
  {Tiwari}}, \bibinfo {author} {\bibfnamefont {M.}~\bibnamefont {Ebersold}}, \
  and\ \bibinfo {author} {\bibfnamefont {E.~Z.}\ \bibnamefont {Hamilton}},\
  }\href {\doibase 10.1103/PhysRevD.104.123024} {\bibfield  {journal} {\bibinfo
   {journal} {Phys. Rev. D}\ }\textbf {\bibinfo {volume} {104}},\ \bibinfo
  {pages} {123024} (\bibinfo {year} {2021})}\BibitemShut {NoStop}%
\bibitem [{\citenamefont {Lopez}\ \emph {et~al.}(2024)\citenamefont {Lopez},
  \citenamefont {Tiwari},\ and\ \citenamefont {Ebersold}}]{Lopez:2023aja}%
  \BibitemOpen
  \bibfield  {author} {\bibinfo {author} {\bibfnamefont {D.}~\bibnamefont
  {Lopez}}, \bibinfo {author} {\bibfnamefont {S.}~\bibnamefont {Tiwari}}, \
  and\ \bibinfo {author} {\bibfnamefont {M.}~\bibnamefont {Ebersold}},\ }\href
  {\doibase 10.1103/PhysRevD.109.043039} {\bibfield  {journal} {\bibinfo
  {journal} {Phys. Rev. D}\ }\textbf {\bibinfo {volume} {109}},\ \bibinfo
  {pages} {043039} (\bibinfo {year} {2024})}\BibitemShut {NoStop}%
\bibitem [{\citenamefont {Goncharov}\ \emph {et~al.}(2023)\citenamefont
  {Goncharov}, \citenamefont {Donnay},\ and\ \citenamefont
  {Harms}}]{Goncharov:2023woe}%
  \BibitemOpen
  \bibfield  {author} {\bibinfo {author} {\bibfnamefont {B.}~\bibnamefont
  {Goncharov}}, \bibinfo {author} {\bibfnamefont {L.}~\bibnamefont {Donnay}}, \
  and\ \bibinfo {author} {\bibfnamefont {J.}~\bibnamefont {Harms}},\
  }\href@noop {} {\  (\bibinfo {year} {2023})}\BibitemShut {NoStop}%
\bibitem [{\citenamefont {Heisenberg}(2023)}]{heisenberg_balance_2023}%
  \BibitemOpen
  \bibfield  {author} {\bibinfo {author} {\bibfnamefont {L.}~\bibnamefont
  {Heisenberg}},\ }\href {\doibase 10.1098/rsta.2023.0086} {\bibfield
  {journal} {\bibinfo  {journal} {Philosophical Transactions of the Royal
  Society A: Mathematical, Physical and Engineering Sciences}\ }\textbf
  {\bibinfo {volume} {382}},\ \bibinfo {pages} {20230086} (\bibinfo {year}
  {2023})},\ \bibinfo {note} {publisher: Royal Society}\BibitemShut {NoStop}%
\bibitem [{\citenamefont {D'Ambrosio}\ \emph {et~al.}(2024)\citenamefont
  {D'Ambrosio}, \citenamefont {Gozzini}, \citenamefont {Heisenberg},
  \citenamefont {Inchauspé}, \citenamefont {Maibach},\ and\ \citenamefont
  {Zosso}}]{dambrosio_testing_2024}%
  \BibitemOpen
  \bibfield  {author} {\bibinfo {author} {\bibfnamefont {F.}~\bibnamefont
  {D'Ambrosio}}, \bibinfo {author} {\bibfnamefont {F.}~\bibnamefont {Gozzini}},
  \bibinfo {author} {\bibfnamefont {L.}~\bibnamefont {Heisenberg}}, \bibinfo
  {author} {\bibfnamefont {H.}~\bibnamefont {Inchauspé}}, \bibinfo {author}
  {\bibfnamefont {D.}~\bibnamefont {Maibach}}, \ and\ \bibinfo {author}
  {\bibfnamefont {J.}~\bibnamefont {Zosso}},\ }\href {\doibase
  10.48550/arXiv.2402.19397} {\enquote {\bibinfo {title} {Testing gravitational
  waveforms in full {General} {Relativity}},}\ } (\bibinfo {year} {2024}),\
  \bibinfo {note} {arXiv:2402.19397 null}\BibitemShut {NoStop}%
\bibitem [{\citenamefont {Du}\ and\ \citenamefont
  {Nishizawa}(2016)}]{Du:2016hww}%
  \BibitemOpen
  \bibfield  {author} {\bibinfo {author} {\bibfnamefont {S.~M.}\ \bibnamefont
  {Du}}\ and\ \bibinfo {author} {\bibfnamefont {A.}~\bibnamefont {Nishizawa}},\
  }\href {\doibase 10.1103/PhysRevD.94.104063} {\bibfield  {journal} {\bibinfo
  {journal} {Phys. Rev. D}\ }\textbf {\bibinfo {volume} {94}},\ \bibinfo
  {pages} {104063} (\bibinfo {year} {2016})}\BibitemShut {NoStop}%
\bibitem [{\citenamefont {Tahura}\ \emph {et~al.}(2021)\citenamefont {Tahura},
  \citenamefont {Nichols},\ and\ \citenamefont {Yagi}}]{Tahura:2021hbk}%
  \BibitemOpen
  \bibfield  {author} {\bibinfo {author} {\bibfnamefont {S.}~\bibnamefont
  {Tahura}}, \bibinfo {author} {\bibfnamefont {D.~A.}\ \bibnamefont {Nichols}},
  \ and\ \bibinfo {author} {\bibfnamefont {K.}~\bibnamefont {Yagi}},\ }\href
  {\doibase 10.1103/PhysRevD.104.104010} {\bibfield  {journal} {\bibinfo
  {journal} {Phys. Rev. D}\ }\textbf {\bibinfo {volume} {104}},\ \bibinfo
  {pages} {104010} (\bibinfo {year} {2021})}\BibitemShut {NoStop}%
\bibitem [{\citenamefont {Heisenberg}\ \emph
  {et~al.}(2023{\natexlab{a}})\citenamefont {Heisenberg}, \citenamefont
  {Yunes},\ and\ \citenamefont {Zosso}}]{Heisenberg:2023prj}%
  \BibitemOpen
  \bibfield  {author} {\bibinfo {author} {\bibfnamefont {L.}~\bibnamefont
  {Heisenberg}}, \bibinfo {author} {\bibfnamefont {N.}~\bibnamefont {Yunes}}, \
  and\ \bibinfo {author} {\bibfnamefont {J.}~\bibnamefont {Zosso}},\ }\href
  {\doibase 10.1103/PhysRevD.108.024010} {\bibfield  {journal} {\bibinfo
  {journal} {Phys. Rev. D}\ }\textbf {\bibinfo {volume} {108}},\ \bibinfo
  {pages} {024010} (\bibinfo {year} {2023}{\natexlab{a}})}\BibitemShut
  {NoStop}%
\bibitem [{\citenamefont {Heisenberg}\ \emph {et~al.}(2024)\citenamefont
  {Heisenberg}, \citenamefont {Xu},\ and\ \citenamefont
  {Zosso}}]{Heisenberg:2024cjk}%
  \BibitemOpen
  \bibfield  {author} {\bibinfo {author} {\bibfnamefont {L.}~\bibnamefont
  {Heisenberg}}, \bibinfo {author} {\bibfnamefont {G.}~\bibnamefont {Xu}}, \
  and\ \bibinfo {author} {\bibfnamefont {J.}~\bibnamefont {Zosso}},\ }\href
  {\doibase 10.1088/1475-7516/2024/05/119} {\bibfield  {journal} {\bibinfo
  {journal} {JCAP}\ }\textbf {\bibinfo {volume} {05}},\ \bibinfo {pages} {119}
  (\bibinfo {year} {2024})},\ \Eprint {http://arxiv.org/abs/2401.05936}
  {arXiv:2401.05936 [gr-qc]} \BibitemShut {NoStop}%
\bibitem [{\citenamefont {McNeill}\ \emph {et~al.}(2017)\citenamefont
  {McNeill}, \citenamefont {Thrane},\ and\ \citenamefont
  {Lasky}}]{McNeill:2017uvq}%
  \BibitemOpen
  \bibfield  {author} {\bibinfo {author} {\bibfnamefont {L.~O.}\ \bibnamefont
  {McNeill}}, \bibinfo {author} {\bibfnamefont {E.}~\bibnamefont {Thrane}}, \
  and\ \bibinfo {author} {\bibfnamefont {P.~D.}\ \bibnamefont {Lasky}},\ }\href
  {\doibase 10.1103/PhysRevLett.118.181103} {\bibfield  {journal} {\bibinfo
  {journal} {Phys. Rev. Lett.}\ }\textbf {\bibinfo {volume} {118}},\ \bibinfo
  {pages} {181103} (\bibinfo {year} {2017})}\BibitemShut {NoStop}%
\bibitem [{\citenamefont {Ghosh}\ \emph {et~al.}(2023)\citenamefont {Ghosh},
  \citenamefont {Weaver}, \citenamefont {Sanjuan}, \citenamefont {Fulda},\ and\
  \citenamefont {Mueller}}]{Ghosh:2023rbe}%
  \BibitemOpen
  \bibfield  {author} {\bibinfo {author} {\bibfnamefont {S.}~\bibnamefont
  {Ghosh}}, \bibinfo {author} {\bibfnamefont {A.}~\bibnamefont {Weaver}},
  \bibinfo {author} {\bibfnamefont {J.}~\bibnamefont {Sanjuan}}, \bibinfo
  {author} {\bibfnamefont {P.}~\bibnamefont {Fulda}}, \ and\ \bibinfo {author}
  {\bibfnamefont {G.}~\bibnamefont {Mueller}},\ }\href {\doibase
  10.1103/PhysRevD.107.084051} {\bibfield  {journal} {\bibinfo  {journal}
  {Phys. Rev. D}\ }\textbf {\bibinfo {volume} {107}},\ \bibinfo {pages}
  {084051} (\bibinfo {year} {2023})}\BibitemShut {NoStop}%
\bibitem [{\citenamefont {H\"ubner}\ \emph {et~al.}(2020)\citenamefont
  {H\"ubner}, \citenamefont {Talbot}, \citenamefont {Lasky},\ and\
  \citenamefont {Thrane}}]{Hubner:2019sly}%
  \BibitemOpen
  \bibfield  {author} {\bibinfo {author} {\bibfnamefont {M.}~\bibnamefont
  {H\"ubner}}, \bibinfo {author} {\bibfnamefont {C.}~\bibnamefont {Talbot}},
  \bibinfo {author} {\bibfnamefont {P.~D.}\ \bibnamefont {Lasky}}, \ and\
  \bibinfo {author} {\bibfnamefont {E.}~\bibnamefont {Thrane}},\ }\href
  {\doibase 10.1103/PhysRevD.101.023011} {\bibfield  {journal} {\bibinfo
  {journal} {Phys. Rev. D}\ }\textbf {\bibinfo {volume} {101}},\ \bibinfo
  {pages} {023011} (\bibinfo {year} {2020})}\BibitemShut {NoStop}%
\bibitem [{\citenamefont {Ebersold}\ and\ \citenamefont
  {Tiwari}(2020)}]{Ebersold:2020zah}%
  \BibitemOpen
  \bibfield  {author} {\bibinfo {author} {\bibfnamefont {M.}~\bibnamefont
  {Ebersold}}\ and\ \bibinfo {author} {\bibfnamefont {S.}~\bibnamefont
  {Tiwari}},\ }\href {\doibase 10.1103/PhysRevD.101.104041} {\bibfield
  {journal} {\bibinfo  {journal} {Phys. Rev. D}\ }\textbf {\bibinfo {volume}
  {101}},\ \bibinfo {pages} {104041} (\bibinfo {year} {2020})}\BibitemShut
  {NoStop}%
\bibitem [{\citenamefont {Zhao}\ \emph {et~al.}(2021)\citenamefont {Zhao},
  \citenamefont {Liu}, \citenamefont {Cao},\ and\ \citenamefont
  {He}}]{Zhao:2021hmx}%
  \BibitemOpen
  \bibfield  {author} {\bibinfo {author} {\bibfnamefont {Z.-C.}\ \bibnamefont
  {Zhao}}, \bibinfo {author} {\bibfnamefont {X.}~\bibnamefont {Liu}}, \bibinfo
  {author} {\bibfnamefont {Z.}~\bibnamefont {Cao}}, \ and\ \bibinfo {author}
  {\bibfnamefont {X.}~\bibnamefont {He}},\ }\href {\doibase
  10.1103/PhysRevD.104.064056} {\bibfield  {journal} {\bibinfo  {journal}
  {Phys. Rev. D}\ }\textbf {\bibinfo {volume} {104}},\ \bibinfo {pages}
  {064056} (\bibinfo {year} {2021})}\BibitemShut {NoStop}%
\bibitem [{\citenamefont {H\"ubner}\ \emph {et~al.}(2021)\citenamefont
  {H\"ubner}, \citenamefont {Lasky},\ and\ \citenamefont
  {Thrane}}]{Hubner:2021amk}%
  \BibitemOpen
  \bibfield  {author} {\bibinfo {author} {\bibfnamefont {M.}~\bibnamefont
  {H\"ubner}}, \bibinfo {author} {\bibfnamefont {P.}~\bibnamefont {Lasky}}, \
  and\ \bibinfo {author} {\bibfnamefont {E.}~\bibnamefont {Thrane}},\ }\href
  {\doibase 10.1103/PhysRevD.104.023004} {\bibfield  {journal} {\bibinfo
  {journal} {Phys. Rev. D}\ }\textbf {\bibinfo {volume} {104}},\ \bibinfo
  {pages} {023004} (\bibinfo {year} {2021})}\BibitemShut {NoStop}%
\bibitem [{\citenamefont {Seto}(2009)}]{Seto:2009nv}%
  \BibitemOpen
  \bibfield  {author} {\bibinfo {author} {\bibfnamefont {N.}~\bibnamefont
  {Seto}},\ }\href {\doibase 10.1111/j.1745-3933.2009.00758.x} {\bibfield
  {journal} {\bibinfo  {journal} {Mon. Not. Roy. Astron. Soc.}\ }\textbf
  {\bibinfo {volume} {400}},\ \bibinfo {pages} {L38} (\bibinfo {year}
  {2009})}\BibitemShut {NoStop}%
\bibitem [{\citenamefont {Cordes}\ and\ \citenamefont
  {Jenet}(2012)}]{Cordes:2012zz}%
  \BibitemOpen
  \bibfield  {author} {\bibinfo {author} {\bibfnamefont {J.~M.}\ \bibnamefont
  {Cordes}}\ and\ \bibinfo {author} {\bibfnamefont {F.~A.}\ \bibnamefont
  {Jenet}},\ }\href {\doibase 10.1088/0004-637X/752/1/54} {\bibfield  {journal}
  {\bibinfo  {journal} {Astrophys. J.}\ }\textbf {\bibinfo {volume} {752}},\
  \bibinfo {pages} {54} (\bibinfo {year} {2012})}\BibitemShut {NoStop}%
\bibitem [{\citenamefont {Agazie}\ \emph {et~al.}(2024)\citenamefont {Agazie}
  \emph {et~al.}}]{NANOGrav:2023vfo}%
  \BibitemOpen
  \bibfield  {author} {\bibinfo {author} {\bibfnamefont {G.}~\bibnamefont
  {Agazie}} \emph {et~al.},\ }\href {\doibase 10.3847/1538-4357/ad0726}
  {\bibfield  {journal} {\bibinfo  {journal} {Astrophys. J.}\ }\textbf
  {\bibinfo {volume} {963}},\ \bibinfo {pages} {61} (\bibinfo {year}
  {2024})}\BibitemShut {NoStop}%
\bibitem [{\citenamefont {Punturo}\ \emph {et~al.}(2010)\citenamefont {Punturo}
  \emph {et~al.}}]{Punturo:2010zz}%
  \BibitemOpen
  \bibfield  {author} {\bibinfo {author} {\bibfnamefont {M.}~\bibnamefont
  {Punturo}} \emph {et~al.},\ }\href {\doibase 10.1088/0264-9381/27/19/194002}
  {\bibfield  {journal} {\bibinfo  {journal} {Class. Quant. Grav.}\ }\textbf
  {\bibinfo {volume} {27}},\ \bibinfo {pages} {194002} (\bibinfo {year}
  {2010})}\BibitemShut {NoStop}%
\bibitem [{\citenamefont {Reitze}\ \emph {et~al.}(2019)\citenamefont {Reitze}
  \emph {et~al.}}]{Reitze:2019iox}%
  \BibitemOpen
  \bibfield  {author} {\bibinfo {author} {\bibfnamefont {D.}~\bibnamefont
  {Reitze}} \emph {et~al.},\ }\href@noop {} {\bibfield  {journal} {\bibinfo
  {journal} {Bull. Am. Astron. Soc.}\ }\textbf {\bibinfo {volume} {51}},\
  \bibinfo {pages} {035} (\bibinfo {year} {2019})}\BibitemShut {NoStop}%
\bibitem [{\citenamefont {Favata}(2009{\natexlab{c}})}]{Favata:2009ii}%
  \BibitemOpen
  \bibfield  {author} {\bibinfo {author} {\bibfnamefont {M.}~\bibnamefont
  {Favata}},\ }\href {\doibase 10.1088/0004-637X/696/2/L159} {\bibfield
  {journal} {\bibinfo  {journal} {Astrophys. J. Lett.}\ }\textbf {\bibinfo
  {volume} {696}},\ \bibinfo {pages} {L159} (\bibinfo {year}
  {2009}{\natexlab{c}})}\BibitemShut {NoStop}%
\bibitem [{\citenamefont {Johnson}\ \emph {et~al.}(2019)\citenamefont
  {Johnson}, \citenamefont {Kapadia}, \citenamefont {Osborne}, \citenamefont
  {Hixon},\ and\ \citenamefont {Kennefick}}]{Johnson:2018xly}%
  \BibitemOpen
  \bibfield  {author} {\bibinfo {author} {\bibfnamefont {A.~D.}\ \bibnamefont
  {Johnson}}, \bibinfo {author} {\bibfnamefont {S.~J.}\ \bibnamefont
  {Kapadia}}, \bibinfo {author} {\bibfnamefont {A.}~\bibnamefont {Osborne}},
  \bibinfo {author} {\bibfnamefont {A.}~\bibnamefont {Hixon}}, \ and\ \bibinfo
  {author} {\bibfnamefont {D.}~\bibnamefont {Kennefick}},\ }\href {\doibase
  10.1103/PhysRevD.99.044045} {\bibfield  {journal} {\bibinfo  {journal} {Phys.
  Rev. D}\ }\textbf {\bibinfo {volume} {99}},\ \bibinfo {pages} {044045}
  (\bibinfo {year} {2019})}\BibitemShut {NoStop}%
\bibitem [{\citenamefont {Islo}\ \emph {et~al.}(2019)\citenamefont {Islo},
  \citenamefont {Simon}, \citenamefont {Burke-Spolaor},\ and\ \citenamefont
  {Siemens}}]{Islo:2019qht}%
  \BibitemOpen
  \bibfield  {author} {\bibinfo {author} {\bibfnamefont {K.}~\bibnamefont
  {Islo}}, \bibinfo {author} {\bibfnamefont {J.}~\bibnamefont {Simon}},
  \bibinfo {author} {\bibfnamefont {S.}~\bibnamefont {Burke-Spolaor}}, \ and\
  \bibinfo {author} {\bibfnamefont {X.}~\bibnamefont {Siemens}},\ }\href@noop
  {} {\  (\bibinfo {year} {2019})}\BibitemShut {NoStop}%
\bibitem [{\citenamefont {Islam}\ \emph {et~al.}(2021)\citenamefont {Islam},
  \citenamefont {Field}, \citenamefont {Khanna},\ and\ \citenamefont
  {Warburton}}]{Islam:2021old}%
  \BibitemOpen
  \bibfield  {author} {\bibinfo {author} {\bibfnamefont {T.}~\bibnamefont
  {Islam}}, \bibinfo {author} {\bibfnamefont {S.~E.}\ \bibnamefont {Field}},
  \bibinfo {author} {\bibfnamefont {G.}~\bibnamefont {Khanna}}, \ and\ \bibinfo
  {author} {\bibfnamefont {N.}~\bibnamefont {Warburton}},\ }\href@noop {} {\
  (\bibinfo {year} {2021})}\BibitemShut {NoStop}%
\bibitem [{\citenamefont {Sun}\ \emph {et~al.}(2023)\citenamefont {Sun},
  \citenamefont {Shi}, \citenamefont {Zhang},\ and\ \citenamefont
  {Mei}}]{Sun:2022pvh}%
  \BibitemOpen
  \bibfield  {author} {\bibinfo {author} {\bibfnamefont {S.}~\bibnamefont
  {Sun}}, \bibinfo {author} {\bibfnamefont {C.}~\bibnamefont {Shi}}, \bibinfo
  {author} {\bibfnamefont {J.-d.}\ \bibnamefont {Zhang}}, \ and\ \bibinfo
  {author} {\bibfnamefont {J.}~\bibnamefont {Mei}},\ }\href {\doibase
  10.1103/PhysRevD.107.044023} {\bibfield  {journal} {\bibinfo  {journal}
  {Phys. Rev. D}\ }\textbf {\bibinfo {volume} {107}},\ \bibinfo {pages}
  {044023} (\bibinfo {year} {2023})},\ \Eprint
  {http://arxiv.org/abs/2207.13009} {arXiv:2207.13009 [gr-qc]} \BibitemShut
  {NoStop}%
\bibitem [{\citenamefont {Sun}\ \emph {et~al.}(2024)\citenamefont {Sun},
  \citenamefont {Shi}, \citenamefont {Zhang},\ and\ \citenamefont
  {Mei}}]{Sun:2024nut}%
  \BibitemOpen
  \bibfield  {author} {\bibinfo {author} {\bibfnamefont {S.}~\bibnamefont
  {Sun}}, \bibinfo {author} {\bibfnamefont {C.}~\bibnamefont {Shi}}, \bibinfo
  {author} {\bibfnamefont {J.-d.}\ \bibnamefont {Zhang}}, \ and\ \bibinfo
  {author} {\bibfnamefont {J.}~\bibnamefont {Mei}},\ }\href@noop {} {\
  (\bibinfo {year} {2024})},\ \Eprint {http://arxiv.org/abs/2401.11416}
  {arXiv:2401.11416 [gr-qc]} \BibitemShut {NoStop}%
\bibitem [{\citenamefont {van Haasteren}\ and\ \citenamefont
  {Levin}(2010)}]{vanHaasteren:2009fy}%
  \BibitemOpen
  \bibfield  {author} {\bibinfo {author} {\bibfnamefont {R.}~\bibnamefont {van
  Haasteren}}\ and\ \bibinfo {author} {\bibfnamefont {Y.}~\bibnamefont
  {Levin}},\ }\href {\doibase 10.1111/j.1365-2966.2009.15885.x} {\bibfield
  {journal} {\bibinfo  {journal} {Mon. Not. Roy. Astron. Soc.}\ }\textbf
  {\bibinfo {volume} {401}},\ \bibinfo {pages} {2372} (\bibinfo {year}
  {2010})}\BibitemShut {NoStop}%
\bibitem [{\citenamefont {Janssen}\ \emph {et~al.}(2015)\citenamefont {Janssen}
  \emph {et~al.}}]{Janssen:2014dka}%
  \BibitemOpen
  \bibfield  {author} {\bibinfo {author} {\bibfnamefont {G.}~\bibnamefont
  {Janssen}} \emph {et~al.},\ }\href {\doibase 10.22323/1.215.0037} {\bibfield
  {journal} {\bibinfo  {journal} {PoS}\ }\textbf {\bibinfo {volume}
  {AASKA14}},\ \bibinfo {pages} {037} (\bibinfo {year} {2015})}\BibitemShut
  {NoStop}%
\bibitem [{\citenamefont {Varma}\ \emph {et~al.}(2019)\citenamefont {Varma},
  \citenamefont {Field}, \citenamefont {Scheel}, \citenamefont {Blackman},
  \citenamefont {Kidder},\ and\ \citenamefont
  {Pfeiffer}}]{varma_surrogate_2019}%
  \BibitemOpen
  \bibfield  {author} {\bibinfo {author} {\bibfnamefont {V.}~\bibnamefont
  {Varma}}, \bibinfo {author} {\bibfnamefont {S.~E.}\ \bibnamefont {Field}},
  \bibinfo {author} {\bibfnamefont {M.~A.}\ \bibnamefont {Scheel}}, \bibinfo
  {author} {\bibfnamefont {J.}~\bibnamefont {Blackman}}, \bibinfo {author}
  {\bibfnamefont {L.~E.}\ \bibnamefont {Kidder}}, \ and\ \bibinfo {author}
  {\bibfnamefont {H.~P.}\ \bibnamefont {Pfeiffer}},\ }\href {\doibase
  10.1103/PhysRevD.99.064045} {\bibfield  {journal} {\bibinfo  {journal}
  {Physical Review D}\ }\textbf {\bibinfo {volume} {99}},\ \bibinfo {pages}
  {064045} (\bibinfo {year} {2019})},\ \bibinfo {note} {publisher: American
  Physical Society}\BibitemShut {NoStop}%
\bibitem [{\citenamefont {Yoo}\ \emph {et~al.}(2023)\citenamefont {Yoo},
  \citenamefont {Mitman}, \citenamefont {Varma}, \citenamefont {Boyle},
  \citenamefont {Field}, \citenamefont {Deppe}, \citenamefont {Hébert},
  \citenamefont {Kidder}, \citenamefont {Moxon}, \citenamefont {Pfeiffer},
  \citenamefont {Scheel}, \citenamefont {Stein}, \citenamefont {Teukolsky},
  \citenamefont {Throwe},\ and\ \citenamefont {Vu}}]{yoo_numerical_2023}%
  \BibitemOpen
  \bibfield  {author} {\bibinfo {author} {\bibfnamefont {J.}~\bibnamefont
  {Yoo}}, \bibinfo {author} {\bibfnamefont {K.}~\bibnamefont {Mitman}},
  \bibinfo {author} {\bibfnamefont {V.}~\bibnamefont {Varma}}, \bibinfo
  {author} {\bibfnamefont {M.}~\bibnamefont {Boyle}}, \bibinfo {author}
  {\bibfnamefont {S.~E.}\ \bibnamefont {Field}}, \bibinfo {author}
  {\bibfnamefont {N.}~\bibnamefont {Deppe}}, \bibinfo {author} {\bibfnamefont
  {F.}~\bibnamefont {Hébert}}, \bibinfo {author} {\bibfnamefont {L.~E.}\
  \bibnamefont {Kidder}}, \bibinfo {author} {\bibfnamefont {J.}~\bibnamefont
  {Moxon}}, \bibinfo {author} {\bibfnamefont {H.~P.}\ \bibnamefont {Pfeiffer}},
  \bibinfo {author} {\bibfnamefont {M.~A.}\ \bibnamefont {Scheel}}, \bibinfo
  {author} {\bibfnamefont {L.~C.}\ \bibnamefont {Stein}}, \bibinfo {author}
  {\bibfnamefont {S.~A.}\ \bibnamefont {Teukolsky}}, \bibinfo {author}
  {\bibfnamefont {W.}~\bibnamefont {Throwe}}, \ and\ \bibinfo {author}
  {\bibfnamefont {N.~L.}\ \bibnamefont {Vu}},\ }\href {\doibase
  10.1103/PhysRevD.108.064027} {\bibfield  {journal} {\bibinfo  {journal}
  {Physical Review D}\ }\textbf {\bibinfo {volume} {108}},\ \bibinfo {pages}
  {064027} (\bibinfo {year} {2023})},\ \bibinfo {note} {publisher: American
  Physical Society}\BibitemShut {NoStop}%
\bibitem [{\citenamefont {Barausse}\ and\ \citenamefont
  {Lapi}(2020)}]{Barausse:2020gbp}%
  \BibitemOpen
  \bibfield  {author} {\bibinfo {author} {\bibfnamefont {E.}~\bibnamefont
  {Barausse}}\ and\ \bibinfo {author} {\bibfnamefont {A.}~\bibnamefont
  {Lapi}},\ }\href@noop {} {\  (\bibinfo {year} {2020})},\ \Eprint
  {http://arxiv.org/abs/2011.01994} {arXiv:2011.01994 [astro-ph.GA]}
  \BibitemShut {NoStop}%
\bibitem [{\citenamefont {Barausse}\ \emph {et~al.}(2020)\citenamefont
  {Barausse}, \citenamefont {Dvorkin}, \citenamefont {Tremmel}, \citenamefont
  {Volonteri},\ and\ \citenamefont {Bonetti}}]{Barausse:2020mdt}%
  \BibitemOpen
  \bibfield  {author} {\bibinfo {author} {\bibfnamefont {E.}~\bibnamefont
  {Barausse}}, \bibinfo {author} {\bibfnamefont {I.}~\bibnamefont {Dvorkin}},
  \bibinfo {author} {\bibfnamefont {M.}~\bibnamefont {Tremmel}}, \bibinfo
  {author} {\bibfnamefont {M.}~\bibnamefont {Volonteri}}, \ and\ \bibinfo
  {author} {\bibfnamefont {M.}~\bibnamefont {Bonetti}},\ }\href {\doibase
  10.3847/1538-4357/abba7f} {\bibfield  {journal} {\bibinfo  {journal}
  {Astrophys. J.}\ }\textbf {\bibinfo {volume} {904}},\ \bibinfo {pages} {16}
  (\bibinfo {year} {2020})}\BibitemShut {NoStop}%
\bibitem [{\citenamefont {Barausse}(2012)}]{EB12}%
  \BibitemOpen
  \bibfield  {author} {\bibinfo {author} {\bibfnamefont {E.}~\bibnamefont
  {Barausse}},\ }\href {\doibase 10.1111/j.1365-2966.2012.21057.x} {\bibfield
  {journal} {\bibinfo  {journal} {Mon. Not. Roy. Astron. Soc.}\ }\textbf
  {\bibinfo {volume} {423}},\ \bibinfo {pages} {2533} (\bibinfo {year}
  {2012})}\BibitemShut {NoStop}%
\bibitem [{\citenamefont {Sesana}\ \emph {et~al.}(2014)\citenamefont {Sesana},
  \citenamefont {Barausse}, \citenamefont {Dotti},\ and\ \citenamefont
  {Rossi}}]{Sesana:2014bea}%
  \BibitemOpen
  \bibfield  {author} {\bibinfo {author} {\bibfnamefont {A.}~\bibnamefont
  {Sesana}}, \bibinfo {author} {\bibfnamefont {E.}~\bibnamefont {Barausse}},
  \bibinfo {author} {\bibfnamefont {M.}~\bibnamefont {Dotti}}, \ and\ \bibinfo
  {author} {\bibfnamefont {E.~M.}\ \bibnamefont {Rossi}},\ }\href {\doibase
  10.1088/0004-637X/794/2/104} {\bibfield  {journal} {\bibinfo  {journal}
  {Astrophys. J.}\ }\textbf {\bibinfo {volume} {794}},\ \bibinfo {pages} {104}
  (\bibinfo {year} {2014})}\BibitemShut {NoStop}%
\bibitem [{\citenamefont {Antonini}\ \emph {et~al.}(2015)\citenamefont
  {Antonini}, \citenamefont {Barausse},\ and\ \citenamefont
  {Silk}}]{Antonini:2015sza}%
  \BibitemOpen
  \bibfield  {author} {\bibinfo {author} {\bibfnamefont {F.}~\bibnamefont
  {Antonini}}, \bibinfo {author} {\bibfnamefont {E.}~\bibnamefont {Barausse}},
  \ and\ \bibinfo {author} {\bibfnamefont {J.}~\bibnamefont {Silk}},\ }\href
  {\doibase 10.1088/0004-637X/812/1/72} {\bibfield  {journal} {\bibinfo
  {journal} {Astrophys. J.}\ }\textbf {\bibinfo {volume} {812}},\ \bibinfo
  {pages} {72} (\bibinfo {year} {2015})}\BibitemShut {NoStop}%
\bibitem [{\citenamefont {Colpi}\ \emph {et~al.}(2024)\citenamefont {Colpi}
  \emph {et~al.}}]{colpi_lisa_2024}%
  \BibitemOpen
  \bibfield  {author} {\bibinfo {author} {\bibfnamefont {M.}~\bibnamefont
  {Colpi}} \emph {et~al.},\ }\href {\doibase 10.48550/arXiv.2402.07571}
  {\enquote {\bibinfo {title} {{LISA} {Definition} {Study} {Report}},}\ }
  (\bibinfo {year} {2024}),\ \bibinfo {note} {arXiv:2402.07571 [astro-ph,
  physics:gr-qc]}\BibitemShut {NoStop}%
\bibitem [{\citenamefont {Bayle}\ \emph
  {et~al.}(2023{\natexlab{a}})\citenamefont {Bayle}, \citenamefont {Baghi},
  \citenamefont {Renzini},\ and\ \citenamefont {Le~Jeune}}]{bayle_lisa_2023}%
  \BibitemOpen
  \bibfield  {author} {\bibinfo {author} {\bibfnamefont {J.-B.}\ \bibnamefont
  {Bayle}}, \bibinfo {author} {\bibfnamefont {Q.}~\bibnamefont {Baghi}},
  \bibinfo {author} {\bibfnamefont {A.}~\bibnamefont {Renzini}}, \ and\
  \bibinfo {author} {\bibfnamefont {M.}~\bibnamefont {Le~Jeune}},\ }\href
  {\doibase 10.5281/zenodo.8321733} {\enquote {\bibinfo {title} {{LISA} {GW}
  {Response}},}\ } (\bibinfo {year} {2023}{\natexlab{a}}),\ \bibinfo {note}
  {language: eng}\BibitemShut {NoStop}%
\bibitem [{\citenamefont {Bayle}\ \emph {et~al.}(2022)\citenamefont {Bayle},
  \citenamefont {Hees}, \citenamefont {Lilley},\ and\ \citenamefont
  {Le~Poncin-Lafitte}}]{bayle_lisa_2022}%
  \BibitemOpen
  \bibfield  {author} {\bibinfo {author} {\bibfnamefont {J.-B.}\ \bibnamefont
  {Bayle}}, \bibinfo {author} {\bibfnamefont {A.}~\bibnamefont {Hees}},
  \bibinfo {author} {\bibfnamefont {M.}~\bibnamefont {Lilley}}, \ and\ \bibinfo
  {author} {\bibfnamefont {C.}~\bibnamefont {Le~Poncin-Lafitte}},\ }\href
  {\doibase 10.5281/zenodo.6412992} {\enquote {\bibinfo {title} {{LISA}
  {Orbits}},}\ } (\bibinfo {year} {2022}),\ \bibinfo {note} {language:
  eng}\BibitemShut {NoStop}%
\bibitem [{\citenamefont {Bayle}\ and\ \citenamefont
  {Hartwig}(2023)}]{bayle_unified_2023}%
  \BibitemOpen
  \bibfield  {author} {\bibinfo {author} {\bibfnamefont {J.-B.}\ \bibnamefont
  {Bayle}}\ and\ \bibinfo {author} {\bibfnamefont {O.}~\bibnamefont
  {Hartwig}},\ }\href {\doibase 10.1103/PhysRevD.107.083019} {\bibfield
  {journal} {\bibinfo  {journal} {Physical Review D}\ }\textbf {\bibinfo
  {volume} {107}},\ \bibinfo {pages} {083019} (\bibinfo {year} {2023})},\
  \bibinfo {note} {publisher: American Physical Society}\BibitemShut {NoStop}%
\bibitem [{\citenamefont {Bayle}\ \emph
  {et~al.}(2023{\natexlab{b}})\citenamefont {Bayle}, \citenamefont {Hartwig},\
  and\ \citenamefont {Staab}}]{lisa_instrument}%
  \BibitemOpen
  \bibfield  {author} {\bibinfo {author} {\bibfnamefont {J.-B.}\ \bibnamefont
  {Bayle}}, \bibinfo {author} {\bibfnamefont {O.}~\bibnamefont {Hartwig}}, \
  and\ \bibinfo {author} {\bibfnamefont {M.}~\bibnamefont {Staab}},\ }\href
  {\doibase 10.5281/zenodo.8329714} {\enquote {\bibinfo {title} {{LISA}
  {Instrument}},}\ } (\bibinfo {year} {2023}{\natexlab{b}})\BibitemShut
  {NoStop}%
\bibitem [{\citenamefont {Staab}\ \emph {et~al.}(2023)\citenamefont {Staab},
  \citenamefont {Bayle},\ and\ \citenamefont {Hartwig}}]{staab_pytdi_2023}%
  \BibitemOpen
  \bibfield  {author} {\bibinfo {author} {\bibfnamefont {M.}~\bibnamefont
  {Staab}}, \bibinfo {author} {\bibfnamefont {J.-B.}\ \bibnamefont {Bayle}}, \
  and\ \bibinfo {author} {\bibfnamefont {O.}~\bibnamefont {Hartwig}},\ }\href
  {\doibase 10.5281/zenodo.7704609} {\enquote {\bibinfo {title} {{PyTDI}},}\ }
  (\bibinfo {year} {2023}),\ \bibinfo {note} {language: eng}\BibitemShut
  {NoStop}%
\bibitem [{\citenamefont {Favata}(2009{\natexlab{d}})}]{Kick_Favata2009}%
  \BibitemOpen
  \bibfield  {author} {\bibinfo {author} {\bibfnamefont {M.}~\bibnamefont
  {Favata}},\ }\href {\doibase 10.1088/1742-6596/154/1/012043} {\bibfield
  {journal} {\bibinfo  {journal} {Journal of Physics: Conference Series}\
  }\textbf {\bibinfo {volume} {154}},\ \bibinfo {pages} {012043} (\bibinfo
  {year} {2009}{\natexlab{d}})}\BibitemShut {NoStop}%
\bibitem [{\citenamefont {Maggiore}(2007)}]{maggiore2008gravitational}%
  \BibitemOpen
  \bibfield  {author} {\bibinfo {author} {\bibfnamefont {M.}~\bibnamefont
  {Maggiore}},\ }\href {\doibase 10.1093/acprof:oso/9780198570745.001.0001}
  {\emph {\bibinfo {title} {Gravitational Waves: Volume 1: Theory and
  Experiments}}}\ (\bibinfo  {publisher} {Oxford University Press},\ \bibinfo
  {year} {2007})\BibitemShut {NoStop}%
\bibitem [{\citenamefont {Barausse}\ \emph {et~al.}(2012)\citenamefont
  {Barausse}, \citenamefont {Morozova},\ and\ \citenamefont
  {Rezzolla}}]{Barausse:2012qz}%
  \BibitemOpen
  \bibfield  {author} {\bibinfo {author} {\bibfnamefont {E.}~\bibnamefont
  {Barausse}}, \bibinfo {author} {\bibfnamefont {V.}~\bibnamefont {Morozova}},
  \ and\ \bibinfo {author} {\bibfnamefont {L.}~\bibnamefont {Rezzolla}},\
  }\href {\doibase 10.1088/0004-637X/758/1/63} {\bibfield  {journal} {\bibinfo
  {journal} {Astrophys. J.}\ }\textbf {\bibinfo {volume} {758}},\ \bibinfo
  {pages} {63} (\bibinfo {year} {2012})},\ \bibinfo {note} {[Erratum:
  Astrophys.J. 786, 76 (2014)]},\ \Eprint {http://arxiv.org/abs/1206.3803}
  {arXiv:1206.3803 [gr-qc]} \BibitemShut {NoStop}%
\bibitem [{\citenamefont {Arnowitt}\ \emph {et~al.}(1961)\citenamefont
  {Arnowitt}, \citenamefont {Deser},\ and\ \citenamefont
  {Misner}}]{PhysRev.121.1556}%
  \BibitemOpen
  \bibfield  {author} {\bibinfo {author} {\bibfnamefont {R.}~\bibnamefont
  {Arnowitt}}, \bibinfo {author} {\bibfnamefont {S.}~\bibnamefont {Deser}}, \
  and\ \bibinfo {author} {\bibfnamefont {C.~W.}\ \bibnamefont {Misner}},\
  }\href {\doibase 10.1103/PhysRev.121.1556} {\bibfield  {journal} {\bibinfo
  {journal} {Phys. Rev.}\ }\textbf {\bibinfo {volume} {121}},\ \bibinfo {pages}
  {1556} (\bibinfo {year} {1961})}\BibitemShut {NoStop}%
\bibitem [{\citenamefont {Isaacson}(1968{\natexlab{a}})}]{PhysRev.166.1272}%
  \BibitemOpen
  \bibfield  {author} {\bibinfo {author} {\bibfnamefont {R.~A.}\ \bibnamefont
  {Isaacson}},\ }\href {\doibase 10.1103/PhysRev.166.1272} {\bibfield
  {journal} {\bibinfo  {journal} {Phys. Rev.}\ }\textbf {\bibinfo {volume}
  {166}},\ \bibinfo {pages} {1272} (\bibinfo {year}
  {1968}{\natexlab{a}})}\BibitemShut {NoStop}%
\bibitem [{\citenamefont {Favata}(2011)}]{Favata:2011qi}%
  \BibitemOpen
  \bibfield  {author} {\bibinfo {author} {\bibfnamefont {M.}~\bibnamefont
  {Favata}},\ }\href {\doibase 10.1103/PhysRevD.84.124013} {\bibfield
  {journal} {\bibinfo  {journal} {Phys. Rev. D}\ }\textbf {\bibinfo {volume}
  {84}},\ \bibinfo {pages} {124013} (\bibinfo {year} {2011})},\ \Eprint
  {http://arxiv.org/abs/1108.3121} {arXiv:1108.3121 [gr-qc]} \BibitemShut
  {NoStop}%
\bibitem [{\citenamefont
  {Isaacson}(1968{\natexlab{b}})}]{Isaacson_PhysRev.166.1263}%
  \BibitemOpen
  \bibfield  {author} {\bibinfo {author} {\bibfnamefont {R.~A.}\ \bibnamefont
  {Isaacson}},\ }\href {\doibase 10.1103/PhysRev.166.1263} {\bibfield
  {journal} {\bibinfo  {journal} {Phys. Rev.}\ }\textbf {\bibinfo {volume}
  {166}},\ \bibinfo {pages} {1263} (\bibinfo {year}
  {1968}{\natexlab{b}})}\BibitemShut {NoStop}%
\bibitem [{\citenamefont
  {Isaacson}(1968{\natexlab{c}})}]{Isaacson_PhysRev.166.1272}%
  \BibitemOpen
  \bibfield  {author} {\bibinfo {author} {\bibfnamefont {R.~A.}\ \bibnamefont
  {Isaacson}},\ }\href {\doibase 10.1103/PhysRev.166.1272} {\bibfield
  {journal} {\bibinfo  {journal} {Phys. Rev.}\ }\textbf {\bibinfo {volume}
  {166}},\ \bibinfo {pages} {1272} (\bibinfo {year}
  {1968}{\natexlab{c}})}\BibitemShut {NoStop}%
\bibitem [{\citenamefont {D'Ambrosio}\ \emph {et~al.}(2022)\citenamefont
  {D'Ambrosio}, \citenamefont {Fell}, \citenamefont {Heisenberg}, \citenamefont
  {Maibach}, \citenamefont {Zentarra},\ and\ \citenamefont
  {Zosso}}]{DAmbrosio:2022clk}%
  \BibitemOpen
  \bibfield  {author} {\bibinfo {author} {\bibfnamefont {F.}~\bibnamefont
  {D'Ambrosio}}, \bibinfo {author} {\bibfnamefont {S.~D.~B.}\ \bibnamefont
  {Fell}}, \bibinfo {author} {\bibfnamefont {L.}~\bibnamefont {Heisenberg}},
  \bibinfo {author} {\bibfnamefont {D.}~\bibnamefont {Maibach}}, \bibinfo
  {author} {\bibfnamefont {S.}~\bibnamefont {Zentarra}}, \ and\ \bibinfo
  {author} {\bibfnamefont {J.}~\bibnamefont {Zosso}},\ }\href@noop {} {\
  (\bibinfo {year} {2022})},\ \Eprint {http://arxiv.org/abs/2201.11634}
  {arXiv:2201.11634 [gr-qc]} \BibitemShut {NoStop}%
\bibitem [{\citenamefont {Talbot}\ \emph
  {et~al.}(2018{\natexlab{a}})\citenamefont {Talbot}, \citenamefont {Thrane},
  \citenamefont {Lasky},\ and\ \citenamefont {Lin}}]{Talbot:2018sgr}%
  \BibitemOpen
  \bibfield  {author} {\bibinfo {author} {\bibfnamefont {C.}~\bibnamefont
  {Talbot}}, \bibinfo {author} {\bibfnamefont {E.}~\bibnamefont {Thrane}},
  \bibinfo {author} {\bibfnamefont {P.~D.}\ \bibnamefont {Lasky}}, \ and\
  \bibinfo {author} {\bibfnamefont {F.}~\bibnamefont {Lin}},\ }\href {\doibase
  10.1103/PhysRevD.98.064031} {\bibfield  {journal} {\bibinfo  {journal} {Phys.
  Rev. D}\ }\textbf {\bibinfo {volume} {98}},\ \bibinfo {pages} {064031}
  (\bibinfo {year} {2018}{\natexlab{a}})}\BibitemShut {NoStop}%
\bibitem [{\citenamefont {Mitman}\ \emph {et~al.}(2020)\citenamefont {Mitman},
  \citenamefont {Moxon}, \citenamefont {Scheel}, \citenamefont {Teukolsky},
  \citenamefont {Boyle}, \citenamefont {Deppe}, \citenamefont {Kidder},\ and\
  \citenamefont {Throwe}}]{Mitman:2020pbt}%
  \BibitemOpen
  \bibfield  {author} {\bibinfo {author} {\bibfnamefont {K.}~\bibnamefont
  {Mitman}}, \bibinfo {author} {\bibfnamefont {J.}~\bibnamefont {Moxon}},
  \bibinfo {author} {\bibfnamefont {M.~A.}\ \bibnamefont {Scheel}}, \bibinfo
  {author} {\bibfnamefont {S.~A.}\ \bibnamefont {Teukolsky}}, \bibinfo {author}
  {\bibfnamefont {M.}~\bibnamefont {Boyle}}, \bibinfo {author} {\bibfnamefont
  {N.}~\bibnamefont {Deppe}}, \bibinfo {author} {\bibfnamefont {L.~E.}\
  \bibnamefont {Kidder}}, \ and\ \bibinfo {author} {\bibfnamefont
  {W.}~\bibnamefont {Throwe}},\ }\href {\doibase 10.1103/PhysRevD.102.104007}
  {\bibfield  {journal} {\bibinfo  {journal} {Phys. Rev. D}\ }\textbf {\bibinfo
  {volume} {102}},\ \bibinfo {pages} {104007} (\bibinfo {year}
  {2020})}\BibitemShut {NoStop}%
\bibitem [{\citenamefont {Rossell\'o-Sastre}\ \emph {et~al.}(2024)\citenamefont
  {Rossell\'o-Sastre}, \citenamefont {Husa},\ and\ \citenamefont
  {Bera}}]{Rossello-Sastre:2024zlr}%
  \BibitemOpen
  \bibfield  {author} {\bibinfo {author} {\bibfnamefont {M.}~\bibnamefont
  {Rossell\'o-Sastre}}, \bibinfo {author} {\bibfnamefont {S.}~\bibnamefont
  {Husa}}, \ and\ \bibinfo {author} {\bibfnamefont {S.}~\bibnamefont {Bera}},\
  }\href@noop {} {\  (\bibinfo {year} {2024})}\BibitemShut {NoStop}%
\bibitem [{\citenamefont {Vallisneri}(2005)}]{vallisneri_geometric_2005}%
  \BibitemOpen
  \bibfield  {author} {\bibinfo {author} {\bibfnamefont {M.}~\bibnamefont
  {Vallisneri}},\ }\href {\doibase 10.1103/PhysRevD.72.042003} {\bibfield
  {journal} {\bibinfo  {journal} {Physical Review D}\ }\textbf {\bibinfo
  {volume} {72}},\ \bibinfo {pages} {042003} (\bibinfo {year} {2005})},\
  \bibinfo {note} {publisher: American Physical Society}\BibitemShut {NoStop}%
\bibitem [{\citenamefont {Baghi}\ \emph {et~al.}(2023)\citenamefont {Baghi},
  \citenamefont {Karnesis}, \citenamefont {Bayle}, \citenamefont {Besançon},\
  and\ \citenamefont {Inchauspé}}]{baghi_uncovering_2023}%
  \BibitemOpen
  \bibfield  {author} {\bibinfo {author} {\bibfnamefont {Q.}~\bibnamefont
  {Baghi}}, \bibinfo {author} {\bibfnamefont {N.}~\bibnamefont {Karnesis}},
  \bibinfo {author} {\bibfnamefont {J.-B.}\ \bibnamefont {Bayle}}, \bibinfo
  {author} {\bibfnamefont {M.}~\bibnamefont {Besançon}}, \ and\ \bibinfo
  {author} {\bibfnamefont {H.}~\bibnamefont {Inchauspé}},\ }\href {\doibase
  10.1088/1475-7516/2023/04/066} {\bibfield  {journal} {\bibinfo  {journal}
  {Journal of Cosmology and Astroparticle Physics}\ }\textbf {\bibinfo {volume}
  {2023}},\ \bibinfo {pages} {066} (\bibinfo {year} {2023})},\ \bibinfo {note}
  {publisher: IOP Publishing}\BibitemShut {NoStop}%
\bibitem [{\citenamefont {Heisenberg}\ \emph
  {et~al.}(2023{\natexlab{b}})\citenamefont {Heisenberg}, \citenamefont
  {Inchauspé}, \citenamefont {Nam}, \citenamefont {Sauter}, \citenamefont
  {Waibel},\ and\ \citenamefont {Wass}}]{lisa_dynamics}%
  \BibitemOpen
  \bibfield  {author} {\bibinfo {author} {\bibfnamefont {L.}~\bibnamefont
  {Heisenberg}}, \bibinfo {author} {\bibfnamefont {H.}~\bibnamefont
  {Inchauspé}}, \bibinfo {author} {\bibfnamefont {D.~Q.}\ \bibnamefont {Nam}},
  \bibinfo {author} {\bibfnamefont {O.}~\bibnamefont {Sauter}}, \bibinfo
  {author} {\bibfnamefont {R.}~\bibnamefont {Waibel}}, \ and\ \bibinfo {author}
  {\bibfnamefont {P.}~\bibnamefont {Wass}},\ }\href {\doibase
  10.1103/PhysRevD.108.122007} {\bibfield  {journal} {\bibinfo  {journal}
  {Physical Review D}\ }\textbf {\bibinfo {volume} {108}},\ \bibinfo {pages}
  {122007} (\bibinfo {year} {2023}{\natexlab{b}})},\ \bibinfo {note}
  {publisher: American Physical Society}\BibitemShut {NoStop}%
\bibitem [{\citenamefont {Tinto}\ and\ \citenamefont
  {Dhurandhar}(2020)}]{tinto_time-delay_2020}%
  \BibitemOpen
  \bibfield  {author} {\bibinfo {author} {\bibfnamefont {M.}~\bibnamefont
  {Tinto}}\ and\ \bibinfo {author} {\bibfnamefont {S.~V.}\ \bibnamefont
  {Dhurandhar}},\ }\href {\doibase 10.1007/s41114-020-00029-6} {\bibfield
  {journal} {\bibinfo  {journal} {Living Reviews in Relativity}\ }\textbf
  {\bibinfo {volume} {24}},\ \bibinfo {pages} {1} (\bibinfo {year}
  {2020})}\BibitemShut {NoStop}%
\bibitem [{\citenamefont {Team}(2018)}]{lisa_scird}%
  \BibitemOpen
  \bibfield  {author} {\bibinfo {author} {\bibfnamefont {L.~S.~S.}\
  \bibnamefont {Team}},\ }\href
  {https://www.cosmos.esa.int/documents/678316/1700384/
  SciRD.pdf/25831f6b-3c01-e215-5916-4ac6e4b306fb? t=1526479841000} {\emph
  {\bibinfo {title} {Science {Requirement} {Document}}}},\ \bibinfo {type}
  {Tech. Rep.}\ \bibinfo {number} {ESA-L3-EST-SCI-RS-001}\ (\bibinfo {year}
  {2018})\BibitemShut {NoStop}%
\bibitem [{\citenamefont {Babak}\ \emph
  {et~al.}(2021{\natexlab{a}})\citenamefont {Babak}, \citenamefont {Hewitson},\
  and\ \citenamefont {Petiteau}}]{babak_lisa_2021}%
  \BibitemOpen
  \bibfield  {author} {\bibinfo {author} {\bibfnamefont {S.}~\bibnamefont
  {Babak}}, \bibinfo {author} {\bibfnamefont {M.}~\bibnamefont {Hewitson}}, \
  and\ \bibinfo {author} {\bibfnamefont {A.}~\bibnamefont {Petiteau}},\ }\href
  {\doibase 10.48550/arXiv.2108.01167} {\enquote {\bibinfo {title} {{LISA}
  {Sensitivity} and {SNR} {Calculations}},}\ } (\bibinfo {year}
  {2021}{\natexlab{a}}),\ \bibinfo {note} {arXiv:2108.01167 [astro-ph,
  physics:gr-qc]}\BibitemShut {NoStop}%
\bibitem [{\citenamefont {Boyle}\ \emph {et~al.}(2019)\citenamefont {Boyle}
  \emph {et~al.}}]{boyle_sxs_2019}%
  \BibitemOpen
  \bibfield  {author} {\bibinfo {author} {\bibfnamefont {M.}~\bibnamefont
  {Boyle}} \emph {et~al.},\ }\href {\doibase 10.1088/1361-6382/ab34e2}
  {\bibfield  {journal} {\bibinfo  {journal} {Classical and Quantum Gravity}\
  }\textbf {\bibinfo {volume} {36}},\ \bibinfo {pages} {195006} (\bibinfo
  {year} {2019})},\ \bibinfo {note} {publisher: IOP Publishing}\BibitemShut
  {NoStop}%
\bibitem [{\citenamefont {McKechan}\ \emph {et~al.}(2010)\citenamefont
  {McKechan}, \citenamefont {Robinson},\ and\ \citenamefont
  {Sathyaprakash}}]{mckechan_tapering_2010}%
  \BibitemOpen
  \bibfield  {author} {\bibinfo {author} {\bibfnamefont {D.~J.~A.}\
  \bibnamefont {McKechan}}, \bibinfo {author} {\bibfnamefont {C.}~\bibnamefont
  {Robinson}}, \ and\ \bibinfo {author} {\bibfnamefont {B.~S.}\ \bibnamefont
  {Sathyaprakash}},\ }\href {\doibase 10.1088/0264-9381/27/8/084020} {\bibfield
   {journal} {\bibinfo  {journal} {Classical and Quantum Gravity}\ }\textbf
  {\bibinfo {volume} {27}},\ \bibinfo {pages} {084020} (\bibinfo {year}
  {2010})}\BibitemShut {NoStop}%
\bibitem [{\citenamefont {Quang~Nam}\ \emph {et~al.}(2023)\citenamefont
  {Quang~Nam}, \citenamefont {Martino}, \citenamefont {Lemière}, \citenamefont
  {Petiteau}, \citenamefont {Bayle}, \citenamefont {Hartwig},\ and\
  \citenamefont {Staab}}]{quang_nam_time-delay_2023}%
  \BibitemOpen
  \bibfield  {author} {\bibinfo {author} {\bibfnamefont {D.}~\bibnamefont
  {Quang~Nam}}, \bibinfo {author} {\bibfnamefont {J.}~\bibnamefont {Martino}},
  \bibinfo {author} {\bibfnamefont {Y.}~\bibnamefont {Lemière}}, \bibinfo
  {author} {\bibfnamefont {A.}~\bibnamefont {Petiteau}}, \bibinfo {author}
  {\bibfnamefont {J.-B.}\ \bibnamefont {Bayle}}, \bibinfo {author}
  {\bibfnamefont {O.}~\bibnamefont {Hartwig}}, \ and\ \bibinfo {author}
  {\bibfnamefont {M.}~\bibnamefont {Staab}},\ }\href {\doibase
  10.1103/PhysRevD.108.082004} {\bibfield  {journal} {\bibinfo  {journal}
  {Physical Review D}\ }\textbf {\bibinfo {volume} {108}},\ \bibinfo {pages}
  {082004} (\bibinfo {year} {2023})},\ \bibinfo {note} {publisher: American
  Physical Society}\BibitemShut {NoStop}%
\bibitem [{\citenamefont {Pitte}\ \emph
  {et~al.}(2023{\natexlab{a}})\citenamefont {Pitte}, \citenamefont {Baghi},
  \citenamefont {Marsat}, \citenamefont {Besançon},\ and\ \citenamefont
  {Petiteau}}]{pitte_detectability_2023}%
  \BibitemOpen
  \bibfield  {author} {\bibinfo {author} {\bibfnamefont {C.}~\bibnamefont
  {Pitte}}, \bibinfo {author} {\bibfnamefont {Q.}~\bibnamefont {Baghi}},
  \bibinfo {author} {\bibfnamefont {S.}~\bibnamefont {Marsat}}, \bibinfo
  {author} {\bibfnamefont {M.}~\bibnamefont {Besançon}}, \ and\ \bibinfo
  {author} {\bibfnamefont {A.}~\bibnamefont {Petiteau}},\ }\href {\doibase
  10.1103/PhysRevD.108.044053} {\bibfield  {journal} {\bibinfo  {journal}
  {Physical Review D}\ }\textbf {\bibinfo {volume} {108}},\ \bibinfo {pages}
  {044053} (\bibinfo {year} {2023}{\natexlab{a}})},\ \bibinfo {note}
  {publisher: American Physical Society}\BibitemShut {NoStop}%
\bibitem [{\citenamefont {Babak}\ \emph
  {et~al.}(2021{\natexlab{b}})\citenamefont {Babak}, \citenamefont {Petiteau},\
  and\ \citenamefont {Hewitson}}]{Babak:2021mhe}%
  \BibitemOpen
  \bibfield  {author} {\bibinfo {author} {\bibfnamefont {S.}~\bibnamefont
  {Babak}}, \bibinfo {author} {\bibfnamefont {A.}~\bibnamefont {Petiteau}}, \
  and\ \bibinfo {author} {\bibfnamefont {M.}~\bibnamefont {Hewitson}},\
  }\href@noop {} {\  (\bibinfo {year} {2021}{\natexlab{b}})}\BibitemShut
  {NoStop}%
\bibitem [{\citenamefont {Buonanno}\ \emph {et~al.}(2007)\citenamefont
  {Buonanno}, \citenamefont {Cook},\ and\ \citenamefont
  {Pretorius}}]{Buonanno:2006ui}%
  \BibitemOpen
  \bibfield  {author} {\bibinfo {author} {\bibfnamefont {A.}~\bibnamefont
  {Buonanno}}, \bibinfo {author} {\bibfnamefont {G.~B.}\ \bibnamefont {Cook}},
  \ and\ \bibinfo {author} {\bibfnamefont {F.}~\bibnamefont {Pretorius}},\
  }\href {\doibase 10.1103/PhysRevD.75.124018} {\bibfield  {journal} {\bibinfo
  {journal} {Phys. Rev. D}\ }\textbf {\bibinfo {volume} {75}},\ \bibinfo
  {pages} {124018} (\bibinfo {year} {2007})}\BibitemShut {NoStop}%
\bibitem [{\citenamefont {Chen}\ \emph {et~al.}(2024)\citenamefont {Chen} \emph
  {et~al.}}]{Chen:2024ieh}%
  \BibitemOpen
  \bibfield  {author} {\bibinfo {author} {\bibfnamefont {Y.}~\bibnamefont
  {Chen}} \emph {et~al.},\ }\href@noop {} {\  (\bibinfo {year}
  {2024})}\BibitemShut {NoStop}%
\bibitem [{\citenamefont {Zonca}\ \emph {et~al.}(2019)\citenamefont {Zonca},
  \citenamefont {Singer}, \citenamefont {Lenz}, \citenamefont {Reinecke},
  \citenamefont {Rosset}, \citenamefont {Hivon},\ and\ \citenamefont
  {Gorski}}]{healpy}%
  \BibitemOpen
  \bibfield  {author} {\bibinfo {author} {\bibfnamefont {A.}~\bibnamefont
  {Zonca}}, \bibinfo {author} {\bibfnamefont {L.~P.}\ \bibnamefont {Singer}},
  \bibinfo {author} {\bibfnamefont {D.}~\bibnamefont {Lenz}}, \bibinfo {author}
  {\bibfnamefont {M.}~\bibnamefont {Reinecke}}, \bibinfo {author}
  {\bibfnamefont {C.}~\bibnamefont {Rosset}}, \bibinfo {author} {\bibfnamefont
  {E.}~\bibnamefont {Hivon}}, \ and\ \bibinfo {author} {\bibfnamefont {K.~M.}\
  \bibnamefont {Gorski}},\ }\href {\doibase 10.21105/joss.01298} {\bibfield
  {journal} {\bibinfo  {journal} {Journal of Open Source Software}\ }\textbf
  {\bibinfo {volume} {4}},\ \bibinfo {pages} {1298} (\bibinfo {year}
  {2019})}\BibitemShut {NoStop}%
\bibitem [{\citenamefont {Robson}\ \emph {et~al.}(2019)\citenamefont {Robson},
  \citenamefont {Cornish},\ and\ \citenamefont {Liu}}]{Robson:2018ifk}%
  \BibitemOpen
  \bibfield  {author} {\bibinfo {author} {\bibfnamefont {T.}~\bibnamefont
  {Robson}}, \bibinfo {author} {\bibfnamefont {N.~J.}\ \bibnamefont {Cornish}},
  \ and\ \bibinfo {author} {\bibfnamefont {C.}~\bibnamefont {Liu}},\ }\href
  {\doibase 10.1088/1361-6382/ab1101} {\bibfield  {journal} {\bibinfo
  {journal} {Class. Quant. Grav.}\ }\textbf {\bibinfo {volume} {36}},\ \bibinfo
  {pages} {105011} (\bibinfo {year} {2019})},\ \Eprint
  {http://arxiv.org/abs/1803.01944} {arXiv:1803.01944 [astro-ph.HE]}
  \BibitemShut {NoStop}%
\bibitem [{\citenamefont {Górski}\ \emph {et~al.}(2005)\citenamefont
  {Górski}, \citenamefont {Hivon}, \citenamefont {Banday}, \citenamefont
  {Wandelt}, \citenamefont {Hansen}, \citenamefont {Reinecke},\ and\
  \citenamefont {Bartelmann}}]{healpix}%
  \BibitemOpen
  \bibfield  {author} {\bibinfo {author} {\bibfnamefont {K.~M.}\ \bibnamefont
  {Górski}}, \bibinfo {author} {\bibfnamefont {E.}~\bibnamefont {Hivon}},
  \bibinfo {author} {\bibfnamefont {A.~J.}\ \bibnamefont {Banday}}, \bibinfo
  {author} {\bibfnamefont {B.~D.}\ \bibnamefont {Wandelt}}, \bibinfo {author}
  {\bibfnamefont {F.~K.}\ \bibnamefont {Hansen}}, \bibinfo {author}
  {\bibfnamefont {M.}~\bibnamefont {Reinecke}}, \ and\ \bibinfo {author}
  {\bibfnamefont {M.}~\bibnamefont {Bartelmann}},\ }\href {\doibase
  10.1086/427976} {\bibfield  {journal} {\bibinfo  {journal} {The Astrophysical
  Journal}\ }\textbf {\bibinfo {volume} {622}},\ \bibinfo {pages} {759}
  (\bibinfo {year} {2005})},\ \bibinfo {note} {publisher: IOP
  Publishing}\BibitemShut {NoStop}%
\bibitem [{\citenamefont {Pitte}\ \emph
  {et~al.}(2023{\natexlab{b}})\citenamefont {Pitte}, \citenamefont {Baghi},
  \citenamefont {Marsat}, \citenamefont {Besan\c{c}on},\ and\ \citenamefont
  {Petiteau}}]{Pitte:2023ltw}%
  \BibitemOpen
  \bibfield  {author} {\bibinfo {author} {\bibfnamefont {C.}~\bibnamefont
  {Pitte}}, \bibinfo {author} {\bibfnamefont {Q.}~\bibnamefont {Baghi}},
  \bibinfo {author} {\bibfnamefont {S.}~\bibnamefont {Marsat}}, \bibinfo
  {author} {\bibfnamefont {M.}~\bibnamefont {Besan\c{c}on}}, \ and\ \bibinfo
  {author} {\bibfnamefont {A.}~\bibnamefont {Petiteau}},\ }\href {\doibase
  10.1103/PhysRevD.108.044053} {\bibfield  {journal} {\bibinfo  {journal}
  {Phys. Rev. D}\ }\textbf {\bibinfo {volume} {108}},\ \bibinfo {pages}
  {044053} (\bibinfo {year} {2023}{\natexlab{b}})}\BibitemShut {NoStop}%
\bibitem [{\citenamefont {{Liu}}\ \emph {et~al.}(2021)\citenamefont {{Liu}},
  \citenamefont {{He}},\ and\ \citenamefont {{Cao}}}]{2021PhRvD.103d3005L}%
  \BibitemOpen
  \bibfield  {author} {\bibinfo {author} {\bibfnamefont {X.}~\bibnamefont
  {{Liu}}}, \bibinfo {author} {\bibfnamefont {X.}~\bibnamefont {{He}}}, \ and\
  \bibinfo {author} {\bibfnamefont {Z.}~\bibnamefont {{Cao}}},\ }\href
  {\doibase 10.1103/PhysRevD.103.043005} {\bibfield  {journal} {\bibinfo
  {journal} {\prd}\ }\textbf {\bibinfo {volume} {103}},\ \bibinfo {eid}
  {043005} (\bibinfo {year} {2021})}\BibitemShut {NoStop}%
\bibitem [{\citenamefont {Cao}\ and\ \citenamefont {Han}(2016)}]{Cao_2016}%
  \BibitemOpen
  \bibfield  {author} {\bibinfo {author} {\bibfnamefont {Z.}~\bibnamefont
  {Cao}}\ and\ \bibinfo {author} {\bibfnamefont {W.-B.}\ \bibnamefont {Han}},\
  }\href {\doibase 10.1088/0264-9381/33/15/155011} {\bibfield  {journal}
  {\bibinfo  {journal} {Classical and Quantum Gravity}\ }\textbf {\bibinfo
  {volume} {33}},\ \bibinfo {pages} {155011} (\bibinfo {year}
  {2016})}\BibitemShut {NoStop}%
\bibitem [{\citenamefont {Pollney}\ and\ \citenamefont
  {Reisswig}(2011)}]{Pollney:2010hs}%
  \BibitemOpen
  \bibfield  {author} {\bibinfo {author} {\bibfnamefont {D.}~\bibnamefont
  {Pollney}}\ and\ \bibinfo {author} {\bibfnamefont {C.}~\bibnamefont
  {Reisswig}},\ }\href {\doibase 10.1088/2041-8205/732/1/L13} {\bibfield
  {journal} {\bibinfo  {journal} {Astrophys. J. Lett.}\ }\textbf {\bibinfo
  {volume} {732}},\ \bibinfo {pages} {L13} (\bibinfo {year}
  {2011})}\BibitemShut {NoStop}%
\bibitem [{\citenamefont {Barausse}\ \emph {et~al.}(2023)\citenamefont
  {Barausse}, \citenamefont {Dey}, \citenamefont {Crisostomi}, \citenamefont
  {Panayada}, \citenamefont {Marsat},\ and\ \citenamefont
  {Basak}}]{Barausse:2023yrx}%
  \BibitemOpen
  \bibfield  {author} {\bibinfo {author} {\bibfnamefont {E.}~\bibnamefont
  {Barausse}}, \bibinfo {author} {\bibfnamefont {K.}~\bibnamefont {Dey}},
  \bibinfo {author} {\bibfnamefont {M.}~\bibnamefont {Crisostomi}}, \bibinfo
  {author} {\bibfnamefont {A.}~\bibnamefont {Panayada}}, \bibinfo {author}
  {\bibfnamefont {S.}~\bibnamefont {Marsat}}, \ and\ \bibinfo {author}
  {\bibfnamefont {S.}~\bibnamefont {Basak}},\ }\href {\doibase
  10.1103/PhysRevD.108.103034} {\bibfield  {journal} {\bibinfo  {journal}
  {Phys. Rev. D}\ }\textbf {\bibinfo {volume} {108}},\ \bibinfo {pages}
  {103034} (\bibinfo {year} {2023})}\BibitemShut {NoStop}%
\bibitem [{\citenamefont {García-Quirós}\ \emph {et~al.}(2020)\citenamefont
  {García-Quirós}, \citenamefont {Colleoni}, \citenamefont {Husa},
  \citenamefont {Estellés}, \citenamefont {Pratten}, \citenamefont
  {Ramos-Buades}, \citenamefont {Mateu-Lucena},\ and\ \citenamefont
  {Jaume}}]{garcia-quiros_multimode_2020}%
  \BibitemOpen
  \bibfield  {author} {\bibinfo {author} {\bibfnamefont {C.}~\bibnamefont
  {García-Quirós}}, \bibinfo {author} {\bibfnamefont {M.}~\bibnamefont
  {Colleoni}}, \bibinfo {author} {\bibfnamefont {S.}~\bibnamefont {Husa}},
  \bibinfo {author} {\bibfnamefont {H.}~\bibnamefont {Estellés}}, \bibinfo
  {author} {\bibfnamefont {G.}~\bibnamefont {Pratten}}, \bibinfo {author}
  {\bibfnamefont {A.}~\bibnamefont {Ramos-Buades}}, \bibinfo {author}
  {\bibfnamefont {M.}~\bibnamefont {Mateu-Lucena}}, \ and\ \bibinfo {author}
  {\bibfnamefont {R.}~\bibnamefont {Jaume}},\ }\href {\doibase
  10.1103/PhysRevD.102.064002} {\bibfield  {journal} {\bibinfo  {journal}
  {Physical Review D}\ }\textbf {\bibinfo {volume} {102}},\ \bibinfo {pages}
  {064002} (\bibinfo {year} {2020})},\ \bibinfo {note} {publisher: American
  Physical Society}\BibitemShut {NoStop}%
\bibitem [{\citenamefont {Caldarola}\ \emph {et~al.}(2024)\citenamefont
  {Caldarola}, \citenamefont {Kuroyanagi}, \citenamefont {Nesseris},\ and\
  \citenamefont {Garcia-Bellido}}]{Caldarola:2023ipo}%
  \BibitemOpen
  \bibfield  {author} {\bibinfo {author} {\bibfnamefont {M.}~\bibnamefont
  {Caldarola}}, \bibinfo {author} {\bibfnamefont {S.}~\bibnamefont
  {Kuroyanagi}}, \bibinfo {author} {\bibfnamefont {S.}~\bibnamefont
  {Nesseris}}, \ and\ \bibinfo {author} {\bibfnamefont {J.}~\bibnamefont
  {Garcia-Bellido}},\ }\href {\doibase 10.1103/PhysRevD.109.064001} {\bibfield
  {journal} {\bibinfo  {journal} {Phys. Rev. D}\ }\textbf {\bibinfo {volume}
  {109}},\ \bibinfo {pages} {064001} (\bibinfo {year} {2024})}\BibitemShut
  {NoStop}%
\bibitem [{\citenamefont {Gr\"obner}\ \emph {et~al.}(2020)\citenamefont
  {Gr\"obner}, \citenamefont {Ishibashi}, \citenamefont {Tiwari}, \citenamefont
  {Haney},\ and\ \citenamefont {Jetzer}}]{Grobner:2020drr}%
  \BibitemOpen
  \bibfield  {author} {\bibinfo {author} {\bibfnamefont {M.}~\bibnamefont
  {Gr\"obner}}, \bibinfo {author} {\bibfnamefont {W.}~\bibnamefont
  {Ishibashi}}, \bibinfo {author} {\bibfnamefont {S.}~\bibnamefont {Tiwari}},
  \bibinfo {author} {\bibfnamefont {M.}~\bibnamefont {Haney}}, \ and\ \bibinfo
  {author} {\bibfnamefont {P.}~\bibnamefont {Jetzer}},\ }\href {\doibase
  10.1051/0004-6361/202037681} {\bibfield  {journal} {\bibinfo  {journal}
  {Astron. Astrophys.}\ }\textbf {\bibinfo {volume} {638}},\ \bibinfo {pages}
  {A119} (\bibinfo {year} {2020})},\ \Eprint {http://arxiv.org/abs/2005.03571}
  {arXiv:2005.03571 [astro-ph.GA]} \BibitemShut {NoStop}%
\bibitem [{\citenamefont {Dandapat}\ \emph {et~al.}(2023)\citenamefont
  {Dandapat}, \citenamefont {Ebersold}, \citenamefont {Susobhanan},
  \citenamefont {Rana}, \citenamefont {Gopakumar}, \citenamefont {Tiwari},
  \citenamefont {Haney}, \citenamefont {Lee},\ and\ \citenamefont
  {Kolhe}}]{Dandapat:2023zzn}%
  \BibitemOpen
  \bibfield  {author} {\bibinfo {author} {\bibfnamefont {S.}~\bibnamefont
  {Dandapat}}, \bibinfo {author} {\bibfnamefont {M.}~\bibnamefont {Ebersold}},
  \bibinfo {author} {\bibfnamefont {A.}~\bibnamefont {Susobhanan}}, \bibinfo
  {author} {\bibfnamefont {P.}~\bibnamefont {Rana}}, \bibinfo {author}
  {\bibfnamefont {A.}~\bibnamefont {Gopakumar}}, \bibinfo {author}
  {\bibfnamefont {S.}~\bibnamefont {Tiwari}}, \bibinfo {author} {\bibfnamefont
  {M.}~\bibnamefont {Haney}}, \bibinfo {author} {\bibfnamefont {H.~M.}\
  \bibnamefont {Lee}}, \ and\ \bibinfo {author} {\bibfnamefont
  {N.}~\bibnamefont {Kolhe}},\ }\href {\doibase 10.1103/PhysRevD.108.024013}
  {\bibfield  {journal} {\bibinfo  {journal} {Phys. Rev. D}\ }\textbf {\bibinfo
  {volume} {108}},\ \bibinfo {pages} {024013} (\bibinfo {year} {2023})},\
  \Eprint {http://arxiv.org/abs/2305.19318} {arXiv:2305.19318 [gr-qc]}
  \BibitemShut {NoStop}%
\bibitem [{\citenamefont {Khera}\ \emph {et~al.}(2021)\citenamefont {Khera},
  \citenamefont {Krishnan}, \citenamefont {Ashtekar},\ and\ \citenamefont
  {De~Lorenzo}}]{Khera:2020mcz}%
  \BibitemOpen
  \bibfield  {author} {\bibinfo {author} {\bibfnamefont {N.}~\bibnamefont
  {Khera}}, \bibinfo {author} {\bibfnamefont {B.}~\bibnamefont {Krishnan}},
  \bibinfo {author} {\bibfnamefont {A.}~\bibnamefont {Ashtekar}}, \ and\
  \bibinfo {author} {\bibfnamefont {T.}~\bibnamefont {De~Lorenzo}},\ }\href
  {\doibase 10.1103/PhysRevD.103.044012} {\bibfield  {journal} {\bibinfo
  {journal} {Phys. Rev. D}\ }\textbf {\bibinfo {volume} {103}},\ \bibinfo
  {pages} {044012} (\bibinfo {year} {2021})}\BibitemShut {NoStop}%
\bibitem [{\citenamefont {Mitman}\ \emph {et~al.}(2021)\citenamefont {Mitman}
  \emph {et~al.}}]{Mitman:2020bjf}%
  \BibitemOpen
  \bibfield  {author} {\bibinfo {author} {\bibfnamefont {K.}~\bibnamefont
  {Mitman}} \emph {et~al.},\ }\href {\doibase 10.1103/PhysRevD.103.024031}
  {\bibfield  {journal} {\bibinfo  {journal} {Phys. Rev. D}\ }\textbf {\bibinfo
  {volume} {103}},\ \bibinfo {pages} {024031} (\bibinfo {year}
  {2021})}\BibitemShut {NoStop}%
\bibitem [{\citenamefont {Talbot}\ \emph
  {et~al.}(2018{\natexlab{b}})\citenamefont {Talbot}, \citenamefont {Thrane},
  \citenamefont {Lasky},\ and\ \citenamefont
  {Lin}}]{talbot_gravitational-wave_2018}%
  \BibitemOpen
  \bibfield  {author} {\bibinfo {author} {\bibfnamefont {C.}~\bibnamefont
  {Talbot}}, \bibinfo {author} {\bibfnamefont {E.}~\bibnamefont {Thrane}},
  \bibinfo {author} {\bibfnamefont {P.~D.}\ \bibnamefont {Lasky}}, \ and\
  \bibinfo {author} {\bibfnamefont {F.}~\bibnamefont {Lin}},\ }\href {\doibase
  10.1103/PhysRevD.98.064031} {\bibfield  {journal} {\bibinfo  {journal}
  {Physical Review D}\ }\textbf {\bibinfo {volume} {98}},\ \bibinfo {pages}
  {064031} (\bibinfo {year} {2018}{\natexlab{b}})},\ \bibinfo {note}
  {publisher: American Physical Society}\BibitemShut {NoStop}%
\end{thebibliography}%

\end{document}